\documentclass[aps,prd,a4paper,amsfonts,amssymb,nofootinbib,onecolumn,longbibliography,preprintnumbers]{revtex4-2}
\usepackage[a4paper,
            bindingoffset=0.2in,
            left=1.5cm,
            right=1.5cm,
            top=1in,
            bottom=1.5cm,
            footskip=.25in]{geometry}

\usepackage{amsmath}
\usepackage{amsthm}
\usepackage{amssymb}
\usepackage{amsfonts}
\usepackage{crimson}
\usepackage[utf8]{inputenc}
\usepackage{hyperref}
\hypersetup{
    colorlinks,
    citecolor=blue,
    filecolor=blue,
    linkcolor=blue,
    urlcolor=blue
}

\usepackage{ulem}
\usepackage{epsfig}
\usepackage{braket}
\usepackage{graphicx}
\usepackage{paralist}
\usepackage{slashed}
\usepackage{color}
\usepackage{tikz} 
\usetikzlibrary{arrows} 
\usetikzlibrary{shapes,arrows,positioning,automata,backgrounds,calc,er,patterns}
\usepackage{tikz-feynman}

\newcommand{\bp}{\boldsymbol{p}}

\begin{document}

\preprint{\vbox{\hbox{IPARCOS-UCM-23-138}}}
\title{Spinning Pairs: \\
Supporting $^3P_0$ Quark-Pair Creation from Landau Gauge Green's Functions}
\author{Reinhard Alkofer}
\affiliation{Institute of Physics, University of Graz, NAWI Graz, Universit\"atsplatz 5, 8010 Graz, Austria}

\author{Felipe J. Llanes-Estrada}
\affiliation{Dept. F\'{i}sica Te\'orica and IPARCOS, Univ. Complutense de Madrid, Plaza de las Ciencias 1, 28040 Madrid, Spain}
\author{Alexandre Salas-Bernárdez}
\affiliation{Instituto de Física Corpuscular (IFIC), Universidad de Valencia-CSIC,
E-46980 Valencia, Spain}
\affiliation{Dept. F\'{i}sica Te\'orica and IPARCOS, Univ. Complutense de Madrid, Plaza de las Ciencias 1, 28040 Madrid, Spain}

\date{\today}


\begin{abstract}
Abundant phenomenology suggests that strong decays from relatively low-excitation hadrons into other hadrons proceed by the creation of a light quark-antiquark pair with zero total angular momentum, the so called $^3P_0$ mechanism originating  from a scalar bilinear. 
Yet the Quantum Chromodynamics (QCD) interaction is perturbatively mediated by gluons of spin one, and
QCD presents a chirally symmetric Lagrangian. Such scalar decay term  must be spontaneously generated upon breaking chiral symmetry. 
We attempt to reproduce this  with the help of the quark-gluon vertex in Landau gauge, whose nonperturbative structure has been reasonably elucidated in the last years, and 
insertions of a uniform, constant chromoelectric field. This is akin to Schwinger pair production in Quantum Electrodynamics (QED), 
 and we provide a comparison with its two field-insertions diagram. 
We find that, the symmetry being cylindrical, the adequate quantum numbers to discuss the production are rather $^3\Sigma_0$, $^3\Sigma_1$ and $^3\Pi_0$ as in diatomic molecules, and we indeed find a sizeable contribution of the third decay mechanism, which may give a rationale for the $^3P_0$ phenomenology, as long as the momentum of the produced pair is at or below the scale of the bare or dynamically generated fermion mass.
On the other hand, ultrarelativistic fermions are rather ejected with $^3\Sigma_1$ quantum numbers. In QED,
our results suggest that $^3\Sigma_0$ dominates, whereas the 
constraint of producing a color singlet in QCD  leads to $^3\Pi_0$ dominance at sub-GeV momenta.

\end{abstract}

\maketitle
\tableofcontents

\section{Introduction}

The production of a fermion-antifermion pair in an intense field, producing a dielectric breakdown of the ``vacuum'' dates back to the 1930s~\cite{Sauter:1931zz}, with~\cite{Schwinger:1951nm} providing a recognised calculation of the pair production rate in quantum field theory. For a recent review on ultra-strong field QED we refer to \cite{Fedotov:2022ely}.

While experimental detection of this phenomenon in Electrodynamics has been elusive, {\it cf.} \cite{Fedotov:2022ely} and references therein, the analog phenomenon in Chromodynamics is run of the mill: the production of valence quark-antiquark pairs in hadron decays (excited baryon to a baryon-meson pair, or a meson decaying to two mesons) has been extensively studied in experiment and phenomenology of the strong interactions. 

This asymmetry in the experimental exploration of pair production translates into a like asymmetry of our knowledge of the circumstances surrounding the produced pair. 
In this article we focus on the angular momentum quantum numbers of such pair. The current situation is that not much information is available on spin in the ultra-strong field QED production
(see, however, refs.~\cite{Kohlfurst:2018kxg,Kohlfurst:2022edl} and~\cite{Alvarez-Dominguez:2023ten}) and the Chromodynamics one is treated with phenomenological models, where some level of understanding has been reached, but without a serious theory background. An exception is the computation of~\cite{Copinger:2022gfz}, whose basic assertion is that the angular momentum of the fermion-antifermion pair reflects a preexisting virtual condensate from which they are extracted. This is a picture with simple physical interpretation, and in line with the traditional $^3P_0$ approach, but we would like to also give some orientation on the ratio of different spin components for future non-perturbative computations, which might be based on Poincar\'e covariant phase space formulations (Wigner
formalism, see \textit{f.e.} \cite{Kohlfurst:2022vwf}), functional methods in continuum quantum field theory, lattice calculations,
and/or a combination
thereof.

We here attempt to put a stepping stone between phenomenological and more advanced treatments based on Chromodynamics, and in doing so, we also 
extract a first theoretical glimpse at the behavior of the electron-positron angular momentum production in a uniform field.

Quark-model phenomenology~\cite{LeYaouanc:1988fx,Roberts:1992esl,Close:2005se,Swanson:2006st} would have hadrons split apart in two-body reactions $A\to  B+C$ by extracting a constituent quark-antiquark pair from the underlying color fields according to the so called $^3P_0$ mechanism~\cite{Micu:1968mk,LeYaouanc:1988fx}, that seems to be in reasonable agreement with strong quarkonium decays. This we briefly recall in section~\ref{sec:decays}.

In the notation of~\cite{Segovia:2012cd}, the effective interaction Hamiltonian that generates the $^3P_0$ transition (see section~\ref{sec:transition}) is the simplest
\begin{equation} \label{3P0model}
H_{^3P_0}  = \sqrt{3} g_s \int d^3\boldsymbol{x} \bar{\psi}(\boldsymbol{x}) \psi(\boldsymbol{x})
\end{equation}
where the constant $g_s$ has dimensions of energy. In a non-relativistic reduction of the spinors, the pair creation is controlled~\cite{Ackleh:1996yt} by the dimensionless parameter 
$\gamma = {g_s}/{2m}$ with $m$ being the constituent quark mass.

While the model of Eq.~(\ref{3P0model}) is very popular and yields reasonable agreement with decay data, it raises evident  theoretical questions. Quark-antiquark pairs are created in a color singlet, while QCD's interaction at leading order produces a color octet, and both singlet and octet configurations are needed, for example, in pNRQCD~\cite{Brambilla:1999xf}. 

Another perhaps not expected property of this $H_{^3P_0}$ is that it is chiral-symmetry breaking (it has the form of a fermion mass term), as can easily be seen by taken its commutator with the chiral charge, $Q_5$,
\begin{equation}
\left[Q_5,H_{^3P_0}\right] = \left[\int d^3 \boldsymbol{x} \psi^\dagger(\boldsymbol{x})  \gamma_5 \psi(\boldsymbol{x}),
H_{^3P_0}\right]\neq 0 \ .
\end{equation}
This is in contrast with the perturbative structure of gauge theories, in which, in the massless fermion limit, all propagators and vertices are proportional to $\gamma^\mu$, and $\left[\gamma^\mu,\gamma^5\right]=0$. Thus, at all orders of perturbation theory, chiral symmetry is respected.
Therefore, the $^3P_0$ vertex must arise from nonperturbative phenomena. 
Here we attempt to connect this effective Hamiltonian with the full (non-perturbative) Landau gauge Green's functions,
that are commonly explored with lattice gauge theory and functional approaches, see, {\it e.g.} ~\cite{Williams:2015cvx,Cyrol:2016tym,Ferreira:2023fva,Gao:2021wun} and references therein.

Because it is chiral-symmetry breaking, the effective $H_{^3P_0}$ has to arise at a low scale: therefore, it should not be applicable to decays where the emitted $q\bar{q}$ have large individual momenta respect to the emitting hadron center of mass (very energetic decays of highly excited states). This phenomenon would be a manifestation of the insensitivity to chiral symmetry breaking in the higher spectrum~\cite{Bicudo:2016eeu}. This is indeed the behavior that we find for the pair creation in this work.

Beyond spectroscopy,  the $^3P_0$ effective decay Hamiltonian has found applications, for example, in the hadronization of high-energy jets~\cite{Kerbizi:2021gos,Artru:2022vqf} where $q\bar{q}$ pairs are extracted from the vacuum in the last steps of the process when hadrons are fragmenting and perturbative QCD reasoning might be insufficient.
As another example, it has also been recently deployed to model polarized hyperon production~\footnote{G. Goldstein, S. Liuti and D. Sivers, 
\href{https://indico.jlab.org/event/667/contributions/12351/}
{Communication to the Xth Workshop of the APS topical group on hadron physics}, Minneapolis, April 12th-14th 2023.}.

Studies of the Schwinger mechanism~\cite{Schwinger:1951nm} for pair creation in a constant field continue to this day, whether in electrodynamics  \cite{Fedotov:2022ely} or in chromodynamics~\cite{Cao:2015dya}, but it appears that the studies center on the dielectric breakdown (``vacuum decay probability''), and eventually on phase transitions such as the chiral one. The community is  hardly paying attention to the spin and orbital angular momentum quantum numbers of the electron-positron
{\it viz.} quark-antiquark pair. Similar comments apply to the Breit-Wheeler process in laser fields \cite{Mahlin:2023aui} and with propagating transverse photons~\cite{Roshchupkin:2023hbt,Cabral:2023rgs}.

Among these aspects, we  explore here the transition between the perturbative, chirally-symmetric emission mediated by one gluon, and the chiral-symmetry breaking regime. We will work with the nonperturbative quark-gluon vertex, so that we will remain at the level of an octet emission mechanism $\int d^3x \bar{\psi}T^a \Gamma^\mu \psi$, but use the skeleton expansion~\cite{Dyson:1949ha,Lu:1992eq} to obtain the color singlet with one more insertion.

We have organized the balance of the article as follows.
Section~\ref{sec:decays} reviews some of the standing evidence that suggested to us that this investigation was worth pursuing, as hadron decays in phenomenologically successful quark models 
seem to have required that pair production was controlled by a chiral-symmetry breaking term which is not in the QCD Lagrangian. 

In section~\ref{sec:QED} we employ a perturbative template inserting a uniform field into a fermion line to ascertain the produced spin of the pair as function of the fermion's  back to back momenta.  This is a much simplified computation but already starts shedding light into the more complicated QCD production. 

Before delving in the fermion structure of Chromodynamics, we devote section~\ref{sec:Afield} to discussing the constant chromoelectric field flux tube joining quarks in hadrons and that locally provides the background homogeneous field equivalent to the electric field in QED.

Section~\ref{sec:Greensfunctions} is then devoted to the preliminaries needed to discuss quarks themselves: for a skeleton expansion of the production kernel, we recall the full quark propagator and quark-gluon vertex, parametrized in Euclidean space from work on the Dyson-Schwinger Equations (DSEs), and an extrapolation to Minkowski space. This extrapolation is 
by far not numerically stable, but fortunately the qualitative results reported later in sections~\ref{sec:transition} and~\ref{sec:singlet} about the pair production with that simple skeleton kernel are not dependent on the precise poles of that continuation to Minkowski space.   The spin-triplet scalar production with orbital angular-momentum projection over the field's direction, $\Lambda=1$, dominates for low momentum; 
the projection with $\Lambda=0$ is instead dominant for momenta sufficiently larger than the fermion mass $m$. 

The discussion is wrapped up in section~\ref{sec:conclusions}.

\section{Phenomenological consistency of the $^3P_0$ mechanism in hadron decays}
\label{sec:decays}
In this section we briefly discuss some of the well-known evidence from hadron decays 
suggesting that pair production triggering them contains an important chiral symmetry breaking
component.

\begin{figure}
    \centering
\includegraphics[width=0.4\columnwidth]{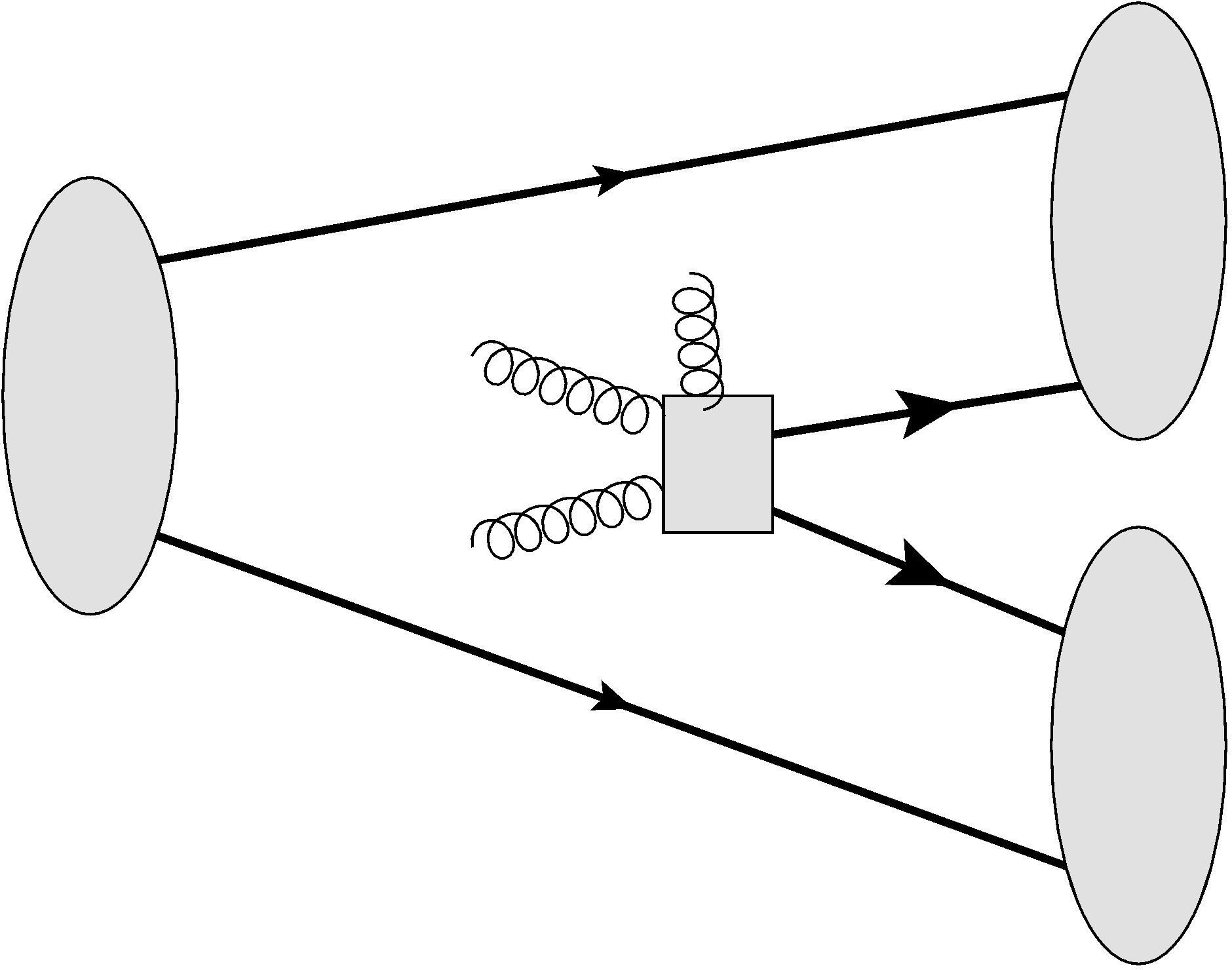}
    \caption{Depiction of an OZI-allowed decay by the creation of a valence $q\bar{q}$ pair via a multilegged transition amplitude.}
    \label{fig:OZI}
\end{figure}

Okubo-Zweig-Iliuka (OZI)-allowed meson decays proceed by the separation of the two valence quarks that provide the quantum numbers of an ordinary meson and the creation of an additional valence $q\bar{q}$ pair that splits
among the two daughter mesons, as reflected in figure~\ref{fig:OZI}.

In the figure, the square box with several gluon legs attached represents the creation of a valence $q\bar{q}$ pair. The angular momentum of the created pair can be decomposed in the usual spectroscopic Russell-Saunders basis $^{2S+1}L_J$. This gives rise to numerous possible combinations, 
{\it e.g.} $^1S_0, ^1P_1, ^3S_1, \boxed{^3P_0}, ^3P_1, ^3P_2\dots$

The most general transition amplitude can be expanded in a basis of operators with all those quantum numbers. However, it has long been known~\cite{Micu:1968mk,LeYaouanc:1972vsx}
that the $^3P_0$ component is compatible with a breadth of decay data, while the other terms, though they may be present, fail to predict sizeable amounts of conspicuous meson decays if employed alone.

Consider as a first example the quark-model spins in the vector to two pseudoscalar mesons decay,
$\rho(\uparrow\uparrow) \to 
\pi(\uparrow\textcolor{blue}{\downarrow}) \pi(\uparrow\textcolor{blue}{\downarrow})     
$. 

The $\rho$-meson  being dominantly in the ground state $^3S_1$ with the two fermion spins aligned, and the two daughter mesons having zero spin each, the newly created pair must be produced with $|m_S|=1$. 
Consequently, the transition amplitude cannot create them with $S=0$. This prominent decay, with $\Gamma\sim 150$ MeV, can therefore not be mediated by either of the spin singlets $^1S_0$, $^1P_1$, etc. 

Likewise, as all three mesons are dominantly $s$-wave and 
thus the relative orbital angular momentum between the two pions is $l=1$, we see that the transition amplitude must carry a $p$-wave, so that it must be of the $^3P_J$ form.

As a second example, consider now the decay of the scalar mesons
$f_0(1370)\to \pi\pi$ and $f_0(1500)\to KK$.
Now, the final state mesons are produced in an $s$-wave. But since the initial-state meson (largely $q\bar{q}$~\cite{Llanes-Estrada:2021evz}) is in a $^3P_0$ configuration, the decay cannot be caused by any transition amplitude with $J\neq 0$.

The only combination of the six quantum-number combinations proposed for the transition amplitude that can appear in both examples is precisely the $^3P_0$.
Continuing studies along these lines show that in different decays, different transition modes may be active; but $^3P_0$ is a very common combination that can mediate an important number of transitions.

The converse line of reasoning, trying to find meson decays that are prominent and would exclude the possibility of a transition matrix element carrying $^3P_0$ quantum numbers, is not easy. That is, there are no good rejection tests of the $^3P_0$ mechanism with light quarks.

A famous selection rule in this direction is
\begin{equation}
A(S=0) \nrightarrow B(S=0) + C(S=0) 
\end{equation}
that tests the ``$^3$'' part of $^3P_0$, {\it i.e.,} whether the produced valence pair is in a spin-1 state. The list of possible mesons at hand that, having $S=0$, can be used to reject that triplet state, is given in Table~\ref{tab:swaves}.

\begin{table}\centering
\caption{List of $S=0$ allowed quantum numbers and example mesons that carry them.\label{tab:swaves}}
\begin{tabular}{|c|c|c|}\hline
$L$   & $J^{P(C)}$ & \\ \hline
0     & $0^{-(+)}$ & $\pi$, $\eta$, $K$\dots \\ \hline
1     & $1^{+(-)}$ & $h_1$, $b_1$, \dots \\ \hline
2     & $2^{-(+)}$ & $\pi_2$, $\eta_2$ \dots \\ \hline
3     & $3^{+(-)}$ & $h_3$, $b_3$, \dots \\
\dots & & \\ \hline
\end{tabular}
\end{table}

The simplest possible decay would be  $0^{-+}\to 0^{-+}\ 0^{-+}$; but this is forbidden by the global $J^{PC}$ quantum numbers, so it cannot possibly test any internal mechanisms; this also applies to $1^{+-}$ decaying to the same final-state mesons.
As a second possibility one could consider
$1^{+-}\to 1^{+-} 0^{-+}$; but we only have at hand the ground state $h_1$, $b_1$ mesons at 1170 and 1235 MeV, respectively, and no excitations thereof; so the would-be left hand side of the decay is unknown. As for open-flavor mesons, the $K_1$'s $^1P_1$ and $^3P_1$ states are mixed with an angle of order of magnitude $\sim 20^{\rm o}$ \cite{Abreu:2019adi} and so the test cannot be conducted because none of the two $K_1$ is a pure $S=0$ meson.
Further, the $2^{-+}\to 1^{+-}\ 1^{+-}$ does not have enough phase-space to proceed, because the mesons with a quark-model $d$-wave, namely the $\pi_2(1670)$ and the slightly more massive $\eta_2$ are not heavy enough to open the phase space for the decay channel. 
In conclusion, no direct rejection tests of the $^3P_0$ mechanism can be proposed with the current knowledge of the light-meson spectrum: there is plenty of room for discovery in the 2 GeV region, at accelerator facilities such as Jefferson Lab.

{Reasonable evidence is available in a different regime:} Close and Swanson found the $^3P_0$ mechanism to be compatible, with only one parameter and within about one standard deviation~\cite{Close:2005se} with a whole 32 modes of heavy-light ($D$ or $D_s$) strong meson decays. We believe that there is enough phenomenological work justifying a more detailed study from the point of view of recent progress in Quantum Chromodynamics.

From our results below it will also become apparent that this
$^3P_0$ model cannot hold for very excited states in the spectrum: when the produced $q\bar{q}$ pair has a momentum much larger than the QCD chiral symmetry breaking mass scale, in addition to $SU(3)$ symmetry being restored in the decay~\cite{Estevez:2020vsm}, the spin quantum numbers of the produced $q\bar{q}$ need to revert to $^3S_1$ (or rather, $^3\Sigma_1$).

\section{Spin of emitted pair in Quantum Electrodynamics
} \label{sec:QED}
\begin{figure}[ht!]
    \centering
    \begin{tikzpicture}[>=stealth,scale=1.2]
  \draw[-latex,thick] (0,.25)--(.75,.7);
  \draw[thick] (0,.25)--(1.25,1);
   \draw [dashed] (-1.25,.25) -- (0,.25);
  \draw[-latex,thick] (1.25,-.5)--(.5,-0.05);
  \draw[thick] (1.25,-.5)--(0,0.25);
    \draw (1,1.3) node {$\bar{u}(p)$};
    \draw (-1,0) node {$\tilde{A}^0$};
    \draw (1.1,-0.9) node {$v(q)$};
\end{tikzpicture}
\hspace{2cm}
\begin{tikzpicture}[>=stealth,scale=1.2]
  \draw [dashed] (-1.25,1) -- (0,1);
  \draw[-latex,thick] (0,1)--(.75,1);
  \draw[thick] (0,1)--(1.25,1);
   \draw [dashed] (-1.25,-.5) -- (0,-.5);
  \draw[-latex,thick] (1.25,-.5)--(.5,-.5);
  \draw[thick] (0,-.5)--(1.25,-.5);
  \draw[-latex,thick] (0,-0.5)--(0,0.35);
    \draw[thick] (0,-.5)--(0,1);
    \draw (-1,1.3) node {$\tilde{A}^0$};
    \draw (1,1.3) node {$\bar{u}(p)$};
    \draw (-1,1.3-1.5) node {$\tilde{A}^0$};
    \draw (1,1.3-1.5) node {$v(q)$};
    \draw[->,very thin]  (0.25,1.3-2)--(1,1.3-2);
     \draw[->,very thin]  (0.25,1.3-.5)--(1,1.3-.5);
     \draw[->,very thin]  (-0.25,1.3-1.5)--(-0.25,1.3-.7);
      \draw (-0.4,1.3-1) node {$t$};
      \draw (.6,1.3-2.2) node {$q$};
      \draw (.6,1.3-.7) node {$p$};
\end{tikzpicture}
    \caption{Minimum $A_0$ field insertions coupling to a fermion pair in Quantum Electrodynamics.} 
    \label{fig:QEDdiagram}
\end{figure}
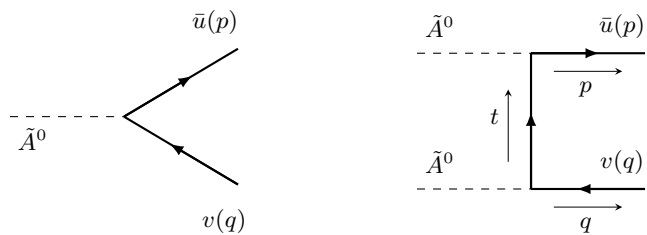
In this section we quickly show, as a perturbative template, the extraction of the fermion-antifermion spin state 
with pure $e\gamma^\mu$ QED vertices and no color.

 Tunnelling in a uniform electric field is traditionally
called the Schwinger effect~\cite{Schwinger:1951nm}. Schwinger's calculation, a prowess of non-perturbative physics, yields the pair production rate summed over spins, and a generalization that can extract the angular momentum composition of the pair requires a quite involved numerical effort.
{\it E.g.,} in ref.\ \cite{Kohlfurst:2022edl} an adaption of the 
Dirac-Heisenberg-Wigner formalism to the Schwinger effect \cite{Hebenstreit:2010vz} has been used. 

Pair production from two photons extracted from a time-dependent transverse field, on the other hand, is a staple of perturbation theory that constitutes the Breit-Wheeler process.
Spin can easily be examined therein, but the computation does not seem  to be easily generalizable to homogeneous, static fields, {\it cf.}, 
\cite{Titov:2020taw,Borysov:2022cwc,Golub:2022cvd} and references therein.

An intermediate-difficulty way of proceeding is to use the Breit-Wheeler perturbative template to investigate the angular momentum quantum numbers, but with the homogeneous electric field insertion instead of the dynamic transverse field. This can then be extended by resummation of Feynman diagrams such as the rainbow approximation for propagators in QCD, and the ladder approximation where each rung is a field insertion, in an attempt at getting closer to true non-perturbative physics. This can of course also later be pursued  with numerical  methods.

We will find, from the lowest order left diagram of figure~\ref{fig:QEDdiagram}, that $^3\Sigma_0$ is dominant (but this is not active in QCD color-singlet production, which requires a fermion skeleton such as the right diagram of the figure).

\subsection{Simplified computation without transversality condition}

The numerator of the two-insertion Feynman production kernel $\mathcal{K}$, reading off the right diagram in figure~\ref{fig:QEDdiagram}, for a conventionally inserted electrostatic field $A^\mu\to A^0$ {\it that does not extend over large distances} (we will lift this restriction shortly in Eq.~(\ref{QEDwithuniformE})), is then 
\begin{equation}\label{QEDzero}
 \bar{u}^s(\bp)\gamma^0(\slashed p + m)\gamma^0{v}^{s'}(-\bp)
 =0
 \ .
\end{equation}
(This is a conventional computation where we made use of the Dirac equation on the well-defined momentum spinor for a fermion of mass $m$, namely  $(\slashed p-m){u}^s(\bp)=0$, and orthogonality among spinors 
.)

One can easily see that two cancelling $^3P_0$ terms  are generated  in QED  (with two field insertions  from the $A_0$ component of the gauge field)  by the electron mass contained in the spinors
\begin{equation}
\bar{u}^s(\bp) \gamma^0 \slashed{p} \gamma^0 {v}^{s'}(-\boldsymbol{p})=2m\bp\cdot \boldsymbol{\sigma}^{s s'} 
\end{equation}
and the second by the explicit mass term from the propagator.
The two contributions exactly cancel each other to yield the zero in Eq.~(\ref{QEDzero}). If successive insertions of the field beyond the two of our kernel are then brought in, each propagator entails a numerator which will contribute a $2E_p$ scalar factor that does not change the spin counting.

\begin{figure}\centering
\includegraphics[width=0.15\textwidth,angle=90]{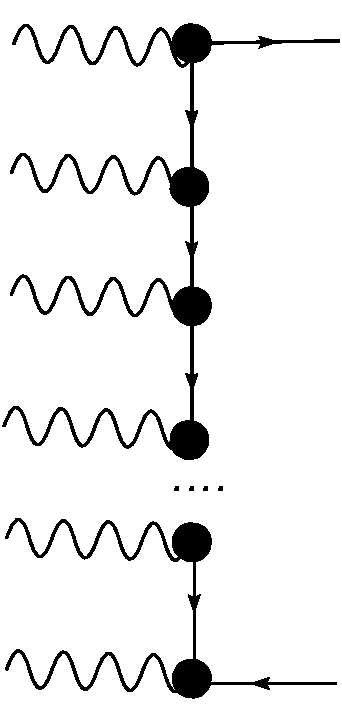}
\caption{\label{Resummation}  A resummation of all ladder diagrams with electric field insertions in QED continues reproducing the zero in Eq.~(\ref{QEDzero})}.
\end{figure}

That zero then persists at every order of perturbation theory, and moreover it also survives a ladder resummation of the type displayed in figure~\ref{Resummation} .
The relevant computation is here
\begin{equation}
\bar{u}^s(\bp) \left( 
\mathbf{1} - \frac{eA^0}{p^2-m^2} \gamma^0(\slashed{p}+m) 
\right)^{-1} {v}^{s'}(-\boldsymbol{p}) = 
\bar{u}^s(\bp) \left(
\mathbf{1} -a \slashed{p}\gamma^0 +a m \gamma^0
\right)b\, {v}^{s'}(-\boldsymbol{p}) 
\end{equation}
with $a$ and $b$ constants; the Dirac $\gamma$ matrix structure between the spinors, upon employing Dirac's equation on the conjugate $u^\dagger {\bf p}\cdot \vec{\gamma} = u^\dagger (m-p^0\gamma^0)$  vanishes once more. 
(This ladder resummation is not exact for indistinguishable electric field insertions; a factor $n!$  affects each term if the insertions can be permuted, and the series is no more geometric, nor summable. It could perhaps be analytically extended via a Borel resummation.)

We now proceed to a constant and uniform electric field that extends over a large distance. When computing the amplitude with an insertion of such  field (that yields the derivative of a delta function in momentum space, see the explanations around Eq.~(\ref{Tmatrixoctet}) and~(\ref{eq:amplitudeoctet}) below) coupled at the vertex, the relevant amplitude's numerator contains
\begin{equation}
 \label{QEDwithuniformE}\mathcal{A}^{ss'}_{\text{QED}}(\bp)\propto 2  \Big[ \frac{\partial}{\partial p^3}\frac{\partial}{\partial q^3} \Big(\bar{u}^s(\bp)\gamma^0\frac{\slashed t + m}{t^2-m^2}\gamma^0{v}^{s'}(\boldsymbol{q})\Big)\Big]\Big|_{\boldsymbol{t}=-\boldsymbol{q}=\bp}\;.
\end{equation}

The derivatives finally remove the zero in some channels.
Explicitly computing  $\mathcal{A}^{ss'}_{\text{QED}}(\bp)$, we find that its spin trace still vanishes, $\sum_s \mathcal{A}^{ss}_{\text{QED}}(\bp)=0$, so that the spin-singlet components of this amplitude are still zero. 
The spin structure of the amplitude can then have nonzero terms of the form
\begin{equation}
\mathcal{A}= A \sigma_\perp\cdot {\bf p}_\perp + B \sigma_z p_z + C  \sigma_\perp + D\sigma_z
\end{equation}
with $A\dots D$ different functions of mass and energy. This means that spherical symmetry in the produced pair is broken (which is natural, given the constant electric field pointing in the $OZ$ direction, manifest in Eq.~(\ref{QEDwithuniformE}) by the $\partial/\partial p_3$ and $\partial/\partial q_3$ 
derivatives). Thus, the atomic term
$^{2S+1}L_J$ including various total angular momenta is not appropriate, as those are  quantum numbers apt for a spherically symmetric problem, and we should instead think of the molecular term notation,
$^{2S+1}\Lambda_{j_z}$ in terms of the components along the $OZ$ axis, with the spherical symmetry reduced down to cylindrical symmetry.

Because the spin-singlet production vanishes, at least for fermions with opposite momenta,
 among the four angular momentum projections analysed below in the more general nonabelian case (Eq.~\ref{QCDprojections}), only the spin-triplet ones remain,  $^3\Pi$ and $^3\Sigma$,  that come with the relative weights (so that $D$ is actually zero)
\begin{align}  
{\mathcal{A}^{^3\Sigma_1}_{\text{QED}}}(|\bp|)&\propto 
-\frac{\pi}{2 |\bp| E_{\bp}^4}  \left(4 |\bp|^2 E_{\bp}-3 m |\bp|^2+2 m^2(E_{\bp}-m)\right)
\xrightarrow[m\to 0]{} -\frac{2\pi}{|{\bf p}|^2}
\\
{\mathcal{A}^{^3\Sigma_0}_{\text{QED}}}(|\bp|)&=0
\\
\mathcal{A}_{\text{QED}}^{^3\Pi_0}(|\bp|)&\propto
-\frac{8}{3|{\bf p}| E^5_{\bp}}\left(
2|{\bf p}|^4-|{\bf p}|^2m(3E_{\bp}-4m)-2m^3(E_{\bp}-m)\right)
\xrightarrow[m\to 0]{} - \frac{16}{3}\frac{1}{|{\bf p}|^2}
\end{align}
where the chiral limit $m\to 0$ has been taken, showing that the ratio of $^3\Sigma_1$ to $^3\Pi_0$ is of order 1 with the $\Sigma$ production slightly dominant, whereas in the opposite, zero momentum limit, all contributions vanish with the leading order for $^3\Sigma_1$ and $^3\Pi_0$
being a power law, {\it i.e.},
$\propto |{\bf p}|^3/m^5$.

If, in the same approximation, we give the production from the one-field insertion on the left diagram of figure~\ref{fig:QEDdiagram}, we would simply expel the pair with the quantum numbers of the zeroth component of a four-vector,
\begin{equation}
\label{QEDonefield}\mathcal{A}^{ss'}_{\text{one-field QED}}(\bp)\propto \Big[ \frac{\partial}{\partial p^3} \Big(\bar{u}^s(\bp)\gamma^0{v}^{s'}(\boldsymbol{q})\Big)\Big]\Big|_{\boldsymbol{q}=-\bp}\;.
\end{equation}
Producing the following spin projected amplitudes:
\begin{align}  \label{onefieldamplitudes}
{\mathcal{A}^{^3\Sigma_1}_{\text{one-field QED}}}(|\bp|)&=0
\\
{\mathcal{A}^{^3\Sigma_0}_{\text{one-field QED}}}(|\bp|)&=-\frac{4 m}{3 E_{\bp}}-\frac{8}{3}\\
\mathcal{A}_{\text{one-field QED}}^{^3\Pi_0}(|\bp|)&=0\ .
\end{align}

\subsection{Incorporating Landau gauge's transversality condition} \label{subsec:QEDwithtransverse}

The computation presented this far in Eq.~(\ref{QEDwithuniformE}) and following is a simplification, to avoid the added difficulty of the Landau gauge transversality condition; 
We now move on and proceed to the full computation in Landau gauge  (which is the one for which we have most information in the QCD case, and a fixed point of the renormalization group in perturbation theory).
To proceed with the estimate of the spin distribution of \emph{back to back leptons with ${\bf p}=-{\bf q}$}, we perform the simple following replacement in the fermion-gauge boson vertex: $\gamma^\mu\to \gamma^\mu - \hat{k}^\mu\hat{\slashed{k}}$ (with $k^\mu$ being the four-momentum extracted from the field, and a caret denoting normalization). In this way, reading from Fig. \ref{fig:QEDdiagram}, we find the following expression for the two-field-insertion kernel, 
\begin{eqnarray}
 \mathcal{A}^{ss'}_{\text{QED}}(\bp) &\propto & 2\Big[ \frac{\partial}{\partial p^3}\frac{\partial}{\partial q^3} \Big(\bar{u}^s(\bp)\left(\gamma^0-\frac{(p^0-t^0)(\slashed p -\slashed t)}{(p-t)^2}\right)\frac{\slashed t + m}{t^2-m^2}\left(\gamma^0-\frac{(q^0+t^0)(\slashed q +\slashed t)}{(q+t)^2}\right){v}^{s'}(\boldsymbol{q})\Big)\Big]\Big|_{\boldsymbol{t}=-\boldsymbol{q}=\bp}\nonumber \\ 
 &=&2\frac{\partial}{\partial p^3}\frac{\partial}{\partial q^3}\Big[  \bar{u}^s(\bp)\left(\gamma^0-\frac{(\slashed p -\slashed t)}{p^0}\right)\frac{\slashed t + m}{t^2-m^2}\left(\gamma^0-\frac{(\slashed q +\slashed t)}{q^0}\right){v}^{s'}(\boldsymbol{q})\Big]\Big|_{\boldsymbol{t}=-\boldsymbol{q}=\bp} \ .
\end{eqnarray}
To pass on to the second line we have noted that each of the two gauge bosons introduces a respective factor $E_{\bp}$ or $E_{\boldsymbol{q}}$ with the momentum of the outgoing fermion to which they are attached. 
Basically, this amounts to $(p-t)^2=p_0^2=E_{\bp}^2$ and $(q+t)^2=q_0^2=E_{\boldsymbol{q}}^2$. With this choice, the energy transferred through the fermion propagator vanishes, $t^0=0$. At least in this kinematic section this will lead to  $^3\Pi_0$ dominance, as can be seen after once more projecting over angular momentum components,
\begin{align}   
{\mathcal{A}^{^3\Sigma_1}_{\text{QED}}}(|\bp|)&\propto -2\pi  |\bp| 
\left(
\frac{E_{\bp}-m}{E^4_{\bp}}
\right)
\label{3S1QED}
\\
{\mathcal{A}^{^3\Sigma_0}_{\text{QED}}}(|\bp|)&=0
\\
\mathcal{A}_{\text{QED}}^{^3\Pi_0}(|\bp|)&\propto\frac{32 m |\bp|}{3 E^4_{\bp}}\ .
\label{3P0QED}
\end{align}

\begin{figure}[ht!]
    \centering    \includegraphics[width=0.48\columnwidth]{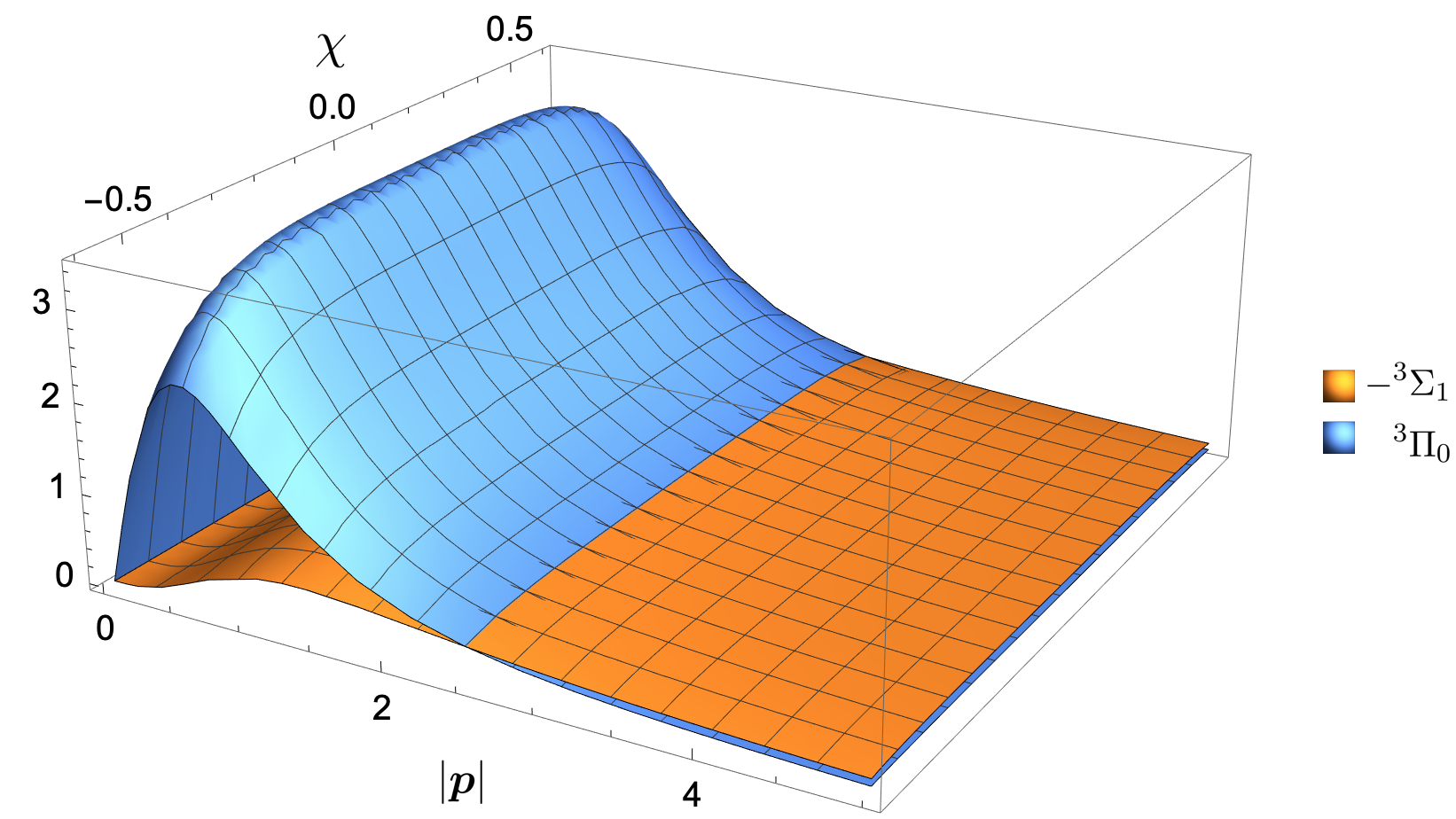}
\includegraphics[width=0.48\columnwidth]{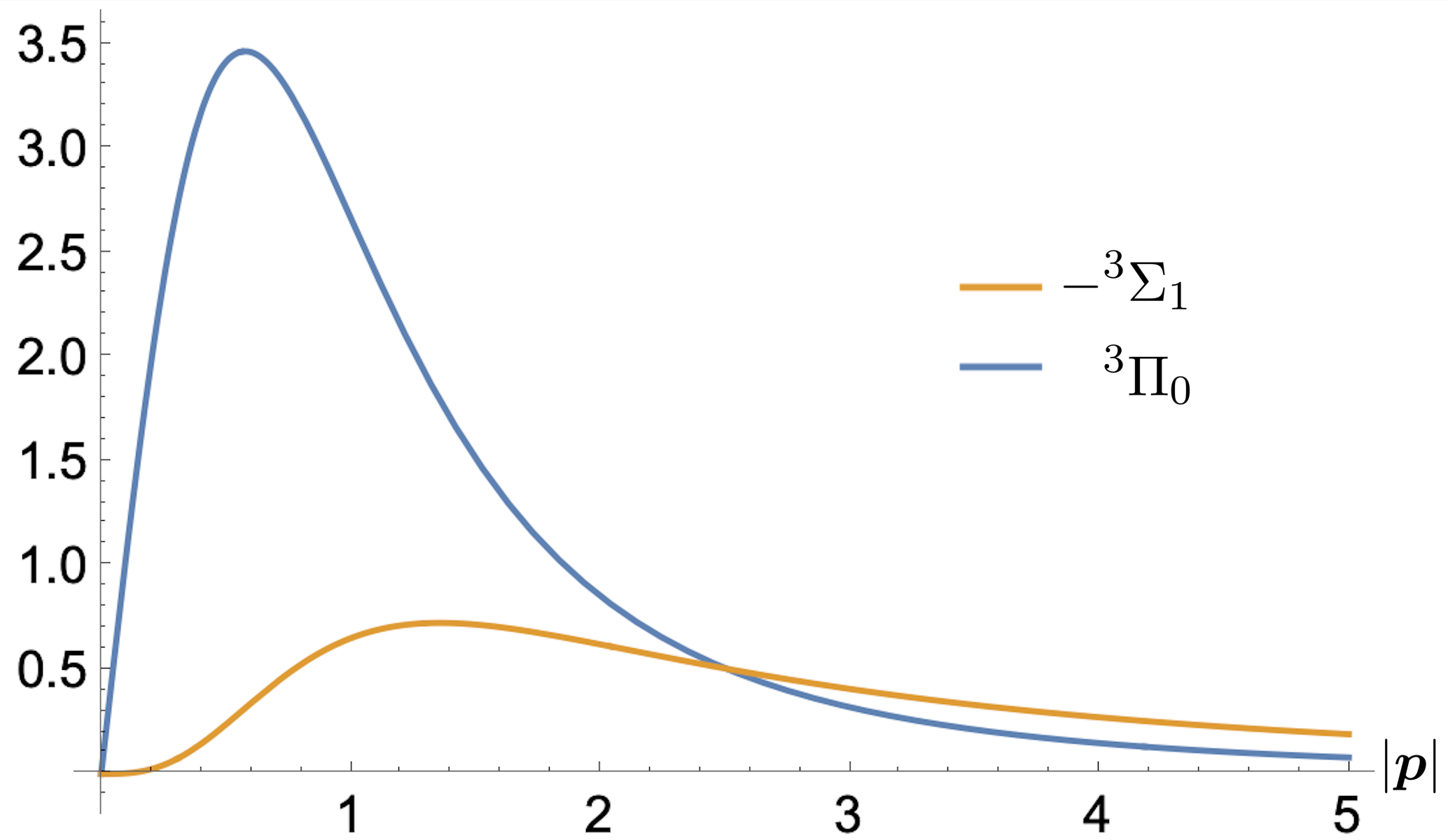}
    \caption{Comparison of the spin-triplet orbital-scalar vs. spin-triplet orbital-vector QED pair production in the uniform electric field. Left: three-dimensional rendering comparing $^3\Pi_0$ and $^3\Sigma_1$ production against the electron-positron momentum and the collinear fraction $\chi$ that the first field insertion provides. Right:  Bidimensional projection with $\chi=0$ as function of the fermion pair momentum. In both cases, the momentum ($x$-axis) is in units of the fermion mass, that is, $m=1$.}
    \label{fig:comparison}
\end{figure}

The limit $|{\bf p}|\to 0$ is not problematic.
In this low-momentum regime the electron mass scale is dominant and we see that the chiral-symmetry breaking piece is leading by two orders in the McLaurin expansion around $|{\bf p}|=0$, 
\begin{equation}
\frac{\mathcal{A}_{\text{QED}}^{^3\Pi_0}}{\mathcal{A}^{^3\Sigma_0}_{\text{QED}}}=O\left(\frac{m^2}{|{\bf p}|^2}\right)\ ,
\end{equation}
that is, while more sophisticated computations should follow this up, we believe that there is a reasonable case that the near-threshold production of an electron-positron pair in an intense homogeneous electric field will see a suppression of $^3\Sigma_1$ versus $^3\Pi_0$ quantum 
numbers.\footnote{This is in agreement with the recently obtained results for the spin-dependent 
amplitudes in multi-photon pair production \cite{Kohlfurst:2022edl}.} The amplitudes (\ref{3S1QED})
and (\ref{3P0QED}), in units of the fermion mass $m$, are depicted in the right plot of Fig. \ref{fig:comparison}.


In it, we can see that the production with one unit of orbital angular momentum is dominant at momenta up to about twice the electron mass, respectively, the mass of the produced pair. For higher energies, this channel falls faster, with the $\Sigma_1$-wave becoming more important; but because the electric breakdown of the Schwinger effect leads to the formation of initially 
slow $e^-e^+$ 
pairs ({\it cf., e.g.,} \cite{Diez:2022ywi} and references therein), we can quite firmly predict that back-to-back leptons will appear with $^3\Pi_0$ spin quantum numbers more often than with $^3\Sigma_1$ ones.  

Next, we allow for a slightly more general kinematic section, in which the two field insertions  contribute differently to the energy balance of the back-to-back lepton pair, that is, $t^0\neq 0$ flows in the fermion propagator of the Feynman diagram in figure~\ref{fig:QEDdiagram}. For this purpose we  introduce an asymmetry parameter $\chi$, such that one of the field insertions provides a fraction $(1-\chi)p^0$ of the electron's energy, and the other field insertion a fraction $(1+\chi)p^0$. We then reobtain the relative weight of the angular momentum components to check for the robustness of the $ 3\Sigma_1\ /\ ^3\Pi_0$ suppression at low momenta, by addressing this more general kinematic section. 

The unprojected amplitude is then
\begin{align}
 \mathcal{A}^{ss'}_{\text{QED}}(\bp)
 \propto&  \Big[ \frac{\partial}{\partial p^3}\frac{\partial}{\partial q^3} \Big(\bar{u}^s(\bp)\left(\gamma^0-\frac{(p^0-t^0)(\slashed p -\slashed t)}{(1-\chi)^2p_0^2}\right)\frac{\slashed t + m}{t^2-m^2}\left(\gamma^0-\frac{(q^0+t^0)(\slashed q +\slashed t)}{(1+\chi)^2q_0^2}\right){v}^{s'}(\boldsymbol{q})\Big)\Big]\Big|_{\substack{\boldsymbol{t}=-\boldsymbol{q}=\bp\\
 t_0=\chi p_0}}\nonumber\\
& + (\chi\to -\chi)
\end{align}
and the corresponding projected amplitudes become
\begin{align}   
{\mathcal{A}^{^3\Sigma_1}_{\text{QED}}}(|\bp|)&\propto \frac{2 \pi  \left(2 \chi^2-1\right) |\bp| \left(E_{\bp}-m\right)}{\left(\chi^2-1\right)^2 E_{\bp}^4}
\\
{\mathcal{A}^{^3\Sigma_0}_{\text{QED}}}(|\bp|)&=0
\\
\mathcal{A}_{\text{QED}}^{^3\Pi_0}(|\bp|)&\propto\frac{32 \left(1-2 \chi^2\right) m |\bp|}{3 \left(\chi^2-1\right)^2 E_{\bp}^4}\ .\end{align}
In consequence, we see once more that among the sub-leading QED amplitudes,  $^3\Pi_0$ dominates over $^3\Sigma_1$ at low-momentum, and remarkably, the resulting ratio among spin components is independent of the $\chi$ asymmetry between the two field insertions,
\begin{equation}
     \frac{
     \mathcal{A}^{^3\Sigma_1}_{\rm QED}(|\bp|)
     }{
     \mathcal{A}_{\rm QED}^{^3\Pi_0}(|\bp|)
     }=\ -3\pi \frac{\left(E_{\bp}-m\right)}{16 m }
     \xrightarrow[|{\bf p}|\to 0]{} \frac{-3\pi}{32} \frac{|{\bf p}|^2}{m^2}
     \; \ \ \ \ \forall \chi\ .
\end{equation}
We should finally note the different sign between the $\Sigma$ and $\Pi$ waves, that might give rise to interference if measurements could be carried out at a high-enough momentum so that both are comparable, for electron momenta of order $2m\simeq 1$ MeV. 

The left panel of figure~\ref{fig:comparison} shows the three-dimensional plot of both nonvanishing amplitudes as a function of the two independent variables, the modulus of the lepton momentum and the asymmetry among the two field insertions $x$.

For completeness, we now note the result with only one field insertion as in the left diagram of figure~\ref{fig:QEDdiagram} but with the transversality condition (Landau gauge) incorporated, that will therefore modify the numerical value of Eq.~(\ref{onefieldamplitudes}) but not the spin structure. This is 
\begin{equation}
 \label{QEDonetransversefield}\mathcal{A}^{ss'}_{\text{one-field QED}}(\bp)\propto \Big[ \frac{\partial}{\partial p^3} \Big(\bar{u}^s(\bp)(\gamma^0-\hat{k}^0\hat{\slashed{k}}){v}^{s'}(\boldsymbol{q})\Big)\Big]\Big|_{\boldsymbol{q}=-\bp}\;.
\end{equation}
with $k^2=(2E_{\bp})^2$,
\begin{align}  
{\mathcal{A}^{^3\Sigma_1}_{\text{one-field QED}}}(|\bp|)&=0
\\
{\mathcal{A}^{^3\Sigma_0}_{\text{one-field QED}}}(|\bp|)&=\frac{4 \left(2 m E_{\bp}+m^2+3 |\bp|^2\right)}{3 E_{\bp}}\\
\mathcal{A}_{\text{one-field QED}}^{^3\Pi_0}(|\bp|)&=0\ .
\end{align}

Thus, we see that the emission with one field insertion corresponds to a wave $^3\Sigma_0$ with cylindrical symmetry around the axis given by the electric field, here chosen according the usual convention in the $z$-direction. This is in contrast with the two-field insertion case, where the $^3\Pi_0$ wave dominates at low momenta. The interplay between both in the full non-perturbative Schwinger effect is difficult to ascertain from our skeleton calculations; but in QCD below, because the one-insertion emission is not possible for a color-singlet fermion pair, and two insertions are needed, we will be able to stake a tentative claim that the $^3\Pi_0$ contribution is possibly dominant for color-singlet $q\overline{q}$ pairs.

\subsection{Corrections to the QED vertex}
\label{subsec:1loopQED}
 
 In QED, the electron-photon vertex at one-loop takes a triangle-diagram correction, yielding 
 ({\it viz.} sect.\ 6.33 of \cite{Peskin:1995ev} or sect.\ 7-1-3 of \cite{Itzykson:1980rh}):
 \begin{equation} \label{QEDvertex}
\gamma^\mu \to \Gamma^\mu = \gamma^\mu F_1(k^2) +\frac{i}{2m} \sigma^{\mu\nu}k_\nu F_2(k^2)\ .
 \end{equation}
The second term, {\it i.e.}, the Pauli term, carries a manifest spin-orbit coupling that may couple the field 
to  $e^-e^+$ pairs directly in the $^3\Pi_0$ state, {\it cf.} the representation given in 
Eq.~(\ref{Lorentzgenerators}) for the generators of the Lorentz boosts $\sigma^{0i}$ 
in terms of the spin Pauli matrices.

{With} respect to the mechanism of figure~\ref{fig:QEDdiagram}, this vertex correction is suppressed by one power of $e$ at the amplitude level (due to the three vertices in the triangle diagram instead of two in the Breit-Wheeler one) so we do not add it to the computation and think of it as a correction, should it not vanish due to other reasons as does the equivalent QCD vertex of Eq.~(\ref{eq:amplitudeoctet}) below whose color factor is zero for a color singlet.
In the case here, the static electric field with back-to-back leptons directly yields zero production (as is expected from such magnetic term).

It is instructive, however, to discuss how the chiral-symmetry breaking nature of this vertex arises. The Pauli form factor can be expressed as an integral over Feynman parameters,
\begin{equation}
    F_2(k^2) = m^2 \frac{\alpha_{\rm QED}}{2\pi}
    \int_0^1 dxdydz \delta(x+y+z-1)
    \frac{2z(1-z)}{m^2(1-z)^2-k^2xy} \, , \quad \alpha_{\rm QED} = \frac{e^2}{4\pi} \, ,
\end{equation}
so that the entire Pauli term of the vertex itself is directly proportional to $m$, upon accounting for the $i/(2m)$ prefactor, and up to logarithmic functions of $m/k$ coming from the integral piece.
Despite the logarithms, it can be easily verified that the Pauli term vanishes in the massless 
limit for all $k^2\not= 0$.

The exception is the limit of zero photon virtuality, where the electron anomalous magnetic moment appears (the leptons are on the mass-shell)
 $F_2(0)= \frac{\alpha_{\rm QED}}{2\pi}$
 and the $m$-dependence has apparently dropped out. This is however a very specific kinematic section relevant for scattering only, as it cannot directly produce a lepton pair, $k^2=s\geq 4m^2$ being far from zero.
 
In Wick-rotated Euclidean space, the Pauli form factor can be rewritten employing an 
implicit transcendental expression \cite{Itzykson:1980rh}
\begin{equation}
F_2(k_E^2) = \frac{\alpha_{\rm QED}}{2\pi} \frac{\Theta}{{\sinh}(\Theta)}  \, , \quad
\frac{k_E^2}{m^2} ={4\ {\sinh}^2\left( \frac{\Theta}{2}\right)} \, ,
\label{F2implicit}
\end{equation}
which, besides making evident that the Pauli form factor is a function of $k_E^2/m^2$ only,
can be used for a straightforward numerical evaluation. 

We can benefit from these well-known formulae because they provide an important check for the phase 
of the quark-gluon vertex form factors below in subsection~\ref{passtoMinkowski}. 
To this end we note that there is an analogue term to the Pauli form factor 
in the quark-gluon vertex, perturbatively and non-perturbatively.  
In our QCD setup, with conventions inherited from earlier Euclidean space Landau gauge computations, 
we will have $-i\gamma^\mu_T$ {\it en lieu} of $\gamma^\mu$. 
(NB: The non-transversality of the usual representation (\ref{QEDvertex}) is 
irrelevant in this context.) 
Then, choosing for convenience the $0$-component of the four-vector $\Gamma^\mu$, {\it i.e.}, $\Gamma^0$, the respective term $i\sigma^{0\nu} (k_\nu/(2m)) F_2$
will be replaced  by $-i(2i \sigma^{0\nu}k_\nu g_3)$, with $g_3$ being 
the corresponding vertex form factor accompanying the vertex structure
$\rho_3^0=\hat{\slashed{k}}(\gamma^0-\hat{k}^0\hat{\slashed{k}})=\hat{k}_i\gamma^i \gamma^0=2i\sigma^{0i}\hat{k}_i$ therein. A comparison between the expression (\ref{F2implicit})
and the numerical results for the form factor $g_3$ (the interested reader my peak ahead to
figure~\ref{fig:vertexparametrization}) allows to fix the correct phase for 
the analytic continuation back to Minkowski space.  
This procedure will be described in detail in the paragraph below Eq.~(\ref{rhotransform}).

 \goodbreak
 
\section{Constant chromoelectric field flux-tube in hadrons} \label{sec:Afield}
We now proceed to our main goal of this study, the spin and orbital angular momentum nature of the pair-production vertex in the strong interactions.
First, it should be noted that constant, homogeneous, infinite chromoelectric fields are not realized in nature. The closest one that comes to this idealization is at the flux tube linking two color sources in a hadron. In this section we discuss this chromoelectric field configuration.

It is known from old that the approximate Regge trajectories on which hadrons sit can be explained by a string-like behavior of mesons. This behavior is heuristically described via the creation of a chromoelectric flux tube between the two constituent quarks (see, {\it e.g.}, \cite{Alkofer:2006fu} and references therein) and is obtainable from Coulomb-gauge variational approaches~\cite{Bowman:2004xd} to QCD. 
Lattice gauge theory has also long indicated~\cite{Bali:1997cp} that this is a physically correct picture, famously exposing a large action density stretching between two static sources.
More sophisticated parametrization of the interquark color potential have been proposed, {\it e.g.} in \cite{Nayak:2012in}.
But to discuss the basic phenomenon, it will suffice to employ the elementary linear one.

Such a field configuration makes the potential rise linearly with the distance between the sources and is the motivation for the well-known flux tube model~\cite{Isgur:1983wj} of hadrons.  Such a linearly rising potential entails a constant, uniform chromoelectric field $\boldsymbol{E}=E\boldsymbol{\hat{z}}$ inside a tubular structure.

\begin{figure}[ht!]
\begin{minipage}{0.45\textwidth}
\centering
\includegraphics[width=65mm]{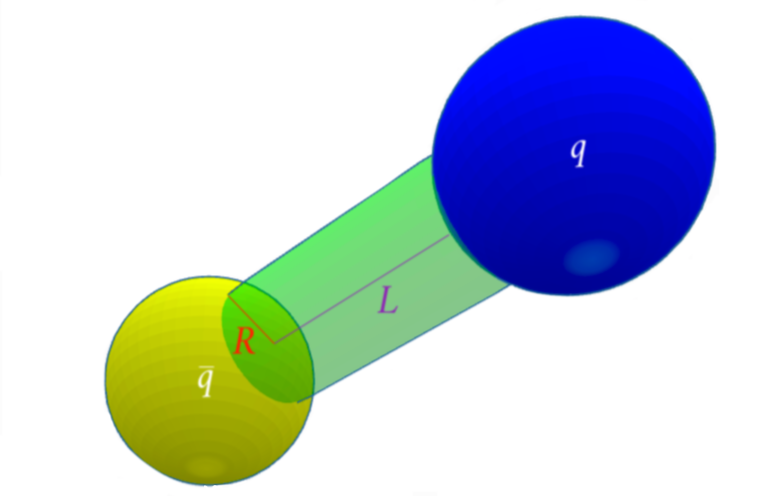}
\end{minipage}
\begin{minipage}{0.53\textwidth}
\caption{Representation of a meson and its chromoelectric flux tube, the blue cylinder (of radius $R$ and length $L$) depicts the region where the electric field can be approximated as constant. We will study the creation of quark pairs in this cylindrical region.} \label{representation}
\end{minipage}
\end{figure}

Since we want to study the creation of dynamical quarks
 inside one of these flux tubes (a topic of interest from old~\cite{Casher:1978wy,Andersson:1983ia}), we will couple the quarks to a constant electric field background (see Fig. \ref{representation}) and study in which spin state they are prone to be created. 

In natural units and cylindrical coordinates, the Landau gauge condition $\partial_\mu A^{\mu\ a} =0 $ for the chromo-electromagnetic potential 
$A^a_\mu(t,\boldsymbol x)$ reads
\begin{equation}
    \partial_t A^a_0(t,\boldsymbol x)=- \frac{1}{\rho}\Big( A^a_\rho(t,\boldsymbol x)+ \partial_\theta A^a_\theta(t,\boldsymbol x)\Big)- \partial_\rho A^a_\rho(t,\boldsymbol x)- \partial_z A^a_z(t,\boldsymbol x)\;.\label{gaugecond}
\end{equation}
Now we set all color components of the the potential, except one, to zero: $A^a(t,\boldsymbol x)=0$ for $a\neq 1$. 
This is not unnatural in a quark-model context where the ends of the tube are charges in the fundamental representation that the tube links, carrying an adjoint index $a$. 
Hence, from now on we omit the color index. Choosing a constant electric field along the $z$ axis of intensity $E$ yields the field equations 
 \begin{align}
  \left\{
                \begin{array}{ll}
                  \partial_\rho A_0(t,\boldsymbol x)=-\partial_t A_\rho(t,\boldsymbol x)\\
                    \partial_\theta A_0(t,\boldsymbol x)=-\rho\partial_t A_\theta(t,\boldsymbol x)\\
                    \partial_z A_0(t,\boldsymbol x)=-\partial_t A_z(t,\boldsymbol x)-E
                \end{array}
              \right.\label{fieldeq}
 \end{align}
Next we assume azimuthal independence and that the components of $A$ perpendicular to the 
meson symmetry axis
(the axis running along the flux-tube, as non-exotic mesons are reasonably well represented by a quark and anti-quark joined by a chromoelectric flux tube) 
are invariant under translations along that axis 
 \begin{align}
  \left\{
                \begin{array}{ll}
                  A_\rho=A_\rho(t,\rho,\theta)\\
                    A_\theta=A_\theta(t,\rho,\theta)\\
                    A_z=A_z(t,z)\;.
                \end{array}
              \right.
 \end{align}
 
Using the gauge condition (\ref{gaugecond}) into the field equations (\ref{fieldeq}) and assuming $A_\mu(t,\boldsymbol x)$ are $C^2$ functions we find 
 \begin{align}
  \left\{
                \begin{array}{ll}
                  \partial_{tt} A_\rho(t,\rho,\theta)=\partial_\rho\Big(\frac{1}{\rho}\big(A_\rho(t,\rho,\theta)+\partial_\theta A_\theta(t,\rho,\theta)\big)+\partial_\rho A_\rho(t,\rho,\theta)\Big) \\
                    \partial_{tt} A_\theta(t,\rho,\theta)=\frac{1}{\rho}\partial_\theta\Big(\frac{1}{\rho}\big(A_\rho(t,\rho,\theta)+\partial_\theta A_\theta(t,\rho,\theta)\big)+\partial_\rho A_\rho(t,\rho,\theta)\Big)\\
                    \partial_{tt}A_z(t,z)=\partial_{zz} A_z(t,z)\label{Laplace}\;.
                \end{array}
              \right.
 \end{align}
 The equations in (\ref{Laplace}) are nothing but the wave equation for each component of the vector field in Cartesian coordinates in the variables $t$ and $x_i$. This will help us to construct the desired solutions.
 
 Let us start by finding the simplest constant-field solution, that can be found easily and will be improved upon shortly.
 We discard the dependence on $\theta$ (cylindrical symmetry) and, for the moment, $\rho$ (so we are basically obtaining the solution equivalent to a parallel-plate capacitor in electrodynamics).
 Choosing a form that provides a representation of the time reversal and parity  discrete symmetries, we obtain the following Abelianized standing waves along the $OZ$ axis,
 \begin{align}
  \left\{
                \begin{array}{ll}
                  A_\rho(t,\rho,\theta)=0\\
                    A_\theta(t,\rho,\theta)=0\\
                    A_z(t,z)=-E\,L\sin \frac{t}{L}\cos{\frac{z}{L}}-{Et}\\
                    A_0(t,z)=E\,L\cos \frac{t}{L}\sin{\frac{z}{L}}
                    \end{array}
              \right.\label{enddd11}
 \end{align}
 where we define $L$ as the length of the cylinder where the electric field can be approximated as constant. Notice that this solution leaves the last expression in Eq. (\ref{fieldeq}) invariant under parity and time reversal. \\
 
 These solutions can however be gauge transformed as usual 
 \begin{equation}
 A_z\to A_z+\partial_z f \;\;\;\;A_0\to A_0-\partial_t f\;,
 \end{equation}
 as long as $f(x)$ satisfies the wave equation $\partial_{tt} f(x)=\partial_{zz} f(x)$, these are the residual gauge transformations after fixing the gauge. This means that we may just write, by taking $f(t,z)=E\,L^2\sin \frac{t}{L}\sin{\frac{z}{L}}+Etz$,
 
 \begin{align}
  \left\{
                \begin{array}{ll}
                  A_\rho(t,\rho,\theta)=0\\
                    A_\theta(t,\rho,\theta)=0\\
                    A_z(t,z)=0\\
                    A_0(t,z)=-Ez
                    \end{array}
              \right.  \label{eq:homogeneousfield}
 \end{align}
there is no surprise here as this is the usual result from electrostatics. This should be sufficient for our purposes. More sophisticated field configurations can be obtained as need arises, as we now exemplify.
 
 For a more realistic flux tube, if so wished, we may look for a potential that 
 contains the  electric field (still along $OZ$)  inside a cylinder of radius $R$, where it is almost $\rho$-independent. A good approximation for this configuration is 
 \begin{equation}
     E_z= \frac{E}{2} \left(1-\tanh \left(\kappa \frac{\rho -R}{R}\right)\right)\;,
 \end{equation}
where we introduced the hyperbolic function to soften the Heaviside step function, $\theta(x)$, since $\lim_{\kappa \to \infty}\tanh (\kappa x)=2\theta(x)-1$ pointwise. \\
Notice now that the following chromo-electromagnetic potential
 \begin{align}
  \left\{
                \begin{array}{ll}
                  A_\rho(t,\rho,\theta)=0\\
                    A_\theta(t,\rho,\theta)=0\\
                    A_z(t,z,\rho)=-{Et}\left(1-\tanh \left(\kappa\frac{\rho -R}{R}\right)\right)/2\\
                    A_0(t,z)=0
                    \end{array}
              \right.\label{enddd3}
 \end{align}
fulfills Landau's gauge condition in Eq. (\ref{gaugecond}) and through the field equations generates the desired $E$ field configuration, contained in a cylindrical region. Also, both ${\bf E}$ and the spatial part ${\bf A}=\sum_{i=1}^3 A_i \hat{e}_i$ change sign under parity, as corresponds to spatial vectors. Because this configuration is 
not curl-free, there is an accompanying stationary chromomagnetic $B$-field, circulating clockwise as 
${\bf B}^a = - \hat{\theta} A_{z,\rho}$ (needed to fulfil $\nabla\times\mathbf{E} + \frac{\partial \mathbf{B}}{\partial t} = 0$). In the limit of large $\kappa$ (sudden extinction of the $E$ field outside the cylinder), this is a surface field. We have not employed this more sophisticated field, but it could be useful in extensions of the work.

\begin{figure}
    \centering
    \includegraphics[width=.5\columnwidth]{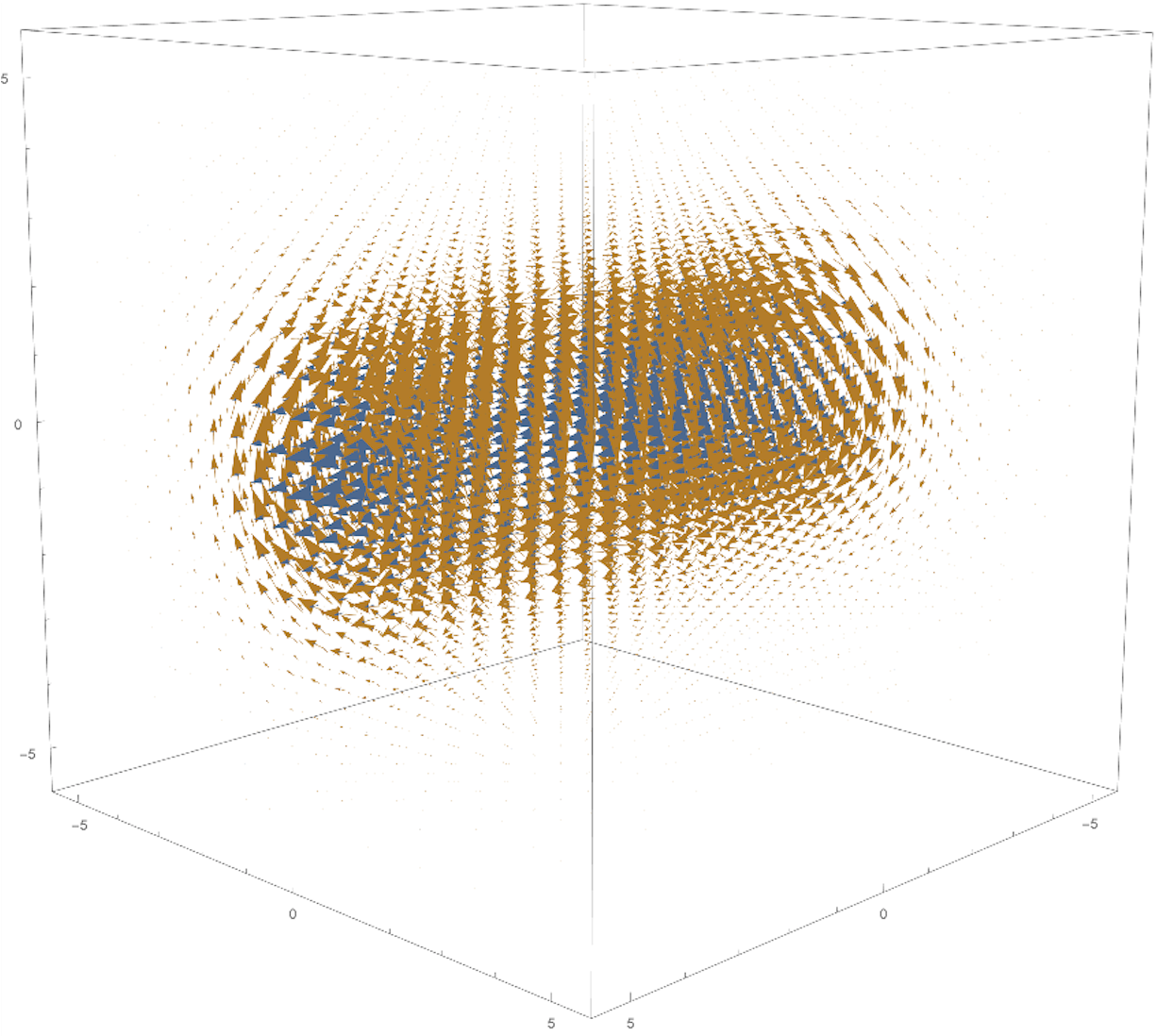}\;\includegraphics[width=.5\columnwidth]{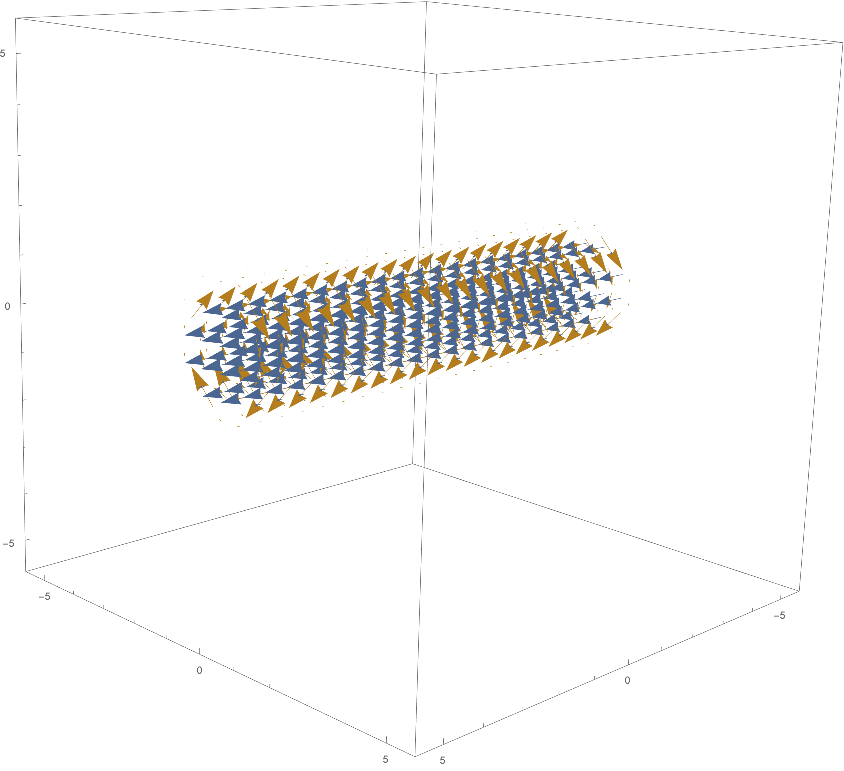}
    \caption{The electric (blue colored) and magnetic field (gold colored) corresponding to the potentials in Eq. (\ref{enddd3}) for $\kappa=1$ (left) and $\kappa=7$ (right). The more intense the fields the bigger the arrowheads of the vectors are.}
    \label{fig:my_label}
\end{figure}

 Much more realistic models for the transverse profile of the chromoelectric field have been studied in lattice gauge field theory. Adopting the values from \cite{Cea:2012qw} we can roughly assume that the flux tube length equals $L=0.76 \text{ fm}$ and has a radius $R=0.46 \text{ fm}$ (since we are looking here for a qualitative explanation of the scalar piece in the decay amplitude, the specific values are not crucial to the analysis). Furthermore we can adopt the rough approximation that the string tension is completely generated by the chromoelectric field, seen as the energy stored into the flux tube per unit length
 \begin{equation}
     \sigma=\frac{1}{2}\int_{\text{flux-tube section}} \boldsymbol{E}^2 dA=\pi R^2 \boldsymbol{E}^2\Longrightarrow E=\sqrt{\frac{2\sigma }{\pi R^2}}\simeq 0.167 \text{ GeV}^2\;,
 \end{equation}
 where we have chosen a typical $\sigma\simeq 0.24 \text{ GeV}^2$ \cite{Buisseret:2004wm}, and where a meson mass is of order $M\sim \sigma L + 2m_q$.

With this field value we can estimate Schwinger's~\cite{Schwinger:1951nm} tunnelling probability per unit time and volume for producing a fermion pair in a constant chromoelectric field, 
\begin{equation}
\label{tunnel}
\frac{dP}{dtdV} = \frac{(gE)^2}{8\pi^2}e^{-\pi m^2/(gE)}\ .
\end{equation}
With $g\sim 3$ so that $\alpha_s\sim 1$ for light quarks,  
this is $(155\ {\rm MeV})^4$.
Multiplying by the volume of the flux tube as described, one gets a tunnelling width
\begin{equation}
\Gamma = \frac{dP}{dt} \sim 0.04 (M-2m_q)\ .
\end{equation}
 For the $\psi(3770)\to D\bar{D}$ decay this predicts about 27 MeV that reasonably compares with the physical width of 32 MeV. 
 For a lighter meson such as $\rho\to \pi\pi$, however, the tunnelling approach of Eq.~(\ref{tunnel}) spectacularly fails, predicting about 7-10 MeV when the physical width is rather of order 150 MeV. 
Since we will limit ourselves to study the ratio of the chiral-symmetry breaking and the chiral-symmetry respecting vertices, these constants are useful only to frame the discussion, but will not affect the result of the article (the relative weight of the two structures) though they can be useful to assess future work, perhaps on the lattice, that may be able to address the absolute value of the effective vertex.
 \section{Primitive Green's functions in Landau-gauge lattice QCD and continuum functional approaches}
\label{sec:Greensfunctions} 
\subsection{The quark propagator and the quark-gluon vertex for space-like momenta}
We now proceed to the fermion part of the effective vertex skeleton construction for the strong interactions, in analogy to the QED discussion of section~\ref{sec:QED}.

Two decades of progress in lattice gauge theory and functional approaches (Dyson-Schwinger equations, Exact Renormalization Group Equations and others) have left us a reasonable, sometimes even quantitatively precise, understanding of the primitive Green's functions of Landau gauge Quantum Chromodynamics. 

These primitive Green's functions are the nonperturbative matrix elements of time-ordered field products at the same point that, if computed to lowest nonvanishing order in perturbation theory, would return one of the pieces of the Landau gauge QCD Lagrangian; or differently put, they are the resummation of all radiative and nonperturbative corrections to each of those pieces in the Lagrangian. 
The gluon and ghost propagators~\cite{Oliveira:2016muq,Cucchieri:2008fc,Aguilar:2007nf}, and the pure Yang-Mills theory interaction vertices (ghost-gluon, three gluon and four gluon vertices)~\cite{Aguilar:2013xqa,Eichmann:2014xya,Mintz:2017qri,Dudal:2012zx,Cyrol:2014kca,Huber:2018ned,Pinto-Gomez:2022brg} have all been reported in the literature. 

\begin{figure}[ht!]
     \centering
     \includegraphics[width=0.5\columnwidth]{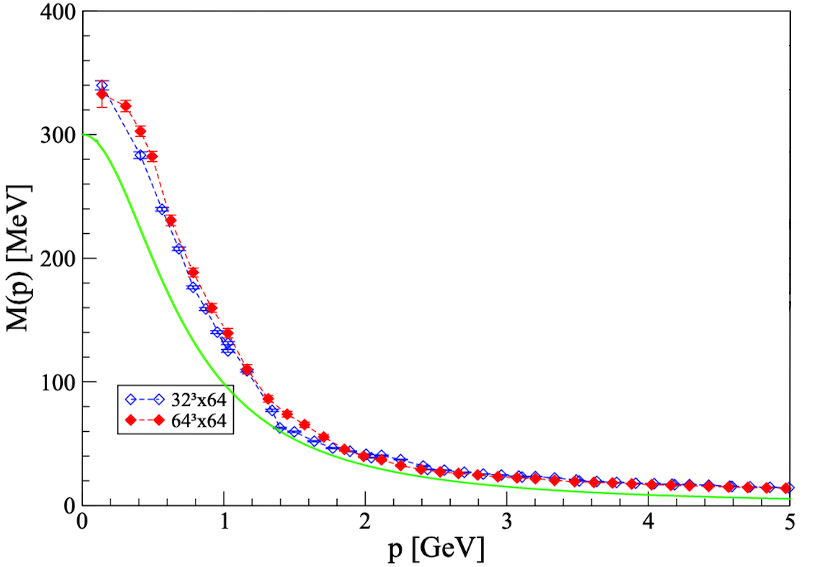}
     \caption{Lattice data for the quark mass function $M(p^2_E)$ from~\cite{Oliveira:2016muq}. 
     The green line is the continuum parametrization of Eq. (\ref{eq:propagatorparametrization}), normalized so that $M(0)=300$ MeV. In this way, it nicely captures the qualitative behavior of the numeric computation. }
     \label{fig:latticevsparametrization}
 \end{figure}

But it is the quark-gluon vertex and the quark propagator that are of highest immediate interest to address the production of quark-antiquark pairs yielding meson decays. 
The lattice data for the quark propagator is now exquisite and we replot part of it, the mass function $M(p^2)$, in Fig.~\ref{fig:latticevsparametrization}.

Chiral-symmetry breaking is manifest as the value of $M$ is commensurate with the QCD scale rather than with the current quark mass; and
while there are not so many points in the infrared region, one can perceive the change of curvature of the function that signals a violation of reflection positivity and thus quark confinement~\cite{Maris:2002mt,Alkofer:2003jj,Leinweber:2022ukj}. 

We then parametrize the full fermion propagator in Minkowski space as
\begin{equation}\label{eq:propagatorparametrization}
    S(p)=\frac{Z_f(p) \,\slashed p+M(p)}{p^2-M^2(p)}\;,\; Z(p)=\frac{p^2-Z_f(0) \Lambda^2_A}{p^2-\Lambda^2_A}\;,\; M(p)=M(0)\frac{\Lambda^2_B}{\Lambda^2_B-p^2}\;,
\end{equation}
with $Z_f(0)=0.80$, $\Lambda_A\simeq 1.7 \text{ GeV}$ and $\Lambda_B= 0.7 \text{ GeV}$.
This captures the overall behavior of the lattice data as well as the many results obtained by functional methods in a simple formula that allows to make qualitative statements about quark lines. 
 
Since the goal of this article is to address the creation of a $q\bar{q}$ pair inside the toy chromoelectric flux tube given by the potential in Eq.~(\ref{enddd3}), we move on to the quark-gluon vertex. Here there is also a trove of lattice work~\cite{Oliveira:2016muq,Kizilersu:2021jen,Kizilersu:2006et} that informs the DSE computations~\cite{Alkofer:2008tt,Aguilar:2023mam,Windisch:2012de,Lessa:2022wqc}. The latter are useful among other things because they allow a parametrization of the entire vertex, as necessary, instead of certain kinematic sections, and because they allow consistency tests among the different Green's functions reported in different lattice computations.

\begin{figure}[ht!]
\centering
\begin{tikzpicture}[>=latex]

\draw[decorate,decoration={coil,segment length=2.4pt}]  (0,0) --(0,-1);
\draw[->]  (0+.3,0-.2) --(0+.3,-1+.2);

\draw[]  (-.8,-1.8) --(0,-1);
\draw[->]  (-.8,-1.8) --(-.35,-1.35);
\draw[->,very thin]  (-.35,-1.35+0.3) -- (-.8,-1.8+0.3);
\draw[->,very thin]  (.35,-1.35+0.3) -- (.8,-1.8+0.3);
\draw[->]  (0,-1) --(.5,-1.5);
\draw[]  (.8,-1.8) --(0,-1);
\draw (0+.5,-.5) node {$k$};
\draw (.7,-1.1) node {$p$};
\draw (-.7,-1.1) node {$q$};

\shade[ball color=blue]  (0,-1) circle (.15cm);
\end{tikzpicture}

\caption{Quark-gluon vertex and our choice of momentum flow.} \label{vertex}
\end{figure}
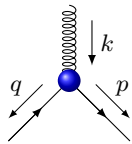

We will address the production of light quark flavors alone,  \textit{up}, \textit{down} and perhaps \textit{strange}, as the heavier ones are closer in scale to pQCD and spontaneous chiral symmetry breaking should affect them less~\cite{Llanes-Estrada:2004edu}. 
This will allow to approximate the value of the $q\bar{q} A$ vertex, $\Gamma(m_1^2,m_2^2)$ for a quark pair with each having very small masses $m_1$ and $m_2$ respectively, with an analytic continuation of the lattice computation in Euclidean space $\Gamma_E$. We will use an original parameterization, with the momentum flow depicted in Fig. \ref{vertex}. 

 With that choice of  flow in Fig. \ref{vertex} we find the transverse part of the vertex, after neglecting the angular dependence, as
 \begin{equation}
     \Gamma_{T}^\mu(q_E,p_E;k_E)=\sum_{i=1}^8 g_i(\bar{p}^2_E)\rho_i^\mu(q_E,p_E)\label{transversevertex}\;,
 \end{equation}
 where $\bar p_E$ is the averaged momentum and ${\rho}_i^\mu(q_E,p_E)$ are all possible Lorentz-invariant vertex components.

 Numerical evidence shows that there are five (or, using equivalences between form factors, three) non-negligible contributions to the vertex
 \cite{Hopfer:2012cnq,Windisch:2014lce,Gao:2021wun}. 
 In Euclidean space, represented as  their respective Padé approximants
 for their model functions (with $x=\bar{p}^2_E/\text{GeV}^2$) \cite{Alkofer:2023lrl}, these are:
 \begin{itemize}
\item The tree-level vertex 
 \begin{equation} \label{vertexpar1}
\rho_{1,E}^\mu=(\delta^{\mu\nu}-\hat{k}^\mu_E\hat{k}^\nu_E)\gamma^\mu_E\equiv \gamma_{T,E}^\mu \;\;\text{ with }\;\; g_1(x)=1+\frac{1.6673+0.2042x}{1+0.6831 x+0.0008509x^2}\;;
 \end{equation}
\item The dynamical chiral-symmetry breaking structures
  \begin{equation}
  \label{vertexpar2}
\rho_{2,E}^\mu=i\hat{s}^\mu_E\;\;\text{ and }\;\; \rho_{3,E}^\mu=i\hat{\slashed{k}}_E\gamma_{T,E}^\mu\;\;\;\text{ with }\;\;g_3(x)=-1.45 g_2(x)=\frac{0.3645 x}{0.01867+0.3530 x+x^2}\;;
 \end{equation}
\item The chirally symmetric structures
  \begin{equation}
\label{vertexpar3}\rho_{4,E}^\mu=\hat{\slashed{k}}_Es^\mu_E\;\;\text{ and }\;\; \rho_{7,E}^\mu=\hat{\slashed{s}}_E\hat{\slashed{k}}_E\gamma_{T,E}^\mu \;\;\text{ with }\;\;g_4(x)=g_7(x)=\frac{2.589x}{0.8587+3.267 x+x^2}\;.
 \end{equation}
\end{itemize}
 Where the caret over the vectors means they are normalized ($\hat{k}_E=k^\mu_E/\sqrt{k^2_E}$), $s^\mu_E =(\delta^{\mu\nu}-\hat{k}^\mu_E\hat{k}^\nu_E)\bar{p}^\nu_E$ and $k^\mu=p^\mu+q^\mu$. A different vertex parametrization can be found in, \textit{e.g.}, \cite{Albino:2018ncl}.

 \subsection{Transformation to Minkowski space} 
\label{passtoMinkowski} 

 To be able to employ this knowledge of the Green's functions in Euclidean space back in the physical  Minkowski space we need the following Wick-rotation rules that define the Euclidean quantities,
 \begin{equation}
   \gamma_E^0\equiv\gamma^0  \;,\;\;\gamma_E^i\equiv+i\gamma^i=-i\gamma_i\;,\;\; \;k_E^i\equiv k_i=-k^i\;,\;\;\;k_E^0\equiv ik^0 \;.
 \end{equation}
 We study how the normalized vectors transform by first noticing that $\sqrt{k^2_E}=\sqrt{(k_E^0)^2+\vec{k}_E^2}=\sqrt{-(k^0)^2+\vec{k}^2}=i\sqrt{k^2}$, hence
 \begin{align}
  \hat{k}^0_E=\frac{k_E^0}{\sqrt{(k_E^0)^2+\vec{k}_E^2}}=\frac{ik^0}{i\sqrt{k^2}}=\hat{k}^0\;\;,\;\;
  \hat{k}^i_E=\frac{k_E^i}{\sqrt{(k_E^0)^2+\vec{k}_E^2}}=\frac{-k^i}{i\sqrt{k^2}}=i\hat{k}^i\;.\label{ktrasform}
 \end{align}
We can then transform the combination 
$$\hat{\slashed{k}}_E=\hat{k}^\mu_E \gamma_E^\mu=\hat{k}^0_E\gamma^0+\hat{k}_E^i\gamma_E^i=\hat{k}^0\gamma^0+(i\hat{k}^i)(i\gamma^i)=\hat{k}^0\gamma^0-\hat{k}^i\gamma^i=\hat{\slashed{k}}\;.$$
We are then ready to transform the components of the vertex starting with
\begin{align}
&\rho_{1, E}^0=i\rho_1^0=\gamma^0_E-\hat{k}^0_E\hat{\slashed{k}}_E=\gamma^0-\hat{k}^0\hat{\slashed{k}}\;\Rightarrow\; \rho_{1}^0=-i(\gamma^0-\hat{k}^0\hat{\slashed{k}})\nonumber \\
&\rho_{1, E}^i=-\rho_1^i=\gamma^i_E-\hat{k}^i_E\hat{\slashed{k}}_E=i\gamma^i-i\hat{k}^i\hat{\slashed{k}}\;\Rightarrow\;\nonumber \rho_{1}^i=-i(\gamma^i-\hat{k}^i\hat{\slashed{k}}) \\
\Longrightarrow\; & \boxed{\rho_{1}^\mu=-i(\gamma^\mu-\hat{k}^\mu \hat{\slashed{k}})} \label{rhotransform}\;.
\end{align}
It is possible to show that the transverse part of the averaged momentum $s_E^\mu$ is related to its Minkowski-space analogue $s^\mu =(\eta^{\mu\nu}-\hat{k}^\mu \hat{k}^\nu)\bar{p}_\nu$ as a usual vector,  $s^0_E=is^0$ and $s^i_E=-s^i$. Also, its normalized version, $\hat{s}^\mu$ obeys an equivalent relation to Eq. (\ref{ktrasform}).

Gathering all this information (and computing the transformation separately for the time and spatial parts of each vertex term analogously to Eq. (\ref{rhotransform})), the dominant contributions to the transverse part of the $qqg$ vertex in Minkowski space come from the following five terms, 
\begin{equation} \label{vertexdefsMinkowski}
\begin{tabular}{lllll}
$\rho_1^\mu =-i(\eta^{\mu\nu}-\hat{k}^\mu\hat{k}^\nu)\gamma_\nu\nonumber$& $\phantom{nothing}$ & 
$\rho_2^\mu =\hat{s}^\mu$& $\phantom{nothing}$ &  \\
$\rho_3^\mu =i\hat{\slashed{k}}\rho_1^\mu$& $\phantom{nothing}$ & $\rho_4^\mu =-i\hat{\slashed{k}}\hat{s}^\mu$& $\phantom{nothing}$& $\rho_7^\mu=\hat{\slashed{s}}\hat{\slashed{k}}\rho_1^\mu$
\end{tabular}\phantom{A}\;.
\end{equation}
In this discussion we have adopted widely used conventions of Euclidean space field theory and directly continued them into Minkowski space at the physically relevant point. The reason is to be able to use the computed parametrizations of the Green's functions.
However, there is a mismatch in conventions that needs to be fixed before proceeding. The full quark-gluon vertex in Minkowski space needs to be real, since it is part of an effective Hamiltonian, by definition Hermitian. This Hermiticity is not directly obtained, and some of the $g_i$ form factors need to be rephased from their conventional Euclidean space usage. But since we do not have a Minkowski space calculation at hand, it is not direct to discern what the correct convention is, respectively, how the correct analytic continuation has to be performed.
We have found expedient to fix this phase by matching to perturbation theory, which can be read off
subsection~\ref{subsec:1loopQED} (the color factor in QCD is common to all vertex structures, so it does not affect the rephasing, and a comparison with the known QED Pauli form factor is possible).  This fixes the phase of $g_3$ and internal consistency then shows what the others are. 
To make every structure real in Minkowski space we then need to rephase $g_3\to -ig_3$, $g_4\to -ig_4$  and $g_7\to -ig_7$ (which are the form factors of the operators carrying a $\hat{\slashed k}$ factor).

~

\paragraph*{\textbf{Analytic structure for time-like momenta}}$\;$\\

The return from Euclidean to physical momentum space is not difficult while using explicit expressions in terms of rational functions, Eq.~(\ref{vertexpar1}), (\ref{vertexpar2}) and~(\ref{vertexpar3}). The concept is understood with the simplest example of such rational functions, 
\begin{equation}
F(P_E^2) \equiv \frac{1}{P_E^2+C^2} = \frac{1}{(p^0_E)^2+E({\bf p})^2+C^2} \label{toyfunction}
\end{equation}
which we imagine known on an Euclidean four-momentum argument as shown.
The analytical back-continuation $p^0_E \to p^0$ is not obstructed because of causality, which means that poles in the $p^0$ plane can be chosen in the first and third quadrants, using the Feynman prescription. The second and fourth quadrants are then free to perform the Wick rotation. (The simple rational functions used here have no cuts to consider.) 

\begin{figure}
    \centering
    \includegraphics[width=0.49\textwidth]{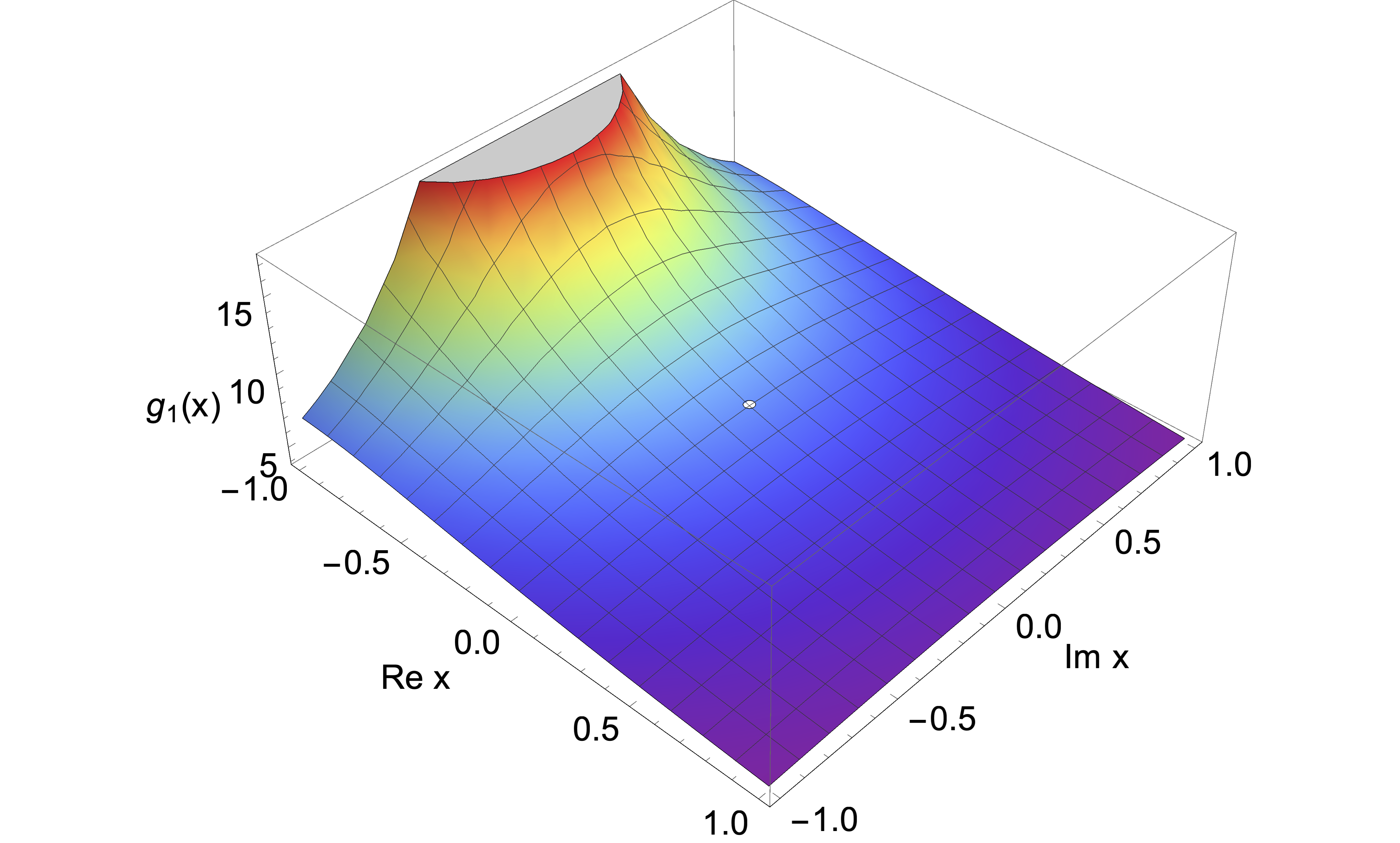} 
    \includegraphics[width=0.49\textwidth]{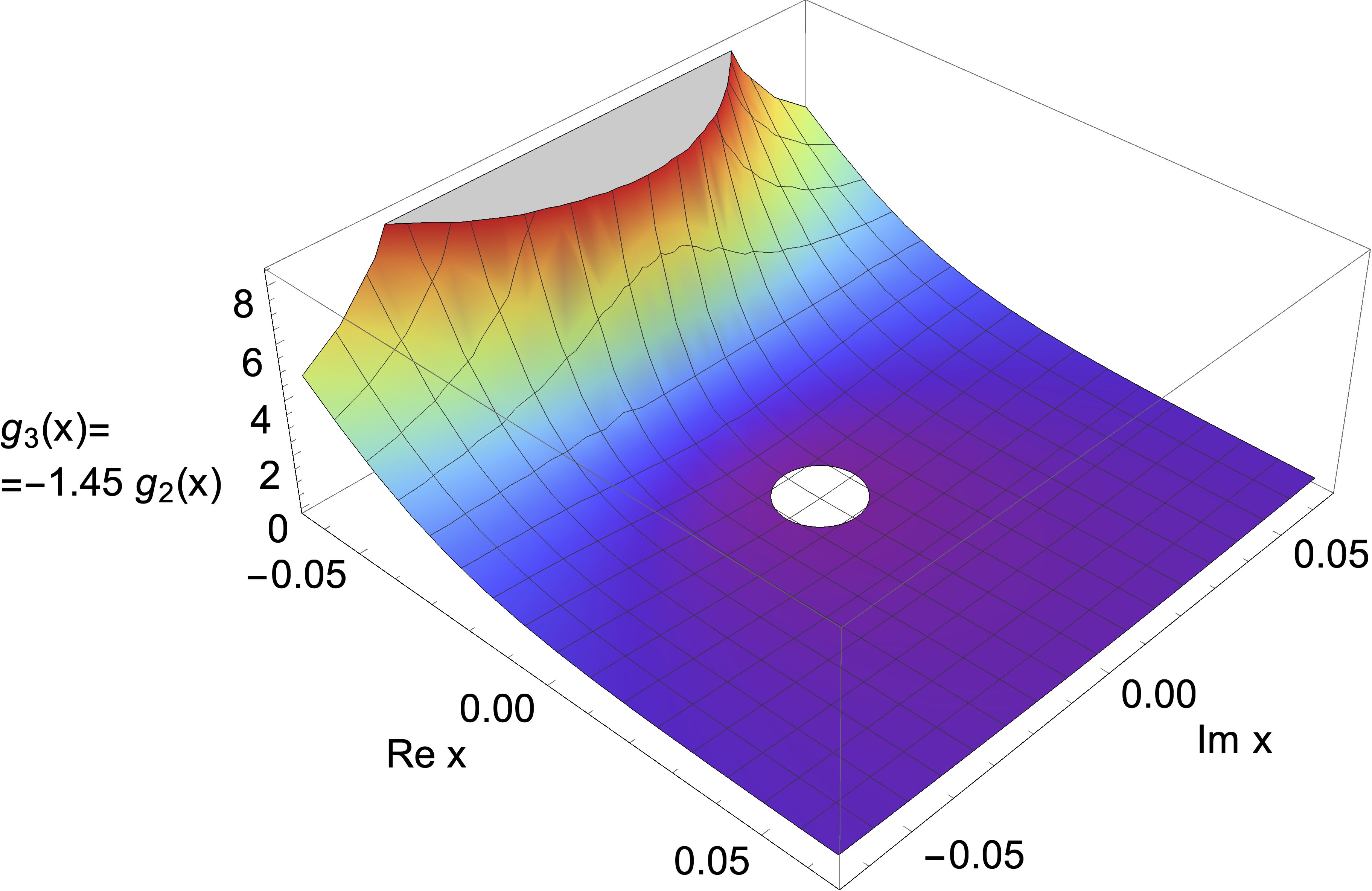} 
    \includegraphics[width=0.42\textwidth]{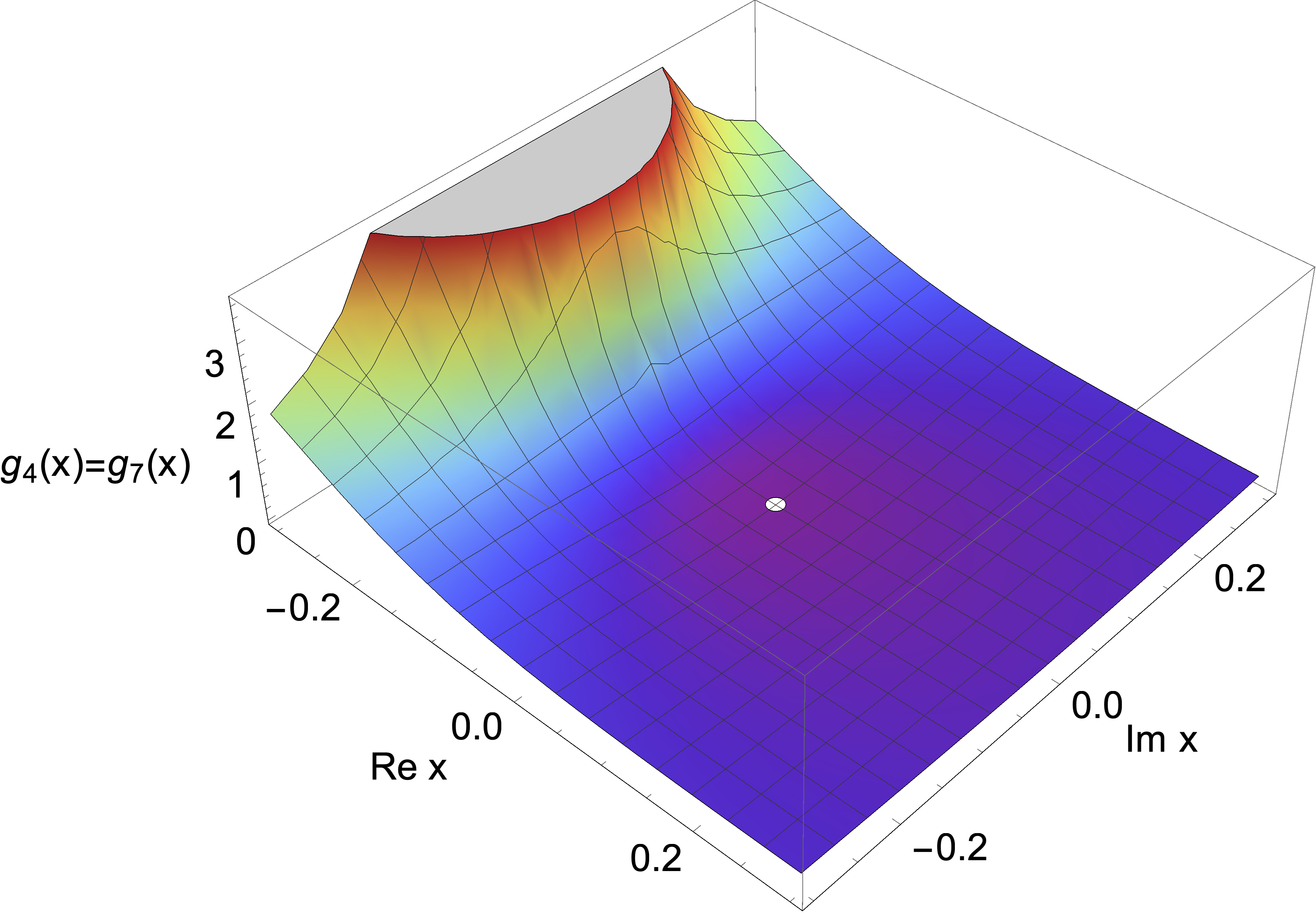}
    \includegraphics[width=0.57\textwidth]{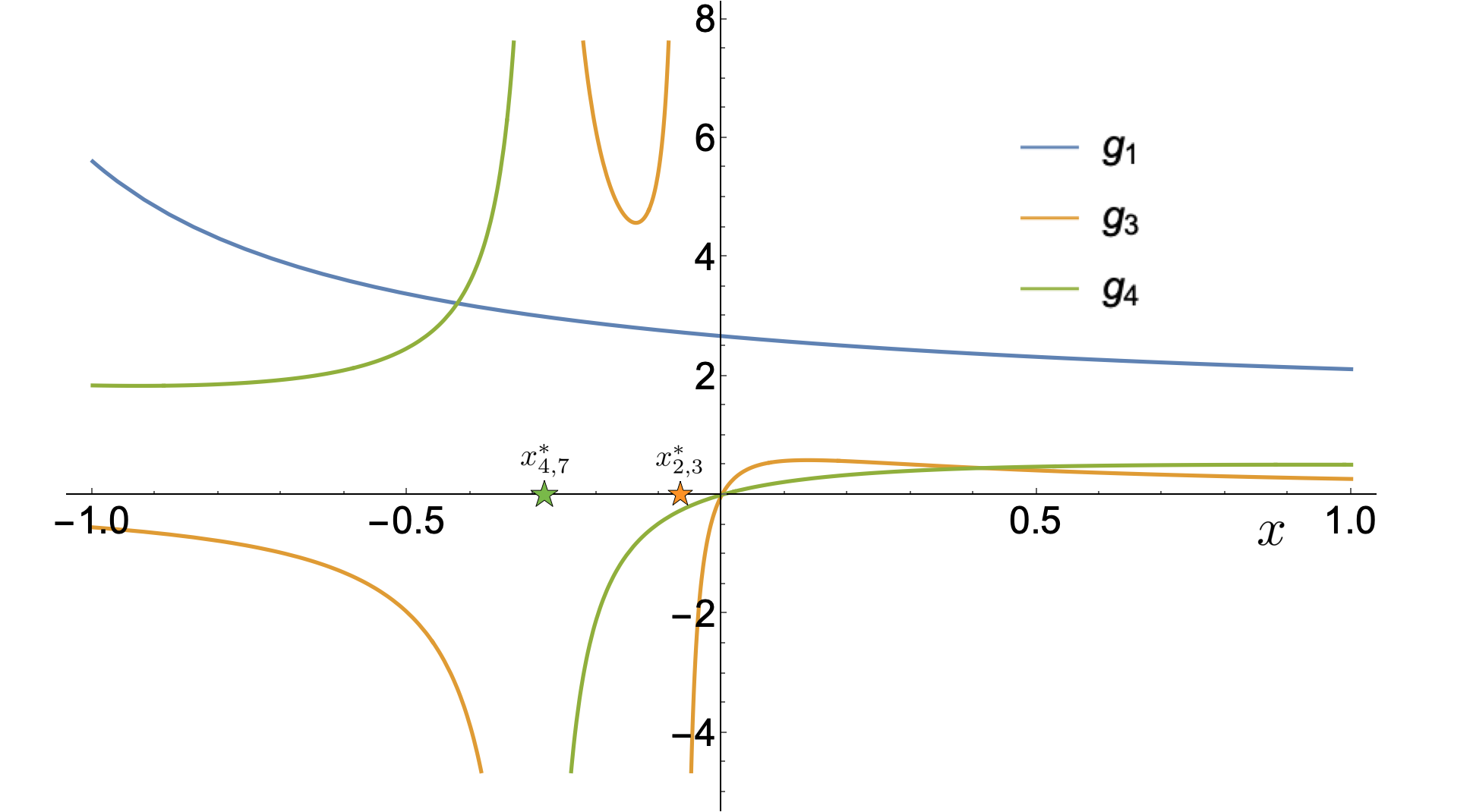}
    \caption{ Evaluation of the absolute values of the vertex model functions $g_i(x)$. These form factors have poles at the value $x_i^\ast$ with $x_1^\ast\simeq -1.46$ (outside the graph and irrelevant for light quark physics), $x_{2,3}^\ast\simeq-6.48\times10^{-2}$ and $x_{4,7}^\ast\simeq-2.88\times10^{-1}$, seen as the mountainlike enhancements in the three-dimensional views, and as the asymptotes in the two-dimensional cut taken along the real axis.}
    \label{fig:vertexparametrization}
\end{figure}

Then, the analytical continuation of $F$ in Eq.~(\ref{toyfunction}) takes us to
\begin{equation}
F(P^2) = \frac{1}{(-(p^0)^2+E({\bf p})^2)+C^2}
= \frac{1}{(-P^2)+C^2}\ .
\end{equation}
Thus, to extend the vertex functions (assuming that their rational approximants are qualitatively reasonable), we only need to work along the real $P^2$ axis and change the argument of the rational function $F$ from positive (the Euclidean side) to negative (the Minkowski side) in the last, two-dimensional plot of Figure~\ref{fig:vertexparametrization}. 

This procedure works because of our explicit model expressions, but is not obvious if one is in possession of just a numerical table on the Euclidean side; but for short-distance continuations where the excursion into Minkowski space wanders not too far from that Euclidean side, one does not need sophisticated dispersion relations that swipe the entire complex plane: one should be able to perform the analytical continuation with only ``regional'' (a small segment suffices~\cite{Gimeno-Segovia:2008fqu}) knowledge of $F$, employing the differential Cauchy-Riemann equations, which are quite unstable but local. They serve to show that it is sound to evaluate the rational function with slightly negative instead of slightly positive argument.

A different issue is to then try to evaluate the systematic uncertainty due to the choice of rational Pad\'e approximant, but that discussion is an entire project in itself, the interested reader can find a worked-out example 
in~\cite{Escudero-Pedrosa:2020rwb}.

We wish to evaluate the vertex functions at physical four-momenta and hence we need to check where the pole positions of the form factors in Eqns.~(\ref{vertexpar1}), (\ref{vertexpar2}, (\ref{vertexpar3}), if any, are encountered, in the region near the physical masses of light quarks. To see this we have prepared Fig. \ref{fig:vertexparametrization}. More robust analytic continuations to the whole complex plane of QCD's correlation functions exist in the literature, see for example \cite{Horak:2021syv,Horak:2023xfb}.

The figure shows three-dimensional plots  of the $g_i$ form factors (as real functions of the $x=p^2_E$ variable extended to the complex plane) and a two-dimensional cut (for real $x$). 

In the three-dimensional plots, small white circles mark the position of the current quark masses in the Lagrangian, that are sufficiently near $x=0$ to not be affected by the form factors. 
The constituent masses from the dressed propagator
are however near those poles. The two dimensional cut shows the precise position of those for $g_3$ and $g_4$: they appear near the constituent quark masses and therefore the relative intensity of the 
form factors for physical, low-scale quark masses is sensitive to model details and future detailed studies are warranted.
(The growth of the primitive $\gamma^\mu$ vertex's form factor $g_1$ is too slow for any pole to appear in the momentum region of interest).

\begin{table}[ht!]
    \centering
    \caption{Actual numerical value of the $g_i$ functions extrapolated to the Minkowski-space side. As the constituent-quark mass value is similar to the value of the pole positions of the $g_i$, we quote the value of the coefficients at two relevant points on both sides of the pole, corresponding to constituent quark masses of 250 MeV and 300 MeV. This larger value is more in line with quark model phenomenology, so the first column marked ``beyond pole'' should be taken as reference. But it is clear that the precise value of the form factor is sensitive to the quark mass, allowing only for qualitative statements.}
    \begin{tabular}{|c|c|c|}
      \hline  & Constituent quark mass $x_c=-0.09$ (beyond pole) & Constituent quark mass $x_c=-0.06$ (just before pole) \\
      \hline 
         $g_1(x)$ & $2.76$ & $2.73$ \\ \hline
         $g_3(x)=-1.45 g_2(x)$ &$6.6$&$-20$ \\ \hline
         $g_4(x)=g_7(x)$ & $-0.41$ & $0.23$ \\
         \hline
    \end{tabular}
    \label{tab:g(x)}
\end{table}

In consequence, we give in Table~\ref{tab:g(x)} the value of the form factors for two typical values of would-be constituent quark masses, taken just above and just below those two poles. 
We see that the chiral-symmetry violating structure has a dominant form factor, $g_3$, in both cases, although we are not able to pinpoint its precise value without going into uncertain model-dependence. One can see on the Euclidean side that $g_3$ and $g_7$ (and therefore also $g_2$ and $g_4$) are rapidly varying and thus the appearance of poles on Minkowski's side is likely, though not assured.\footnote{Higher-order Pad\'e, resp., rational approximations to the form factor $g_3$ 
indicate that the pole closest to the origin is a very robust feature with a quite stable pole position whereas the second pole is an artefact of the employed order of the approximation.}
Happily, the relative weight of the two spin contributions shown below does not seem to be very sensitive to details of this extrapolation.

\section{Pair creation inside the flux tube}\label{sec:transition}

In this section 
we  collect a few intermediate steps concerning the  $q\overline{q}$ spin structure (in pieces that we believe would enter a fuller calculation, and are representative thereof). In first order of business, we establish contact with the quark model formulation of the $^3P_0$ formalism.
We also directly quote the result for a color octet emission. We leave for the next section~\ref{sec:singlet} the treatment of the minimum skeleton diagram that can emit a color singlet pair.

\subsection{For comparison: Effective Hamiltonian in quark model terms}
Let us first establish contact between the Landau gauge DSE and the quark model formalisms, which was the initial intent and motivation of our investigation.
We will slightly modify the treatment of Segovia \textit{et al.} \cite{Segovia:2012cd} so that we can match it to the covariant approach. The transition operator of the $^3 P_0$ model, in relativistic notation, is  
\begin{align} \label{3P0model2}
T_{^3P_0}=\sqrt{3} g_s \int d^3 \boldsymbol x \int \frac{d^3 \boldsymbol{k}_1}{(2\pi)^3\sqrt{2E_{\boldsymbol{k}_1}}} \frac{d^3 \boldsymbol{k}_2}{(2\pi)^3\sqrt{2E_{\boldsymbol{k}_2}}} e^{-i \boldsymbol x\cdot (\boldsymbol{k}_1+\boldsymbol{k}_2)} \sum_{s,s'} {{a}^{s}_{\boldsymbol{k}_1}}^\dagger{{b}^{s'\dagger}_{\boldsymbol{k}_2}}  \bar{u}^s(\boldsymbol{k}_1){v}^{s'}(\boldsymbol{k}_2)
\end{align}
which is nothing but the part of the $^3P_0$ Hamiltonian of Eq. (\ref{3P0model}) that creates a quark-antiquark pair. 
This is missing information about where the energy of the created quark-antiquark pair comes from. In old-fashioned perturbation theory that is not a problem, since the interaction potential acts as the source of that energy. But in covariant perturbation theory, we need to explicitly account for it. 
A possible method that achieves this is to supply an auxiliary field interpolating between the vacuum and the chromoelectric flux tube configuration, 
$\langle E | \Phi  |0 \rangle\neq 0$, so that
\begin{equation} \label{3P0model21}
H_{^3P_0}  \to \sqrt{3} g_s \int d^3\boldsymbol{x} \Phi(x) \bar{\psi}(\boldsymbol{x}) \psi(\boldsymbol{x}).
\end{equation}
The transition operator can be used to compute the corresponding Feynman amplitude, $\mathcal{M}$, by taking its matrix element
\begin{eqnarray}
 \braket{\bp s, \boldsymbol{q} s'|iT_{^3P_0}|E}&=& \int d^4 x\braket{\bp s, \boldsymbol{q} s'|-i\sqrt{3}g_s  \bar{\psi}(x) \psi(x)|0}
 \braket{0|\Phi(x)|E}
 \nonumber \\
 &=&-i\sqrt{3}g_s \int d^4 x e^{-ix(p+q-k)}\bar{u}^s(p){v}^{s'}(q)\;,
\end{eqnarray}
and, after extracting the delta function, we identify
\begin{equation}
    i(2\pi)^4\delta^{(4)}(p+q-k)\mathcal{M}^{ss'}_{^3P_0}(p,q)=\braket{\bp s, \boldsymbol{q} s'|iT_{^3P_0}|0}=-i\sqrt{3}g_s\,(2\pi)^4\delta^{(4)}(p+q-k)\bar{u}^s(p){v}^{s'}(q)\ .  \label{eq:3P0amplitudesegovia}
\end{equation}

Therefore, the Feynman amplitude $\mathcal{M}$ is 
\begin{equation}
\mathcal{M}^{ss'}_{^3P_0}(p,q)=
-\sqrt{3}g_s\,\bar{u}^s(p){v}^{s'}(q)\ .  \label{eq:3P0amplitude}
\end{equation}

It is then direct to show that, defining $\mathcal{A}=\delta^{(3)}(\{\boldsymbol{p}\})\mathcal{M}$, in the CM frame and putting external particles on shell,
\begin{equation}
\mathcal{A}^{ss'}_{^3P_0}(\bp)\propto \bar{u}^s(\bp){v}^{s'}(-\bp)=2\bp\cdot \boldsymbol{\sigma}^{ss'}\;,
\end{equation}
which is of course a purely $^3P_0$ contribution.

\subsection{Setup with QCD Green's functions: color octet production in the skeleton expansion}
Now that we have a clearer idea of what structures we would need to be reproducing from the point of view of the covariant
Landau gauge Green's functions, we write down the pair-production amplitude, first in a color octet state from a single vertex insertion
\begin{align} \label{Tmatrixoctet}
\braket{\bp s, \boldsymbol{q} s'|iT_{\rm octet}|0}&=\braket{\bp s, \boldsymbol{q} s'|-ig \int d^4 x \bar{\psi}_i(x) T^a_{ij} A_\mu^a(x) \Gamma^\mu  \psi_j(x)|0}=\nonumber\\
&=-ig \int d^4 x \int \frac{d^4 k}{(2\pi)^4} e^{-ix(p+q-k)}\tilde{A}_0^a(k)\bar{u}_i^s(p) T^a_{ij}\Gamma^0(p,q,-k){v}^{s'}_j(q)\nonumber=\\
&= -ig \tilde{A}_0^a(p+q)\bar{u}_i^s(p) T^a_{ij}\Gamma^0(p,q){v}^{s'}_j(q)\;,
\end{align}
where $\tilde{A}_0^a(k)$ is the Fourier coefficient of ${A}_0^a(x)$ (this result is analogue to the textbook QED computation in \cite{Peskin:1995ev}).

For the constant (chromo)electric field flux tube, approximated as a uniform (chromo)electric field,  we have that $\tilde{A}_0^a(p+q)\propto\partial_{p_3}\delta^{(4)}(p+q)$ (which can be seen by writing the $x_3=z$ in the Coulomb potential as a derivative respect to the third momentum component before the Fourier transform). The momentum derivative acting on the delta function is passed, upon using Green's theorem for integration by parts, to the fermion kernel, so that we obtain,
\begin{eqnarray}
\braket{\bp s, \boldsymbol{q} s'|iT_{\rm octet}|0}&=&
\braket{\bp s, \boldsymbol{q} s'|-ig \int d^4 x \bar{\psi}_i(x) T^a_{ij} A_\mu^a(x) \Gamma^\mu  \psi_j(x)|0}  \nonumber\\
&=&-ig \int d^4 x e^{ix(p+q)} (-Ex_3)\bar{u}_i^s(p) T^a_{ij}\Gamma^0(p,q){v}^{s'}_j(q)\\
&=& gE(2\pi)^4\delta^{(4)}(p+q) \left[\frac{\partial}{\partial p^3}\big(\bar{u}_i^s(p) T^a_{ij}\Gamma^0(p,q){v}^{s'}_j(q)\big)\right]\;,\label{eq:amplitudeoctet}
\end{eqnarray}
where we assumed that we can integrate by parts in $p$.\footnote{The appearance of a delta function imposing $E_{\bp}+E_{\boldsymbol{q}}=0$ in the amplitudes of eqns. (\ref{eq:3P0amplitude}) and (\ref{eq:amplitudeoctet}) is a consequence of the assumption that there is no time dependence in the interacting Hamiltonian. We can understand that this is an approximation, a more realistic model would implement the time dependence on the flux-tube chromoelectric potential when the flux tube breaks (hence giving the missing excess energy for creating the $q\bar{q}$ pair from the flux tube). These delta distributions are irrelevant for our treatment since we will only compare the amplitudes after extracting those distributions. An example comes by arguing, as would Glendenning and Matsui~\cite{Glendenning:1983qq}, 
that in QCD the created pair would quickly cancel the chromoelectric flux tube, a better approximation would be
$E={\rm constant}\to E=E_0\theta(-t)$, that is, a step function in which the hadron decay at $t=0$ eliminates excess linear field. Then the energy conservation delta rather becomes a finite distribution $\frac{\sin^2(\Delta E t/2)}{(\Delta E t/2)^2}$ that allows for energy redistribution during the time in which the daughter hadrons fly apart. 
However, real-time lattice calculations demonstrate that string-breaking is quite complicated 
with several sub-processes playing an important role \cite{Hebenstreit:2013baa,Kasper:2014uaa} and 
the corresponding time dependence will be quite involved.
In any case, since we are addressing the relative intensity of the various possible spin-orbital
combinations, we will drop the Dirac energy-conservation-$\delta$ distributions and work, preferably, with the Feynman amplitudes $\mathcal{M}$ as soon as they can be provided.}

Therefore, the Feynman amplitude now reads, defining $\mathcal{A}=\delta^{(3)}(\{\boldsymbol{p}\})\mathcal{M}$,
\begin{equation}
\mathcal{A}^{ss'}_{\rm octet}({\bp})=
-i gE \left[\frac{\partial}{\partial p^3}\big(\bar{u}_i^s(p) T^a_{ij}\Gamma^0(p,q){v}^{s'}_j(q)\big)\right]_{{\bf q}=-{\bf p}}\;.
\end{equation}

In comparing this with Eq.~(\ref{eq:3P0amplitude}) we see that, because of the uniform chromoelectric field in the $OZ$ direction, the invariance under rotations is broken and we have an explicit derivative respect to $p_3$.
For the rest, we see that the color-octet production of a quark-antiquark pair falls-off directly from the Landau-gauge quark-gluon vertex. 
\begin{figure}[ht!]
 \centering
    \begin{tikzpicture}[>=stealth,scale=1.2]
  \draw[-latex,thick] (0,.25)--(.75,.7);
  \draw[thick] (0,.25)--(1.25,1);
   \draw [decoration={aspect=0.4, segment length=1mm, amplitude=1mm,coil},decorate] (-1.25,.25) -- (0,.25);
  \draw[-latex,thick] (1.25,-.5)--(.5,-0.05);
  \draw[thick] (1.25,-.5)--(0,0.25);
\end{tikzpicture}
 \begin{tikzpicture}[>=stealth,scale=1.2]
  \draw[-latex,ultra thick, blue] (-.5,0)--(0.375,.0);
  \draw (-.5,-.75) node {$\,$};
\end{tikzpicture}
\begin{tikzpicture}[>=stealth,scale=1.2]
  \draw[-latex,thick] (0,.25)--(.75,.7);
  \draw[thick] (0,.25)--(1.25,1);
   \draw [decoration={aspect=0.4, segment length=1mm, amplitude=1mm,coil},decorate] (-1.25,.25) -- (0,.25);
  \draw[-latex,thick] (1.25,-.5)--(.5,-0.05);
  \draw[thick] (1.25,-.5)--(0,0.25);
          \shade[ball color=blue]  (0,0.25) circle (.15cm);
\end{tikzpicture}
\caption{\label{fig:skeletonoctet} The perturbative octet-channel quark-gluon scattering kernel that gives rise to pair creation is chirally invariant. However, in the skeleton expansion, both the propagator and the two vertices which appear in its skeleton expansion~\cite{Llanes-Estrada:2004hnb} acquire chiral-symmetry breaking parts conveniently generated by their respective Dyson-Schwinger equations assisted by lattice data. }
\end{figure}
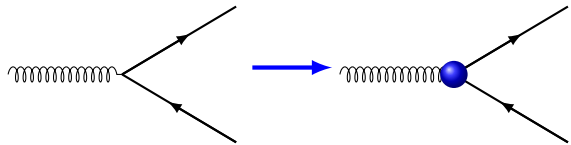

\section{Singlet production: Skeleton expansion of the quark-gluon scattering kernel }
\label{sec:singlet}

The quark-gluon vertex produces a $q\overline{q}$ pair in a color-octet state only. 
To examine color-singlet production, we need to extract two gluons from the background field. 

\begin{figure}[ht!]
 \centering
 \begin{tikzpicture}[>=stealth,scale=1.2]
  \draw [decoration={aspect=0.4, segment length=1mm, amplitude=1mm,coil},decorate] (-1.25,1) -- (0,1);
  \draw[-latex,thick] (0,1)--(.75,1);
  
  \draw[thick] (0,1)--(1.25,1);
   \draw [decoration={aspect=0.4, segment length=1mm, amplitude=1mm,coil},decorate] (-1.25,-.5) -- (0,-.5);
  \draw[-latex,thick] (1.25,-.5)--(.5,-.5);
  \draw[thick] (0,-.5)--(1.25,-.5);
  \draw[-latex,thick] (0,-.5)--(0,0.375);
    \draw[thick] (0,-.5)--(0,1);
\end{tikzpicture}
 \begin{tikzpicture}[>=stealth,scale=1.2]
  \draw[-latex,ultra thick, blue] (-.5,0)--(0.375,.0);
  \draw (-.5,-.75) node {$\,$};
\end{tikzpicture}
 \begin{tikzpicture}[>=stealth,scale=1.2]
  \draw [decoration={aspect=0.4, segment length=1mm, amplitude=1mm,coil},decorate] (-1.25,1) -- (0,1);
  \draw[-latex,thick] (0,1)--(.75,1);
  \draw[thick] (0,1)--(1.25,1);
   \draw [decoration={aspect=0.4, segment length=1mm, amplitude=1mm,coil},decorate] (-1.25,-.5) -- (0,-.5);
  \draw[-latex,thick] (1.25,-.5)--(.5,-.5);
  \draw[thick] (0,-.5)--(1.25,-.5);
  \draw[-latex,thick] (0,-.5)--(0,0);
  \draw[-latex,thick] (0,-.5)--(0,0.75);
    \draw[thick] (0,-.5)--(0,1);
      \shade[ball color=blue] (0,1) circle (.15cm);
        \shade[ball color=blue]  (0,-.5) circle (.15cm);
        \shade[ball color=black]  (0,0.25) circle (.15cm);
\end{tikzpicture}
\caption{\label{fig:skeleton} The perturbative singlet-channel quark-gluon scattering kernel that gives rise to pair creation, analogous to the Breit-Wheeler process, is chirally invariant save for the small current quark mass in the fermion propagator. However, in the skeleton expansion, both the propagator and the two vertices acquire chiral-symmetry breaking parts conveniently generated by their respective Dyson-Schwinger equations assisted by lattice data. }
\end{figure}
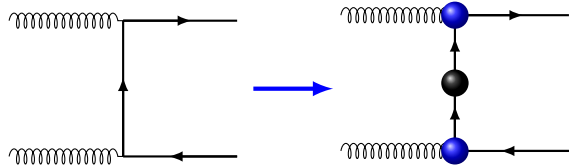

The equivalent of the Breit-Wheeler process depicted in figure~\ref{fig:skeleton} is chirally symmetric in perturbation theory.
However, perturbation theory can be reorganized in the so called ``skeleton expansion'' in which all the primitive Green's functions (those reflecting a term in the renormalizable QCD Lagrangian) are substituted by the fully interacting ones.

We compute the amplitude
\begin{align}
\braket{\bp s, \boldsymbol{q} s'|iT_{\text{singlet}}|0}&=\braket{\bp s, \boldsymbol{q} s'|-\frac{g^2}{2} \int d^4 x \bar{\psi}_i(x) T^a_{ij} A_\mu^a(x) \Gamma^\mu  \psi_j(x)\int d^4 y \bar{\psi}_i(y) T^a_{ij} A_\nu^a(y) \Gamma^\nu  \psi_j(y)|0}=\nonumber\\
&=-g^2 \int d^4 x d^4 y \int \frac{d^4 t}{(2\pi)^4} \tilde{A}_0^a(p-t) \tilde{A}_0^a(q+t) \mathcal{K}_{ab}^{ss'}(p,q,t)\;,
\end{align}
where the skeleton-expanded kernel equals
\begin{equation}
\mathcal{K}_{ab}^{ \,ss'}(p,q,t)\equiv  
\left[\bar{u}_i^s({p})T_{ij}^a\Gamma^0(p,-t)S(t)T_{jk}^b\Gamma^0(q,t){v}_k^{s'}({q})\right]+\text{crossed amplitude}\;,\label{eq:kernel}
\end{equation}
and $S(t)$ is the dressed fermion propagator.

For the case of a constant, classical electric field insertion, as described in section~\ref{sec:Afield}, we get
\begin{align}
\braket{\bp s, \boldsymbol{q} s'|iT_{\text{singlet}}|0}&=\braket{\bp s, \boldsymbol{q} s'|-\frac{g^2}{2} \int d^4 x \bar{\psi}_i(x) T^a_{ij} A_\mu^a(x) \Gamma^\mu  \psi_j(x)\int d^4 y \bar{\psi}_i(y) T^a_{ij} A_\nu^a(y) \Gamma^\nu  \psi_j(y)|0}=\nonumber\\
&=-g^2 \int d^4 x d^4 y \int \frac{d^4 t}{(2\pi)^4} e^{-i(x-y)t}  e^{ixp} e^{iyq}(-Ex_3) (-Ey_3)\mathcal{K}_{ab}^{ss'}(p,q,t)\;.
\end{align}
 Now we assume that we can integrate by parts in $p$ and $q$ to obtain
\begin{align}
\braket{\bp s, \boldsymbol{q} s'|iT_{\text{singlet}}|0}&=-(2\pi)^4\delta^{(4)}(p+q)(gE)^2\left[\frac{\partial }{\partial p^3}\frac{\partial}{\partial q^3}  \mathcal{K}_{ab}^{ \,ss'}(p,q,t)\right]\Big|_{t=-q
}\;.\label{eq:singletamplitude}
\end{align}

\subsection{Comparing the Amplitudes}

Gathering the results from eqns. (\ref{eq:3P0amplitude}), (\ref{eq:amplitudeoctet}) and (\ref{eq:singletamplitude}) we have the following amplitudes, $\mathcal{A}(\{\bp\})=\delta^{(3)}(\{\bp\})\mathcal{M}(\{p\})$, for the case of the homogeneous chromoelectric field of Eq. (\ref{eq:homogeneousfield}):

\begin{table}[ht!]
    \centering
    \begin{tabular}{|c|c|c|c|}
      \hline  & $^3P_0$ Model & Octet channel & Singlet channel \\
      \hline & & & \\
         $\mathcal{A}^{ss'}(\bp)$ & $-\sqrt{3}g_s\,\bar{u}^s(\bp){v}^{s'}(-\bp)$ & $-igE\left[\frac{\partial}{\partial p^3}\big(\bar{u}_i^s(\bp) T^a_{ij}\Gamma^0(\bp,\boldsymbol{q}){v}^{s'}_j(\boldsymbol{q})\big)\right]\Big|_{\boldsymbol{q}=-\bp}$& $i(gE)^2\left[\frac{\partial}{\partial p^3}\frac{\partial}{\partial q^3}  \mathcal{K}_{ab}^{ \,ss'}(p,q,t)\right]\Big|_{\boldsymbol{t}=-\boldsymbol{q}=\bp}$\\
      & & & \\
         \hline
    \end{tabular}
    \label{tab:amplitudes}
\end{table}
where the evaluations after taking momentum derivatives also evaluate the energy components to their on-shell values.

For the sake of comparison, we will extract some of the different angular momentum contributions by projecting the amplitudes $\mathcal{A}^{ss'}(\bp)$ as
\begin{align}    \label{QCDprojections} 
\mathcal{A}^{^1S_0}(|\bp|)&=\sum_{s} \int d\Omega\, \mathcal{A}^{ss}(\bp)\\
    \mathcal{A}^{^3S_1}_i(|\bp|)&=\sum_{s,t} \int d\Omega\, \boldsymbol{\sigma}^{st}_i \mathcal{A}^{ts}(\bp)\\
    \mathcal{A}^{^3P_0}(|\bp|)&=\sum_{s,t} \int d\Omega\, \hat{\bp}\cdot \boldsymbol{\sigma}^{st} \mathcal{A}^{ts}(\bp)\\
\mathcal{A}^{^1P_1}_i(|\bp|)&=\sum_{s} \int d\Omega\, \hat{\bp}_i \,\mathcal{A}^{ss}(\bp)\;.
    \end{align}

Once more, we emphasize that, because of the reduced rotational symmetry in the presence of the homogeneous chromoelectric field, that only the angular momentum component along the said field is a good quantum number, so that we should relabel $S\to \Sigma$, $P\to \Pi$ as in molecular physics.
\subsection{Singlet Channel QCD spin components}
For the full DSE parametrization of the QCD vertex  in Eqs.~(\ref{vertexpar1}), (\ref{vertexpar2}), (\ref{vertexpar3}), 
the amplitude from Eq.~ (\ref{eq:singletamplitude}) 
simplifies if we introduce the following auxiliary quantities with dimensions of squared energy,
in which the various dynamical scales of the propagator parametrization substitute for a mass \begin{eqnarray}
E_{AZ}^2  \equiv |{\bf p}|^2 +Z_0 \Lambda_A^2\,;\;\;\;E_A^2 \equiv |{\bf p}|^2 +\Lambda_A^2\,;\;\;\;E_B^2 \equiv |{\bf p}|^2 +\Lambda_B^2 \,;\;\;\;E_{\bp}^2 \equiv |{\bf p}|^2 + m^2\ .
\end{eqnarray}
The actual mass here is $m=M(0)$ that we also take as approximately equal to $M(M^2)$ in Minkowski space for simplicity of the following expressions, since $M(p^2)$ is smooth for small virtualities, and not distinguish them in the following formulae.

 In them we also take the symmetric kinematic section
$\chi=0$  (setting the asymmetry variable from subsection~\ref{subsec:QEDwithtransverse} which distributes the energy production among the two vertices), that is, both field insertions provide an equal energy $E_p$. Hence,  this is a simplified kinematic case, which still calls for symbolic computing assistance to reduce it before attaining this compact form). Nonetheless, remember that in QED we observed that the ratio $^3\Sigma_1/^3\Pi_0$ of the  amplitudes was independent of the asymmetry $\chi$.

The result becomes
\begin{align}
 {\mathcal{A}^{^3\Sigma_1}_{\text{QCD}}}(|\bp|)\nonumber \propto   &  
\frac{-\pi  }{  E_{\bp}^3 \left(2 E_{B}^2 |\bp|^4+\Lambda_B^4 E_{\bp}^2-|\bp|^6\right)}\frac{E_{B}^2}{E_{A}^2}\Bigg(   
\\ &\phantom{+}\ 
g_1^2\ 2|\bp|  E_{\bp} \Big[E_{A}^2  m (Z_0-2) \Lambda_B^2+E_{AZ}^2 \left(E_{B}^2 E_{\bp}+m |\bp|^2\right)-m |\bp|^2 (Z_0-1) \Lambda_B^2\Big]\nonumber\\
&+ g_2^2 m |\bp| E_{\bp} \Big[E_{A}^2 \left(E_{B}^2 (Z_0+1)-|\bp|^2\right)-E_{B}^2 |\bp|^2 (Z_0-1)\Big]\nonumber\\
&+g_3^2\ 2 |\bp| E_{\bp} \left(2 E_{A}^2 \Lambda_{B}^2 m + E_{AZ}^2 E_{B}^2  m 
+ E_{AZ}^2 E_{B}^2 E_{\bp}
\right) 
\nonumber\\ \nonumber
&+g_4^2 m |\bp| E_{\bp} \left( E_{A}^2 \Lambda_{B}^2   -E_{AZ}^2 E_{B}^2  
\right)
\\ &
+g_7^2 \ 2 m   |\bp| E_{\bp} \left( - E_{A}^2 \Lambda_{B}^2 -2 E_{AZ}^2 E_{B}^2 \right)
\nonumber\\
&+g_1 g_2 \ 2 E_{\bp} \big[\Lambda_{A}^2 E_{B}^2 |\bp|^2 Z_0
+m E_{A}^2 \Lambda_{B}^2(E_{\bp}- m)
+E_{B}^2  |\bp|^4 
\big]\nonumber\\ &
+g_1  g_3 \ 4 |\bp| E_{\bp}E_{AZ}^2 E_{B}^2    \left( m+E_{\bp}\right)
\nonumber\\ &
+  g_1 g_7 \ 2 E_{\bp}  \left[E_{A}^2 \left(m |\bp|^2 \left( m+3 E_{\bp} \right)-E_{B}^2 \left( \left(m^2-|\bp|^2 Z_0\right)+3 m E_{\bp}\right)\right)-E_{B}^2  |\bp|^4 (Z_0-1)\right]
\nonumber \\
&+g_3 g_7 \ 2 E_{\bp} \left( 3 E_{A}^2  m |\bp|^2(E_{\bp}+m)
+3 E_{AZ}^2 E_{B}^2 |\bp|^2 - 3 E_{A}^2 E_{B}^2  m^2 
- m E_{A}^2 E_{B}^2  E_{\bp}
\right)\nonumber\\
&+g_4 g_7 \ 2 m |\bp| E_{\bp} \left( E_{AZ}^2 E_{B}^2 - E_{A}^2 E_{B}^2 + E_{A}^2 |\bp|^2\right)\nonumber\\
&+g_2 g_3 \ 2\left(
-E_{A}^2 \Lambda_{B}^2 m (E_{\bp}-m)
-\Lambda_{A}^2 E_{B}^2 
 |\bp|^2 Z_0
 -E_{B}^2  |\bp|^4
 \right)\nonumber\\
&+ 2g_2 g_7   m |\bp|  E_{\bp}  \left(E_{A}^2 \left(E_{B}^2 (Z_0+1)-|\bp|^2\right)-E_{B}^2 |\bp|^2 (Z_0-1)\right)\Bigg)\\
{\mathcal{A}^{^3\Pi_0}_{\text{QCD}}}(|\bp|)\nonumber \propto
&\frac{-16/3 }{  E_{\bp}^2 \left(-2 E_{B}^2 |\bp|^4-\Lambda_B^4 E_{\bp}^2+|\bp|^6\right)}
\frac{E_{B}^2}{E_{A}^2}\Bigg(
\nonumber\\
& \phantom{+}\ \ g_1^2\  m |\bp| \left(E_{A}^2 \left(E_{B}^2 (Z_0-3)+3 |\bp|^2\right)-E_{B}^2 |\bp|^2 (Z_0-1)\right)\nonumber\\
&+ g_2^2\ m |\bp| \left(E_{A}^2 \left(E_{B}^2 (Z_0+1)-|\bp|^2\right)-E_{B}^2 |\bp|^2 (Z_0-1)\right)\nonumber\\
&+ g_3^2\ m |\bp| \left(E_{A}^2 \left(E_{B}^2 (Z_0+3)-3 |\bp|^2\right)-E_{B}^2 |\bp|^2 (Z_0-1)\right)\nonumber\\
&+ g_4^2\ m |\bp| \left(E_{B}^2 |\bp|^2 (Z_0-1)-E_{A}^2 \left(E_{B}^2 (Z_0-1)+|\bp|^2\right)\right)\nonumber\\
&- g_7^2\ m |\bp| \left(E_{A}^2 \left(3Z_0E_{B}^2+\Lambda_B^2\right)+3 |\bp|^2 E_B^2(1-Z_0)\right)
\nonumber\\
&+ 
g_1 g_2 \ 2 \left(
 -m^2 E_{A}^2\Lambda_{B}^2
+|\bp|^2 \Lambda_{A}^2 E_{B}^2 Z_0
+|\bp|^4 E_{B}^2 
\right) \nonumber\\
& +g_1 g_3\ 2 |\bp| E_{\bp} E_{AZ}^2 E_{B}^2\nonumber\\
&- g_1 g_7\ 4 m E_{\bp} E_{A}^2  \Lambda_{B}^2 \nonumber\\
&- g_2 g_3\ 2 m E_{\bp}  E_{A}^2 \Lambda_{B}^2  \nonumber\\
&+ g_3 g_7\ 4 \left( -m^2E_{A}^2\Lambda_{B}^2
+ |\bp|^2 \Lambda_{A}^2 E_{B}^2  Z_0 
+ |\bp|^4 E_{B}^2 
)\right) \nonumber\\
&+  g_4 g_7 2 m |\bp| \left(\Lambda_{A}^2 E_{B}^2 (Z_0-1)+|\bp|^2 E_{A}^2  \right)\Bigg)\;.
\end{align}
Not all possible structures appear in these equations: for example, the cross product $g_2g_4$ seems to be absent, but this is plausibly a model-dependent feature so we do not make much of it.

\begin{figure}[ht!]
    \centering
    \includegraphics[width=.475\columnwidth]{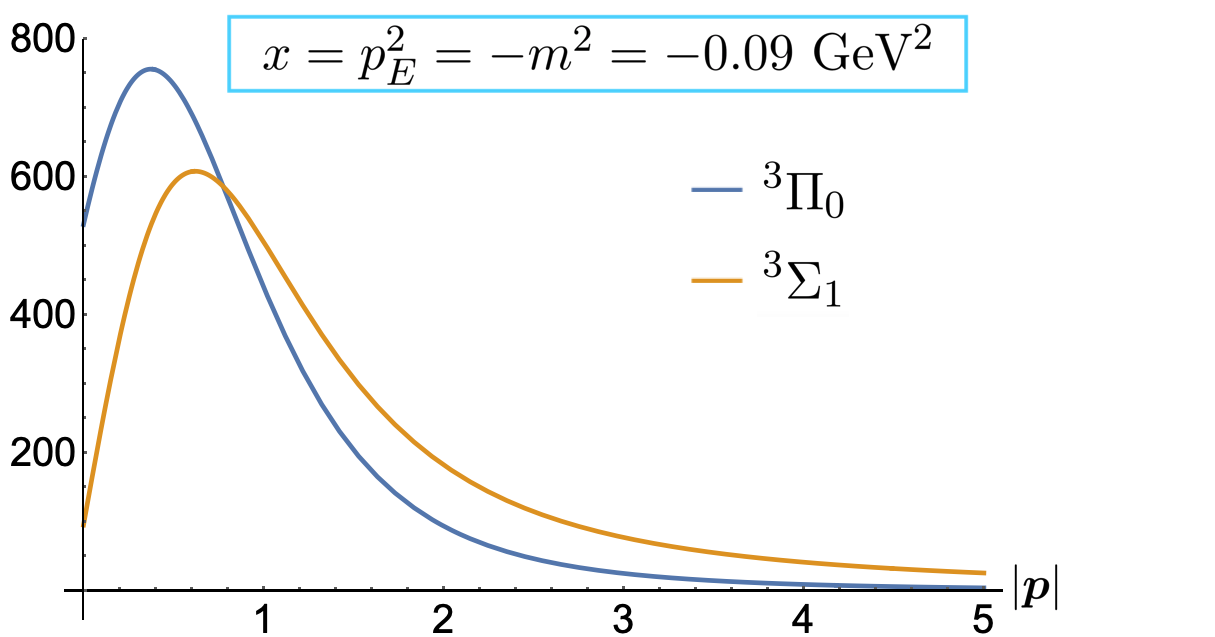}\includegraphics[width=.475\columnwidth]{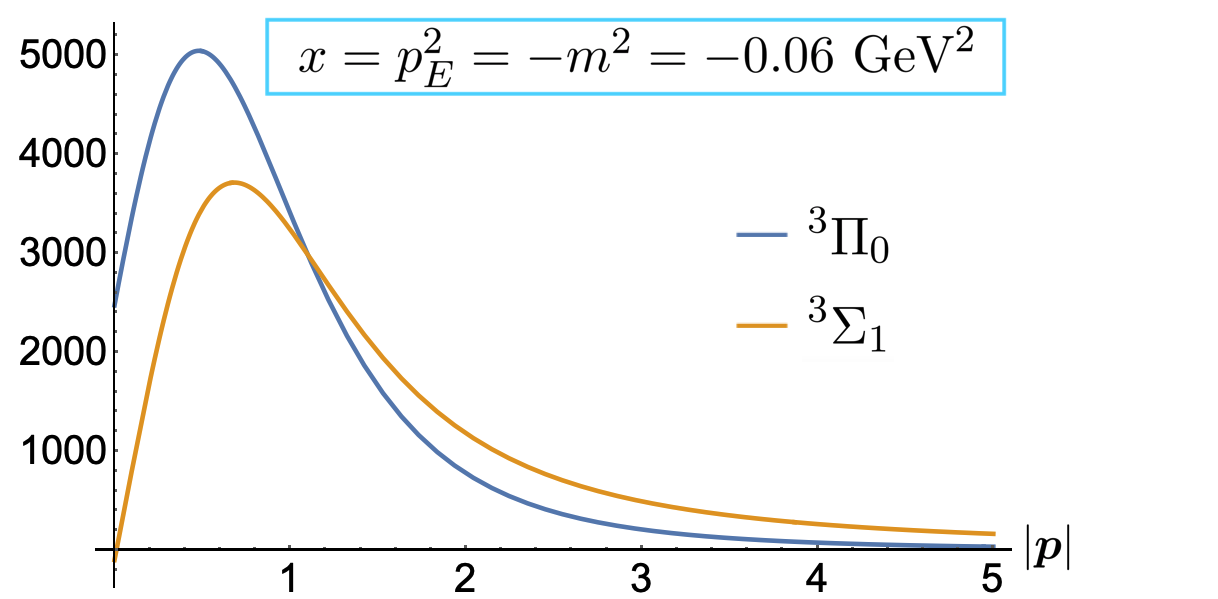}
    \includegraphics[width=.475\columnwidth]{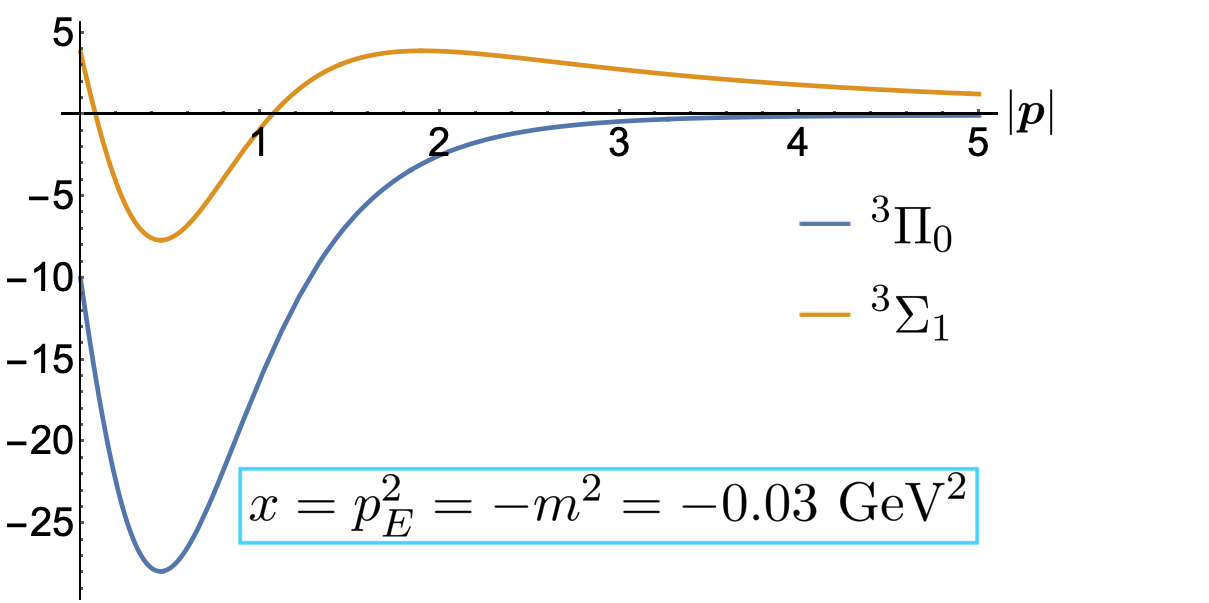}\includegraphics[width=.475\columnwidth]{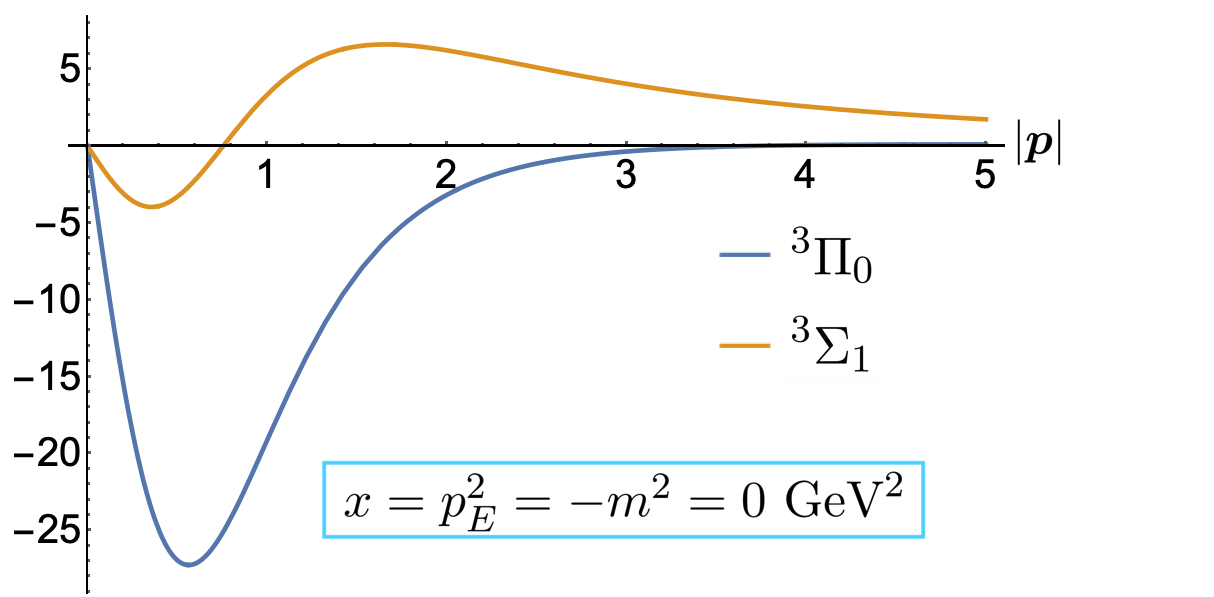}
    \caption{ Comparison of the $^3\Pi_0$ and $^3\Sigma_1$ angular momentum components in the QCD skeleton approximation here developed, in four kinematic sections. The $q\overline{q}$ pair is produced in its center of mass frame. The $^3\Pi_0$ vertex is seen to be dominant for small quark momentum ($OX$ axis, in units of the quark mass) and the $^3\Sigma_1$ dominates for large momenta.  $|\bp|$ is in units of $m$, except for the last plot in which arbitrary units are employed (with the scale ultimately controlled by $\Lambda_A$, $\Lambda_B$, etc. from the Schwinger-Dyson obtained parametrization from section~\ref{sec:Greensfunctions}).
    }
    \label{fig:compareQCD}
\end{figure}

While these explicit expressions are complicated to analyze, it is direct to numerically compute and plot the two spin production mechanisms together to compare their relative intensity as we did above in section~\ref{sec:QED}, and the result is exposed in figure~\ref{fig:compareQCD}.

The result of the figure is clear: for  values of $|\bp|$ below the fermion mass scale $m$, the $^3\Pi_0$ amplitude dominates
independently of the chosen value of the quark mass. 
While the full analytical expressions are difficult to interpret, we can enlighten the discussion by taking two appropriate limits.

\begin{figure}[ht!]
    \centering
\includegraphics[width=.5\columnwidth]{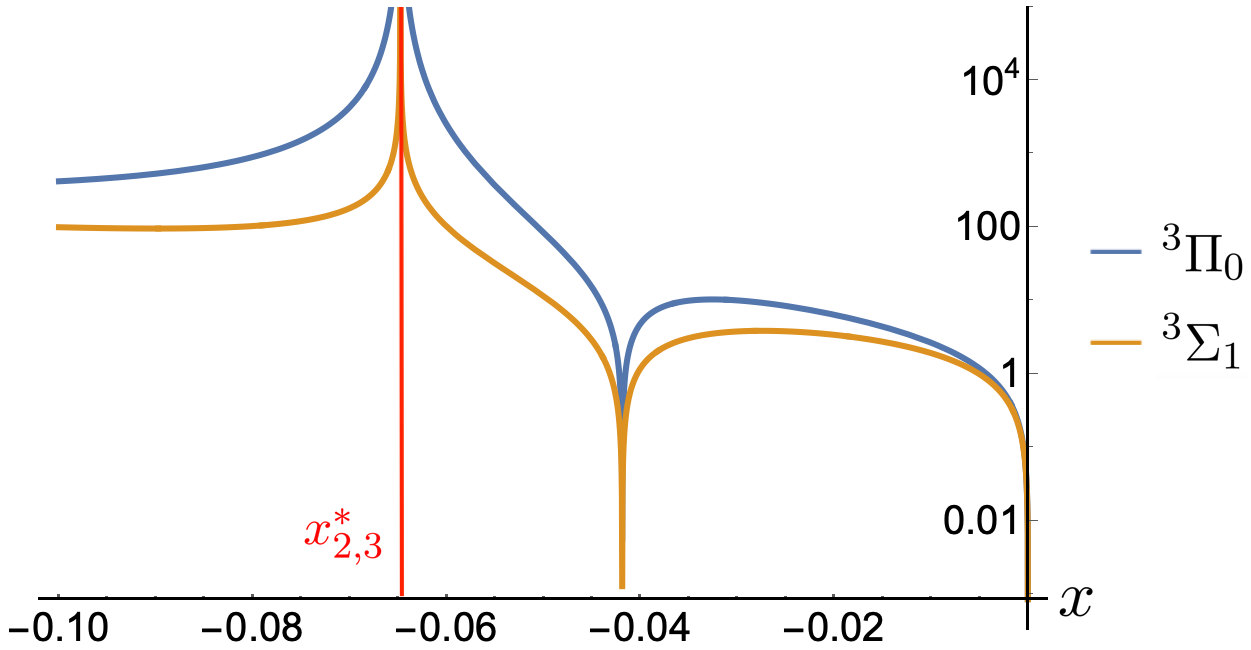}
    \caption{Absolute values (in log scale, and up to factors common to both) of the  $^3\Pi_0$ and $^3\Sigma_1$ contributions
    to our computed pair production kernel at  
    threshold (where $\bp=0$) as function of $m^2$ on the physical Minkowski side. 
    The $^3\Pi_0$ structure dominates
    for all plotted values. This is in spite and independently of the (perhaps artificial) pole of the propagator parametrization 
    (that induces a pole of the amplitudes), when $x_{2,3}^\ast\simeq-6.48\times10^{-2}$. Close to $x=-0.0418857$ both amplitudes almost vanish at the $\bp=0$ threshold.}
    \label{fig:alwaysscalar}
\end{figure}

In the limit $\bp\to {\bf 0}$, we can read off the ratio of the threshold-limit pair production spin amplitudes which, for the two components of interest, equal
\begin{align}
\label{chiralat0}
\lim_{\bp\to 0}{\mathcal{A}^{^3\Sigma_1}_{\text{QCD}}}(|\bp|)&=-\frac{8 \pi  \left(g_1+g_3\right) g_7}{m^2}\\
\lim_{\bp\to 0}{\mathcal{A}^{^3\Pi_0}_{\text{QCD}}}(|\bp|)&=-\frac{32 \left(g_1+g_3\right) \left(g_2+2 g_7\right)}{3 m^2}\ .
\label{nonchiralat0}
\end{align}
Comparing the two expressions shows the reason for the $^3\Pi_0$ production  threshold dominance in the numerical computations of figure~\ref{fig:compareQCD}: the presence of the chiral-symmetry breaking structure $g_2$ in Eq.~(\ref{nonchiralat0}) (since $g_3$ contributes in the combination $g_1+g_3$ in both spin amplitudes). 
Indeed, a glance at Table~\ref{tab:g(x)} above reveals that the chiral-symmetry breaking structures $g_2$, $g_3$ provide much larger vertex contributions than the symmetric $g_4$, $g_7$ ones, for typical kinematic values. 
A lattice-gauge theory computation would here be very interesting to ascertain the degree of model dependence of this assertion. 
All we can do to alleviate it at the present time is to systematically study the dependence with the quark mass on the Minkowski side. There are two poles of the extrapolated  vertex coupling there, which may or may not mean anything depending on the extrapolation used, but the finding, shown in figure~\ref{fig:alwaysscalar} is robust: left or right and at any distance from those two poles, basically everywhere, the $^3\Pi_0$ amplitude is dominant at threshold. 

We also quote the values in the opposite, chiral limit in which $|\bp|>>m$, which turn out to be
\begin{align}
\lim_{m\to 0}{\mathcal{A}^{^3\Sigma_1}_{\text{QCD}}}(|\bp|)&=2 \pi   \frac{E_{AZ}^2}{|\bp|^2 E_A^2}\left(g_1\left(g_1+g_2+2 g_3+g_7\right) +g_3 \left(-g_2+g_3+3 g_7\right)\right)
\\
\lim_{m\to 0}{\mathcal{A}^{^3\Pi_0}_{\text{QCD}}}(|\bp|)&=\frac{32}{3}\frac{E_{AZ}^2}{|\bp|^2 E_A^2}\left(g_1 \left(g_2+g_3\right)+2 g_3 g_7\right)\ .
\end{align}
Here, the $^3\Sigma_1$ amplitude is dominant. 
This is as expected: at large momenta, spontaneous chiral symmetry breaking is less important.


\newpage

\section{Conclusions and Outlook} \label{sec:conclusions}
In this work we set out to ascertain whether the $^3P_0$ production mechanism 
for quark-antiquark pairs in hadron decays could receive support from existing 
QCD Greens functions studies that have accrued in the decades since the mechanism was first proposed. 

In doing so, we have noted that, because of the chromoelectric field providing a net local axis of reference, the total orbital angular momentum is not an adequate quantum number for the produced pair. Hence, as in the theory of diatomic molecules, we need to use its third component, so that $^3\Pi_0$ is a more appropriate term for this vertex. This we do find to be a significant contribution in hadron decays but also in electron-positron production in an intense electric field, a very active line of research in which, surprisingly, spin is not so often addressed. 

This reduced symmetry means that the fermion-antifermion relative orbital angular momentum in the $^3\Pi_0$ decay is parallel (+1) or antiparallel (-1) to the electric (chromoelectric) field, and can also be 0 in the $\Sigma$ waves respecting chiral symmetry. 
This effect is most intense at low momenta $p<m_e$ in Electrodynamics or 
$p<M$, the scale of the constituent quark mass, in Chromodynamics, because it is chiral symmetry breaking. At higher momenta, that make the masses negligible, the production reverts to $^3\Sigma_1$, which is what one expects from a gauge theory treated with a perturbative formalism.

Other decay modes are also possible and we have given (at a very qualitative level)
some idea of their possible relative numerical importance.

We are far from producing a full computation comparable to Schwinger's complete evaluation of the spin-averaged emission rate in QED, but we believe that the skeleton diagrams that we have examined will turn out to be part of any more complete effective action treatment that may be attempted in the future, as they contain the relevant fermion lines and vertices that should enter in such computations. 

With the material at hand, we do not think it is meaningful to attempt an extraction of the total rates in either spin decay channel, but find it instructive to report, as we have, the relative weight of the main spin structures: our results support the quark model lore that the decay mechanism producing quark-antiquark pairs is the $^3P_0$ one, although we qualify this to be the less determined $^3\Pi_0$.

To summarize, our findings would suggest that in QED, $^3\Sigma_0$ is the dominant (lowest order) $e^-e^+$
emission mechanism, followed by $^3\Pi_0$ at threshold, which dominates over $^3\Sigma_1$; this last one, however, takes over at larger momenta of the emitted pair, $p>m_e$.
In QCD, color-singlet $q\overline{q}$ emission requires at least two field insertions and hence $^3\Pi_0$ near threshold (but $^3\Sigma_1$ at large momenta) is obtained. 

This finding lends support to the traditional $^3P_0$ mechanism in the following sense. Since $^3\Pi_0$ requires $m_L=1\implies L\geq 1$, without restricting total $J$, the smallest angular momenta (and thus the smallest energy required to overcome centrifugal barriers and chromomagnetic effects) is the $L=1$, $J=0$ configuration, or $^3P_0$.  
To distinguish $^3P_0$ and $^3\Pi_0$ does not seem possible with basic meson decays, interpreted as $(q\overline{q})\to (q\overline{q})(q\overline{q})$.

To see this, note that for the parent meson, the possible quantum numbers are spin $s_i=0,1$, internal orbital angular momentum $l$, and parity $P_i=(-1)^{l+1}$. 
The two daughter mesons in the final state have $s_f=0,1,2$, internal and relative orbital $l_1$, $l_2$, $L$ and parity (one antiparticle has been produced) $P_f=(-1)^{l_1+l_2+L+2}$, meaning that orbital angular momentum has to change by an odd number of units to preserve parity.

Total angular momentum conservation implies that $\Delta J=0$, so the changes in spin and orbital angular momentum have to compensate each other, 
$|\Delta S|=|\Delta L|$. Because $\Delta S=0,\pm(1,2)$, these are the values that $\Delta L$ can take. Among them, $\Delta L=1$ is common to both $^3P_0$ and $^3\Pi_0$, $\Delta L=0$ to none, and only $\Delta L=2$ could distinguish the two mechanisms. 
But this is an even change in orbital angular momentum, which parity conservation does not allow. So it looks unpromising to try to distinguish both mechanisms, and they are for all purposes undistinguishable in ordinary mesons.

One would hope that future work could shed light on the nuanced distinction between these quantum numbers of the produced pair. Decays of baryon resonances with high spin (Yrast states~\cite{Bicudo:2016eeu}) or multiquark states might offer possibilities for new tests: but the flux tube might then be more complicated and not be well approximated by a uniform field.

We hope that this article will inspire more researchers to address the spin quantum numbers of particle-antiparticle pairs produced in intense fields. Other areas of research where this could be of interest include heavy ion collisions (featuring intense color fields), neutron stars (where magnetic fields are very intense, which do require separate treatment), etc. 

\section*{Acknowledgments}
ASB thanks the theoretical physics group at Graz for its hospitality during an extended visit that facilitated this project.
Support by the City of Graz for this visit is gratefully acknowledged.
Supported likewise by Spanish MICINN under grant numbers PID2019-
108655GB-I00/AEI/10.13039/501100011033 and PID2022-137003NB-I00; EU’s 824093 (STRONG2020);  and Universidad Complutense de Madrid under research group 910309 and the IPARCOS institute. ASB acknowledges the support of the EU's Next Generation funding, grant number CNS2022-135688.

\clearpage
\section{Appendices}

We list here the conventions that, while necessary
for full reproductibility of the work, can be glossed over by most readers.

\subsection{Gamma-matrix and spinor representations}
We follow the conventions of Peskin \& Schroeder\cite{Peskin:1995ev}, Eq.(3.25) and following:
\begin{equation}
\gamma^0 = \begin{pmatrix} 0 & \mathbb{I}_2 \\ \mathbb{I}_2 & 0 \end{pmatrix},\quad \gamma^i = \begin{pmatrix} 0 & \sigma^i \\ -\sigma^i & 0 \end{pmatrix},\quad \gamma^5 = \begin{pmatrix} -\mathbb{I}_2 & 0 \\ 0 & \mathbb{I}_2 \end{pmatrix}.    
\end{equation}
$\sigma=(\mathbb{I}_2,\boldsymbol{\sigma})$ and $\bar{\sigma}=(\mathbb{I}_2,-\boldsymbol{\sigma})$
\begin{equation}
    u^s(p)=\begin{pmatrix}
        \sqrt{p\cdot \sigma} \xi^s\\
        \sqrt{p\cdot \bar{\sigma}} \xi^s
    \end{pmatrix},\;     v^s(p)=\begin{pmatrix}
        \sqrt{p\cdot \sigma} \xi^s\\
        -\sqrt{p\cdot \bar{\sigma}} \xi^s
    \end{pmatrix}
\end{equation}
with 
\begin{equation}
    \xi^+=\begin{pmatrix}
        1\\
        0
    \end{pmatrix},\;     \xi^{-}=\begin{pmatrix}
        0\\
        1
    \end{pmatrix}
\end{equation}
and the spin generators being
\begin{equation} \label{Lorentzgenerators}
\sigma^{0i} = \frac{i}{2} \begin{pmatrix}
-\sigma^i & 0 \\ 0 & \sigma^i
\end{pmatrix}\ .
\end{equation}

\subsection{Color factor for the quark-gluon scattering kernel}
The amputated four-legged function carries $T^a_{ij}T^b_{jk}$ 
that decomposes in an octet piece (that we do not further considered since it is subleading in the skeleton expansion to the three-point function already described at length) and a singlet with a constant $\kappa \delta_{ab}\delta_{ik}$.
This $\kappa$ is determined by projecting with $\frac{\delta^{ab}}{N_c^2-1}$ obtaining
\begin{equation}
\frac{1}{N^2_c -1} T^a_{ij}T^a_{jk}  =  
\kappa \frac{1}{N_c^2-1} \delta^{aa} \delta_{ik}\ .
\end{equation}
The right-hand side is just $\kappa \delta_{ik}$. 
The left-hand side is straightforward to compute from the closure relation
$$
 T^a_{ij}T^a_{jk} = \frac{1}{2} ( \delta_{ik} \delta_{jj}-\frac{1}{N_c} \delta_{ij}\delta_{jk})
$$
so that $\kappa =\frac{1}{2N_c}$.
We then do not need to multiply our skeleton kernel by this factor since 
it does not distinguish the different spin configuration in our computation, it being instead a global factor. It would be necessary if a full model based on the kernel was to be constructed and checked against data. 

\bibliography{references}

\begin{thebibliography}{81}%
\makeatletter
\providecommand \@ifxundefined [1]{%
 \@ifx{#1\undefined}
}%
\providecommand \@ifnum [1]{%
 \ifnum #1\expandafter \@firstoftwo
 \else \expandafter \@secondoftwo
 \fi
}%
\providecommand \@ifx [1]{%
 \ifx #1\expandafter \@firstoftwo
 \else \expandafter \@secondoftwo
 \fi
}%
\providecommand \natexlab [1]{#1}%
\providecommand \enquote  [1]{``#1''}%
\providecommand \bibnamefont  [1]{#1}%
\providecommand \bibfnamefont [1]{#1}%
\providecommand \citenamefont [1]{#1}%
\providecommand \href@noop [0]{\@secondoftwo}%
\providecommand \href [0]{\begingroup \@sanitize@url \@href}%
\providecommand \@href[1]{\@@startlink{#1}\@@href}%
\providecommand \@@href[1]{\endgroup#1\@@endlink}%
\providecommand \@sanitize@url [0]{\catcode `\\12\catcode `\$12\catcode
  `\&12\catcode `\#12\catcode `\^12\catcode `\_12\catcode `\%12\relax}%
\providecommand \@@startlink[1]{}%
\providecommand \@@endlink[0]{}%
\providecommand \url  [0]{\begingroup\@sanitize@url \@url }%
\providecommand \@url [1]{\endgroup\@href {#1}{\urlprefix }}%
\providecommand \urlprefix  [0]{URL }%
\providecommand \Eprint [0]{\href }%
\providecommand \doibase [0]{http://dx.doi.org/}%
\providecommand \selectlanguage [0]{\@gobble}%
\providecommand \bibinfo  [0]{\@secondoftwo}%
\providecommand \bibfield  [0]{\@secondoftwo}%
\providecommand \translation [1]{[#1]}%
\providecommand \BibitemOpen [0]{}%
\providecommand \bibitemStop [0]{}%
\providecommand \bibitemNoStop [0]{.\EOS\space}%
\providecommand \EOS [0]{\spacefactor3000\relax}%
\providecommand \BibitemShut  [1]{\csname bibitem#1\endcsname}%
\let\auto@bib@innerbib\@empty
\bibitem [{\citenamefont {Sauter}(1931)}]{Sauter:1931zz}%
  \BibitemOpen
  \bibfield  {author} {\bibinfo {author} {\bibfnamefont {F.}~\bibnamefont
  {Sauter}},\ }\href {\doibase 10.1007/BF01339461} {\bibfield  {journal}
  {\bibinfo  {journal} {Z. Phys.}\ }\textbf {\bibinfo {volume} {69}},\ \bibinfo
  {pages} {742} (\bibinfo {year} {1931})}\BibitemShut {NoStop}%
\bibitem [{\citenamefont {Schwinger}(1951)}]{Schwinger:1951nm}%
  \BibitemOpen
  \bibfield  {author} {\bibinfo {author} {\bibfnamefont {J.~S.}\ \bibnamefont
  {Schwinger}},\ }\href {\doibase 10.1103/PhysRev.82.664} {\bibfield  {journal}
  {\bibinfo  {journal} {Phys. Rev.}\ }\textbf {\bibinfo {volume} {82}},\
  \bibinfo {pages} {664} (\bibinfo {year} {1951})}\BibitemShut {NoStop}%
\bibitem [{\citenamefont {Fedotov}\ \emph {et~al.}(2023)\citenamefont
  {Fedotov}, \citenamefont {Ilderton}, \citenamefont {Karbstein}, \citenamefont
  {King}, \citenamefont {Seipt}, \citenamefont {Taya},\ and\ \citenamefont
  {Torgrimsson}}]{Fedotov:2022ely}%
  \BibitemOpen
  \bibfield  {author} {\bibinfo {author} {\bibfnamefont {A.}~\bibnamefont
  {Fedotov}}, \bibinfo {author} {\bibfnamefont {A.}~\bibnamefont {Ilderton}},
  \bibinfo {author} {\bibfnamefont {F.}~\bibnamefont {Karbstein}}, \bibinfo
  {author} {\bibfnamefont {B.}~\bibnamefont {King}}, \bibinfo {author}
  {\bibfnamefont {D.}~\bibnamefont {Seipt}}, \bibinfo {author} {\bibfnamefont
  {H.}~\bibnamefont {Taya}}, \ and\ \bibinfo {author} {\bibfnamefont
  {G.}~\bibnamefont {Torgrimsson}},\ }\href {\doibase
  10.1016/j.physrep.2023.01.003} {\bibfield  {journal} {\bibinfo  {journal}
  {Phys. Rept.}\ }\textbf {\bibinfo {volume} {1010}},\ \bibinfo {pages} {1}
  (\bibinfo {year} {2023})},\ \Eprint {http://arxiv.org/abs/2203.00019}
  {arXiv:2203.00019 [hep-ph]} \BibitemShut {NoStop}%
\bibitem [{\citenamefont {Kohlf\"urst}(2019)}]{Kohlfurst:2018kxg}%
  \BibitemOpen
  \bibfield  {author} {\bibinfo {author} {\bibfnamefont {C.}~\bibnamefont
  {Kohlf\"urst}},\ }\href {\doibase 10.1103/PhysRevD.99.096017} {\bibfield
  {journal} {\bibinfo  {journal} {Phys. Rev. D}\ }\textbf {\bibinfo {volume}
  {99}},\ \bibinfo {pages} {096017} (\bibinfo {year} {2019})},\ \Eprint
  {http://arxiv.org/abs/1812.03130} {arXiv:1812.03130 [hep-ph]} \BibitemShut
  {NoStop}%
\bibitem [{\citenamefont
  {Kohlf\"urst}(2022{\natexlab{a}})}]{Kohlfurst:2022edl}%
  \BibitemOpen
  \bibfield  {author} {\bibinfo {author} {\bibfnamefont {C.}~\bibnamefont
  {Kohlf\"urst}},\ }\href@noop {} {\  (\bibinfo {year} {2022}{\natexlab{a}})},\
  \Eprint {http://arxiv.org/abs/2212.03180} {arXiv:2212.03180 [hep-ph]}
  \BibitemShut {NoStop}%
\bibitem [{\citenamefont {\'Alvarez-Dom\'\i{}nguez}\ \emph
  {et~al.}(2023)\citenamefont {\'Alvarez-Dom\'\i{}nguez}, \citenamefont
  {Garay}, \citenamefont {Mart\'\i{}n-Benito},\ and\ \citenamefont
  {Neves}}]{Alvarez-Dominguez:2023ten}%
  \BibitemOpen
  \bibfield  {author} {\bibinfo {author} {\bibfnamefont {A.}~\bibnamefont
  {\'Alvarez-Dom\'\i{}nguez}}, \bibinfo {author} {\bibfnamefont {L.~J.}\
  \bibnamefont {Garay}}, \bibinfo {author} {\bibfnamefont {M.}~\bibnamefont
  {Mart\'\i{}n-Benito}}, \ and\ \bibinfo {author} {\bibfnamefont {R.~B.}\
  \bibnamefont {Neves}},\ }\href {\doibase 10.1007/JHEP06(2023)093} {\bibfield
  {journal} {\bibinfo  {journal} {JHEP}\ }\textbf {\bibinfo {volume} {06}},\
  \bibinfo {pages} {093} (\bibinfo {year} {2023})},\ \Eprint
  {http://arxiv.org/abs/2303.15294} {arXiv:2303.15294 [hep-th]} \BibitemShut
  {NoStop}%
\bibitem [{\citenamefont {Copinger}\ and\ \citenamefont
  {Hidaka}(2022)}]{Copinger:2022gfz}%
  \BibitemOpen
  \bibfield  {author} {\bibinfo {author} {\bibfnamefont {P.}~\bibnamefont
  {Copinger}}\ and\ \bibinfo {author} {\bibfnamefont {Y.}~\bibnamefont
  {Hidaka}},\ }\href@noop {} {\  (\bibinfo {year} {2022})},\ \Eprint
  {http://arxiv.org/abs/2203.10917} {arXiv:2203.10917 [hep-ph]} \BibitemShut
  {NoStop}%
\bibitem [{\citenamefont
  {Kohlf\"urst}(2022{\natexlab{b}})}]{Kohlfurst:2022vwf}%
  \BibitemOpen
  \bibfield  {author} {\bibinfo {author} {\bibfnamefont {C.}~\bibnamefont
  {Kohlf\"urst}},\ }\href@noop {} {\  (\bibinfo {year} {2022}{\natexlab{b}})},\
  \Eprint {http://arxiv.org/abs/2212.06057} {arXiv:2212.06057 [hep-ph]}
  \BibitemShut {NoStop}%
\bibitem [{\citenamefont {Le~Yaouanc}\ \emph {et~al.}(1988)\citenamefont
  {Le~Yaouanc}, \citenamefont {Oliver}, \citenamefont {Pene},\ and\
  \citenamefont {Raynal}}]{LeYaouanc:1988fx}%
  \BibitemOpen
  \bibfield  {author} {\bibinfo {author} {\bibfnamefont {A.}~\bibnamefont
  {Le~Yaouanc}}, \bibinfo {author} {\bibfnamefont {L.}~\bibnamefont {Oliver}},
  \bibinfo {author} {\bibfnamefont {O.}~\bibnamefont {Pene}}, \ and\ \bibinfo
  {author} {\bibfnamefont {J.~C.}\ \bibnamefont {Raynal}},\ }\href@noop {}
  {\emph {\bibinfo {title} {{HADRON TRANSITIONS IN THE QUARK MODEL}}}}\
  (\bibinfo {year} {1988})\BibitemShut {NoStop}%
\bibitem [{\citenamefont {Roberts}\ and\ \citenamefont
  {Silvestre-Brac}(1992)}]{Roberts:1992esl}%
  \BibitemOpen
  \bibfield  {author} {\bibinfo {author} {\bibfnamefont {W.}~\bibnamefont
  {Roberts}}\ and\ \bibinfo {author} {\bibfnamefont {B.}~\bibnamefont
  {Silvestre-Brac}},\ }\href {\doibase 10.1007/bf01641821} {\bibfield
  {journal} {\bibinfo  {journal} {Few Body Syst.}\ }\textbf {\bibinfo {volume}
  {11}},\ \bibinfo {pages} {171} (\bibinfo {year} {1992})}\BibitemShut
  {NoStop}%
\bibitem [{\citenamefont {Close}\ and\ \citenamefont
  {Swanson}(2005)}]{Close:2005se}%
  \BibitemOpen
  \bibfield  {author} {\bibinfo {author} {\bibfnamefont {F.~E.}\ \bibnamefont
  {Close}}\ and\ \bibinfo {author} {\bibfnamefont {E.~S.}\ \bibnamefont
  {Swanson}},\ }\href {\doibase 10.1103/PhysRevD.72.094004} {\bibfield
  {journal} {\bibinfo  {journal} {Phys. Rev. D}\ }\textbf {\bibinfo {volume}
  {72}},\ \bibinfo {pages} {094004} (\bibinfo {year} {2005})},\ \Eprint
  {http://arxiv.org/abs/hep-ph/0505206} {arXiv:hep-ph/0505206} \BibitemShut
  {NoStop}%
\bibitem [{\citenamefont {Swanson}(2006)}]{Swanson:2006st}%
  \BibitemOpen
  \bibfield  {author} {\bibinfo {author} {\bibfnamefont {E.~S.}\ \bibnamefont
  {Swanson}},\ }\href {\doibase 10.1016/j.physrep.2006.04.003} {\bibfield
  {journal} {\bibinfo  {journal} {Phys. Rept.}\ }\textbf {\bibinfo {volume}
  {429}},\ \bibinfo {pages} {243} (\bibinfo {year} {2006})},\ \Eprint
  {http://arxiv.org/abs/hep-ph/0601110} {arXiv:hep-ph/0601110} \BibitemShut
  {NoStop}%
\bibitem [{\citenamefont {Micu}(1969)}]{Micu:1968mk}%
  \BibitemOpen
  \bibfield  {author} {\bibinfo {author} {\bibfnamefont {L.}~\bibnamefont
  {Micu}},\ }\href {\doibase 10.1016/0550-3213(69)90039-X} {\bibfield
  {journal} {\bibinfo  {journal} {Nucl. Phys. B}\ }\textbf {\bibinfo {volume}
  {10}},\ \bibinfo {pages} {521} (\bibinfo {year} {1969})}\BibitemShut
  {NoStop}%
\bibitem [{\citenamefont {Segovia}\ \emph {et~al.}(2012)\citenamefont
  {Segovia}, \citenamefont {Entem},\ and\ \citenamefont
  {Fern\'andez}}]{Segovia:2012cd}%
  \BibitemOpen
  \bibfield  {author} {\bibinfo {author} {\bibfnamefont {J.}~\bibnamefont
  {Segovia}}, \bibinfo {author} {\bibfnamefont {D.~R.}\ \bibnamefont {Entem}},
  \ and\ \bibinfo {author} {\bibfnamefont {F.}~\bibnamefont {Fern\'andez}},\
  }\href {\doibase 10.1016/j.physletb.2012.08.005} {\bibfield  {journal}
  {\bibinfo  {journal} {Phys. Lett. B}\ }\textbf {\bibinfo {volume} {715}},\
  \bibinfo {pages} {322} (\bibinfo {year} {2012})},\ \Eprint
  {http://arxiv.org/abs/1205.2215} {arXiv:1205.2215 [hep-ph]} \BibitemShut
  {NoStop}%
\bibitem [{\citenamefont {Ackleh}\ \emph {et~al.}(1996)\citenamefont {Ackleh},
  \citenamefont {Barnes},\ and\ \citenamefont {Swanson}}]{Ackleh:1996yt}%
  \BibitemOpen
  \bibfield  {author} {\bibinfo {author} {\bibfnamefont {E.~S.}\ \bibnamefont
  {Ackleh}}, \bibinfo {author} {\bibfnamefont {T.}~\bibnamefont {Barnes}}, \
  and\ \bibinfo {author} {\bibfnamefont {E.~S.}\ \bibnamefont {Swanson}},\
  }\href {\doibase 10.1103/PhysRevD.54.6811} {\bibfield  {journal} {\bibinfo
  {journal} {Phys. Rev. D}\ }\textbf {\bibinfo {volume} {54}},\ \bibinfo
  {pages} {6811} (\bibinfo {year} {1996})},\ \Eprint
  {http://arxiv.org/abs/hep-ph/9604355} {arXiv:hep-ph/9604355} \BibitemShut
  {NoStop}%
\bibitem [{\citenamefont {Brambilla}\ \emph {et~al.}(2000)\citenamefont
  {Brambilla}, \citenamefont {Pineda}, \citenamefont {Soto},\ and\
  \citenamefont {Vairo}}]{Brambilla:1999xf}%
  \BibitemOpen
  \bibfield  {author} {\bibinfo {author} {\bibfnamefont {N.}~\bibnamefont
  {Brambilla}}, \bibinfo {author} {\bibfnamefont {A.}~\bibnamefont {Pineda}},
  \bibinfo {author} {\bibfnamefont {J.}~\bibnamefont {Soto}}, \ and\ \bibinfo
  {author} {\bibfnamefont {A.}~\bibnamefont {Vairo}},\ }\href {\doibase
  10.1016/S0550-3213(99)00693-8} {\bibfield  {journal} {\bibinfo  {journal}
  {Nucl. Phys. B}\ }\textbf {\bibinfo {volume} {566}},\ \bibinfo {pages} {275}
  (\bibinfo {year} {2000})},\ \Eprint {http://arxiv.org/abs/hep-ph/9907240}
  {arXiv:hep-ph/9907240} \BibitemShut {NoStop}%
\bibitem [{\citenamefont {Williams}\ \emph {et~al.}(2016)\citenamefont
  {Williams}, \citenamefont {Fischer},\ and\ \citenamefont
  {Heupel}}]{Williams:2015cvx}%
  \BibitemOpen
  \bibfield  {author} {\bibinfo {author} {\bibfnamefont {R.}~\bibnamefont
  {Williams}}, \bibinfo {author} {\bibfnamefont {C.~S.}\ \bibnamefont
  {Fischer}}, \ and\ \bibinfo {author} {\bibfnamefont {W.}~\bibnamefont
  {Heupel}},\ }\href {\doibase 10.1103/PhysRevD.93.034026} {\bibfield
  {journal} {\bibinfo  {journal} {Phys. Rev. D}\ }\textbf {\bibinfo {volume}
  {93}},\ \bibinfo {pages} {034026} (\bibinfo {year} {2016})},\ \Eprint
  {http://arxiv.org/abs/1512.00455} {arXiv:1512.00455 [hep-ph]} \BibitemShut
  {NoStop}%
\bibitem [{\citenamefont {Cyrol}\ \emph {et~al.}(2016)\citenamefont {Cyrol},
  \citenamefont {Fister}, \citenamefont {Mitter}, \citenamefont {Pawlowski},\
  and\ \citenamefont {Strodthoff}}]{Cyrol:2016tym}%
  \BibitemOpen
  \bibfield  {author} {\bibinfo {author} {\bibfnamefont {A.~K.}\ \bibnamefont
  {Cyrol}}, \bibinfo {author} {\bibfnamefont {L.}~\bibnamefont {Fister}},
  \bibinfo {author} {\bibfnamefont {M.}~\bibnamefont {Mitter}}, \bibinfo
  {author} {\bibfnamefont {J.~M.}\ \bibnamefont {Pawlowski}}, \ and\ \bibinfo
  {author} {\bibfnamefont {N.}~\bibnamefont {Strodthoff}},\ }\href {\doibase
  10.1103/PhysRevD.94.054005} {\bibfield  {journal} {\bibinfo  {journal} {Phys.
  Rev. D}\ }\textbf {\bibinfo {volume} {94}},\ \bibinfo {pages} {054005}
  (\bibinfo {year} {2016})},\ \Eprint {http://arxiv.org/abs/1605.01856}
  {arXiv:1605.01856 [hep-ph]} \BibitemShut {NoStop}%
\bibitem [{\citenamefont {Ferreira}\ and\ \citenamefont
  {Papavassiliou}(2023)}]{Ferreira:2023fva}%
  \BibitemOpen
  \bibfield  {author} {\bibinfo {author} {\bibfnamefont {M.~N.}\ \bibnamefont
  {Ferreira}}\ and\ \bibinfo {author} {\bibfnamefont {J.}~\bibnamefont
  {Papavassiliou}},\ }\href {\doibase 10.3390/particles6010017} {\bibfield
  {journal} {\bibinfo  {journal} {Particles}\ }\textbf {\bibinfo {volume}
  {6}},\ \bibinfo {pages} {312} (\bibinfo {year} {2023})},\ \Eprint
  {http://arxiv.org/abs/2301.02314} {arXiv:2301.02314 [hep-ph]} \BibitemShut
  {NoStop}%
\bibitem [{\citenamefont {Gao}\ \emph {et~al.}(2021)\citenamefont {Gao},
  \citenamefont {Papavassiliou},\ and\ \citenamefont
  {Pawlowski}}]{Gao:2021wun}%
  \BibitemOpen
  \bibfield  {author} {\bibinfo {author} {\bibfnamefont {F.}~\bibnamefont
  {Gao}}, \bibinfo {author} {\bibfnamefont {J.}~\bibnamefont {Papavassiliou}},
  \ and\ \bibinfo {author} {\bibfnamefont {J.~M.}\ \bibnamefont {Pawlowski}},\
  }\href {\doibase 10.1103/PhysRevD.103.094013} {\bibfield  {journal} {\bibinfo
   {journal} {Phys. Rev. D}\ }\textbf {\bibinfo {volume} {103}},\ \bibinfo
  {pages} {094013} (\bibinfo {year} {2021})},\ \Eprint
  {http://arxiv.org/abs/2102.13053} {arXiv:2102.13053 [hep-ph]} \BibitemShut
  {NoStop}%
\bibitem [{\citenamefont {Bicudo}\ \emph {et~al.}(2016)\citenamefont {Bicudo},
  \citenamefont {Cardoso}, \citenamefont {Llanes-Estrada},\ and\ \citenamefont
  {Van~Cauteren}}]{Bicudo:2016eeu}%
  \BibitemOpen
  \bibfield  {author} {\bibinfo {author} {\bibfnamefont {P.}~\bibnamefont
  {Bicudo}}, \bibinfo {author} {\bibfnamefont {M.}~\bibnamefont {Cardoso}},
  \bibinfo {author} {\bibfnamefont {F.~J.}\ \bibnamefont {Llanes-Estrada}}, \
  and\ \bibinfo {author} {\bibfnamefont {T.}~\bibnamefont {Van~Cauteren}},\
  }\href {\doibase 10.1103/PhysRevD.94.054006} {\bibfield  {journal} {\bibinfo
  {journal} {Phys. Rev. D}\ }\textbf {\bibinfo {volume} {94}},\ \bibinfo
  {pages} {054006} (\bibinfo {year} {2016})},\ \Eprint
  {http://arxiv.org/abs/1605.05171} {arXiv:1605.05171 [hep-ph]} \BibitemShut
  {NoStop}%
\bibitem [{\citenamefont {Kerbizi}\ \emph {et~al.}(2021)\citenamefont
  {Kerbizi}, \citenamefont {Artru},\ and\ \citenamefont
  {Martin}}]{Kerbizi:2021gos}%
  \BibitemOpen
  \bibfield  {author} {\bibinfo {author} {\bibfnamefont {A.}~\bibnamefont
  {Kerbizi}}, \bibinfo {author} {\bibfnamefont {X.}~\bibnamefont {Artru}}, \
  and\ \bibinfo {author} {\bibfnamefont {A.}~\bibnamefont {Martin}},\ }\href
  {\doibase 10.1103/PhysRevD.104.114038} {\bibfield  {journal} {\bibinfo
  {journal} {Phys. Rev. D}\ }\textbf {\bibinfo {volume} {104}},\ \bibinfo
  {pages} {114038} (\bibinfo {year} {2021})},\ \Eprint
  {http://arxiv.org/abs/2109.06124} {arXiv:2109.06124 [hep-ph]} \BibitemShut
  {NoStop}%
\bibitem [{\citenamefont {Artru}\ and\ \citenamefont
  {Kerbizi}(2022)}]{Artru:2022vqf}%
  \BibitemOpen
  \bibfield  {author} {\bibinfo {author} {\bibfnamefont {X.}~\bibnamefont
  {Artru}}\ and\ \bibinfo {author} {\bibfnamefont {A.}~\bibnamefont
  {Kerbizi}},\ }in\ \href@noop {} {\emph {\bibinfo {booktitle} {{24th
  International Symposium on Spin Physics}}}}\ (\bibinfo {year} {2022})\
  \Eprint {http://arxiv.org/abs/2201.05509} {arXiv:2201.05509 [hep-ph]}
  \BibitemShut {NoStop}%
\bibitem [{\citenamefont {Cao}\ and\ \citenamefont
  {Huang}(2016)}]{Cao:2015dya}%
  \BibitemOpen
  \bibfield  {author} {\bibinfo {author} {\bibfnamefont {G.}~\bibnamefont
  {Cao}}\ and\ \bibinfo {author} {\bibfnamefont {X.-G.}\ \bibnamefont
  {Huang}},\ }\href {\doibase 10.1103/PhysRevD.93.016007} {\bibfield  {journal}
  {\bibinfo  {journal} {Phys. Rev. D}\ }\textbf {\bibinfo {volume} {93}},\
  \bibinfo {pages} {016007} (\bibinfo {year} {2016})},\ \Eprint
  {http://arxiv.org/abs/1510.05125} {arXiv:1510.05125 [nucl-th]} \BibitemShut
  {NoStop}%
\bibitem [{\citenamefont {Mahlin}\ \emph {et~al.}(2023)\citenamefont {Mahlin},
  \citenamefont {Villalba-Ch\'avez},\ and\ \citenamefont
  {M\"uller}}]{Mahlin:2023aui}%
  \BibitemOpen
  \bibfield  {author} {\bibinfo {author} {\bibfnamefont {N.}~\bibnamefont
  {Mahlin}}, \bibinfo {author} {\bibfnamefont {S.}~\bibnamefont
  {Villalba-Ch\'avez}}, \ and\ \bibinfo {author} {\bibfnamefont
  {C.}~\bibnamefont {M\"uller}},\ }\href {\doibase 10.1103/PhysRevD.108.096023}
  {\bibfield  {journal} {\bibinfo  {journal} {Phys. Rev. D}\ }\textbf {\bibinfo
  {volume} {108}},\ \bibinfo {pages} {096023} (\bibinfo {year} {2023})},\
  \Eprint {http://arxiv.org/abs/2306.02735} {arXiv:2306.02735 [hep-ph]}
  \BibitemShut {NoStop}%
\bibitem [{\citenamefont {Roshchupkin}\ \emph {et~al.}(2023)\citenamefont
  {Roshchupkin}, \citenamefont {Serov},\ and\ \citenamefont
  {Dubov}}]{Roshchupkin:2023hbt}%
  \BibitemOpen
  \bibfield  {author} {\bibinfo {author} {\bibfnamefont {S.~P.}\ \bibnamefont
  {Roshchupkin}}, \bibinfo {author} {\bibfnamefont {V.~D.}\ \bibnamefont
  {Serov}}, \ and\ \bibinfo {author} {\bibfnamefont {V.~V.}\ \bibnamefont
  {Dubov}},\ }\href {\doibase 10.3390/sym15101901} {\bibfield  {journal}
  {\bibinfo  {journal} {Symmetry}\ }\textbf {\bibinfo {volume} {15}},\ \bibinfo
  {pages} {1901} (\bibinfo {year} {2023})},\ \Eprint
  {http://arxiv.org/abs/2308.02520} {arXiv:2308.02520 [hep-ph]} \BibitemShut
  {NoStop}%
\bibitem [{\citenamefont {Cabral}\ \emph {et~al.}(2023)\citenamefont {Cabral},
  \citenamefont {Santos},\ and\ \citenamefont {Bufalo}}]{Cabral:2023rgs}%
  \BibitemOpen
  \bibfield  {author} {\bibinfo {author} {\bibfnamefont {D.~S.}\ \bibnamefont
  {Cabral}}, \bibinfo {author} {\bibfnamefont {A.~F.}\ \bibnamefont {Santos}},
  \ and\ \bibinfo {author} {\bibfnamefont {R.}~\bibnamefont {Bufalo}},\ }\href
  {\doibase 10.1140/epjc/s10052-023-12281-5} {\bibfield  {journal} {\bibinfo
  {journal} {Eur. Phys. J. C}\ }\textbf {\bibinfo {volume} {83}},\ \bibinfo
  {pages} {1113} (\bibinfo {year} {2023})},\ \Eprint
  {http://arxiv.org/abs/2311.13376} {arXiv:2311.13376 [hep-ph]} \BibitemShut
  {NoStop}%
\bibitem [{\citenamefont {Dyson}(1949)}]{Dyson:1949ha}%
  \BibitemOpen
  \bibfield  {author} {\bibinfo {author} {\bibfnamefont {F.~J.}\ \bibnamefont
  {Dyson}},\ }\href {\doibase 10.1103/PhysRev.75.1736} {\bibfield  {journal}
  {\bibinfo  {journal} {Phys. Rev.}\ }\textbf {\bibinfo {volume} {75}},\
  \bibinfo {pages} {1736} (\bibinfo {year} {1949})}\BibitemShut {NoStop}%
\bibitem [{\citenamefont {Lu}(1992)}]{Lu:1992eq}%
  \BibitemOpen
  \bibfield  {author} {\bibinfo {author} {\bibfnamefont {H.~J.}\ \bibnamefont
  {Lu}},\ }\emph {\bibinfo {title} {{Dressed skeleton expansion and the
  coupling scale ambiguity problem}}},\ \href@noop {} {Ph.D. thesis},\ \bibinfo
   {school} {Stanford U.} (\bibinfo {year} {1992})\BibitemShut {NoStop}%
\bibitem [{\citenamefont {Le~Yaouanc}\ \emph {et~al.}(1973)\citenamefont
  {Le~Yaouanc}, \citenamefont {Oliver}, \citenamefont {Pene},\ and\
  \citenamefont {Raynal}}]{LeYaouanc:1972vsx}%
  \BibitemOpen
  \bibfield  {author} {\bibinfo {author} {\bibfnamefont {A.}~\bibnamefont
  {Le~Yaouanc}}, \bibinfo {author} {\bibfnamefont {L.}~\bibnamefont {Oliver}},
  \bibinfo {author} {\bibfnamefont {O.}~\bibnamefont {Pene}}, \ and\ \bibinfo
  {author} {\bibfnamefont {J.~C.}\ \bibnamefont {Raynal}},\ }\href {\doibase
  10.1103/PhysRevD.8.2223} {\bibfield  {journal} {\bibinfo  {journal} {Phys.
  Rev. D}\ }\textbf {\bibinfo {volume} {8}},\ \bibinfo {pages} {2223} (\bibinfo
  {year} {1973})}\BibitemShut {NoStop}%
\bibitem [{\citenamefont {Llanes-Estrada}(2021)}]{Llanes-Estrada:2021evz}%
  \BibitemOpen
  \bibfield  {author} {\bibinfo {author} {\bibfnamefont {F.~J.}\ \bibnamefont
  {Llanes-Estrada}},\ }\href {\doibase 10.1140/epjs/s11734-021-00143-8}
  {\bibfield  {journal} {\bibinfo  {journal} {Eur. Phys. J. ST}\ }\textbf
  {\bibinfo {volume} {230}},\ \bibinfo {pages} {1575} (\bibinfo {year}
  {2021})},\ \Eprint {http://arxiv.org/abs/2101.05366} {arXiv:2101.05366
  [hep-ph]} \BibitemShut {NoStop}%
\bibitem [{\citenamefont {Abreu}\ \emph {et~al.}(2019)\citenamefont {Abreu},
  \citenamefont {Favero}, \citenamefont {Llanes-Estrada},\ and\ \citenamefont
  {S\'anchez}}]{Abreu:2019adi}%
  \BibitemOpen
  \bibfield  {author} {\bibinfo {author} {\bibfnamefont {L.~M.}\ \bibnamefont
  {Abreu}}, \bibinfo {author} {\bibfnamefont {A.~G.}\ \bibnamefont {Favero}},
  \bibinfo {author} {\bibfnamefont {F.~J.}\ \bibnamefont {Llanes-Estrada}}, \
  and\ \bibinfo {author} {\bibfnamefont {A.~G.}\ \bibnamefont {S\'anchez}},\
  }\href {\doibase 10.1103/PhysRevD.100.116012} {\bibfield  {journal} {\bibinfo
   {journal} {Phys. Rev. D}\ }\textbf {\bibinfo {volume} {100}},\ \bibinfo
  {pages} {116012} (\bibinfo {year} {2019})},\ \Eprint
  {http://arxiv.org/abs/1908.11154} {arXiv:1908.11154 [hep-ph]} \BibitemShut
  {NoStop}%
\bibitem [{\citenamefont {Est\'evez}\ \emph {et~al.}(2020)\citenamefont
  {Est\'evez} \emph {et~al.}}]{Estevez:2020vsm}%
  \BibitemOpen
  \bibfield  {author} {\bibinfo {author} {\bibfnamefont {J.}~\bibnamefont
  {Est\'evez}} \emph {et~al.},\ }\href {\doibase 10.13140/RG.2.2.36673.38244}
  {\bibfield  {journal} {\bibinfo  {journal} {Phys. Rev. D}\ }\textbf {\bibinfo
  {volume} {102}},\ \bibinfo {pages} {114032} (\bibinfo {year} {2020})},\
  \Eprint {http://arxiv.org/abs/2009.11020} {arXiv:2009.11020 [hep-ph]}
  \BibitemShut {NoStop}%
\bibitem [{\citenamefont {Hebenstreit}\ \emph {et~al.}(2010)\citenamefont
  {Hebenstreit}, \citenamefont {Alkofer},\ and\ \citenamefont
  {Gies}}]{Hebenstreit:2010vz}%
  \BibitemOpen
  \bibfield  {author} {\bibinfo {author} {\bibfnamefont {F.}~\bibnamefont
  {Hebenstreit}}, \bibinfo {author} {\bibfnamefont {R.}~\bibnamefont
  {Alkofer}}, \ and\ \bibinfo {author} {\bibfnamefont {H.}~\bibnamefont
  {Gies}},\ }\href {\doibase 10.1103/PhysRevD.82.105026} {\bibfield  {journal}
  {\bibinfo  {journal} {Phys. Rev. D}\ }\textbf {\bibinfo {volume} {82}},\
  \bibinfo {pages} {105026} (\bibinfo {year} {2010})},\ \Eprint
  {http://arxiv.org/abs/1007.1099} {arXiv:1007.1099 [hep-ph]} \BibitemShut
  {NoStop}%
\bibitem [{\citenamefont {Titov}\ and\ \citenamefont
  {Kampfer}(2020)}]{Titov:2020taw}%
  \BibitemOpen
  \bibfield  {author} {\bibinfo {author} {\bibfnamefont {A.~I.}\ \bibnamefont
  {Titov}}\ and\ \bibinfo {author} {\bibfnamefont {B.}~\bibnamefont
  {Kampfer}},\ }\href {\doibase 10.1140/epjd/e2020-10327-9} {\bibfield
  {journal} {\bibinfo  {journal} {Eur. Phys. J. D}\ }\textbf {\bibinfo {volume}
  {74}},\ \bibinfo {pages} {218} (\bibinfo {year} {2020})},\ \Eprint
  {http://arxiv.org/abs/2006.04496} {arXiv:2006.04496 [hep-ph]} \BibitemShut
  {NoStop}%
\bibitem [{\citenamefont {Borysov}\ \emph {et~al.}(2022)\citenamefont
  {Borysov}, \citenamefont {Heinemann}, \citenamefont {Ilderton}, \citenamefont
  {King},\ and\ \citenamefont {Potylitsyn}}]{Borysov:2022cwc}%
  \BibitemOpen
  \bibfield  {author} {\bibinfo {author} {\bibfnamefont {O.}~\bibnamefont
  {Borysov}}, \bibinfo {author} {\bibfnamefont {B.}~\bibnamefont {Heinemann}},
  \bibinfo {author} {\bibfnamefont {A.}~\bibnamefont {Ilderton}}, \bibinfo
  {author} {\bibfnamefont {B.}~\bibnamefont {King}}, \ and\ \bibinfo {author}
  {\bibfnamefont {A.}~\bibnamefont {Potylitsyn}},\ }\href {\doibase
  10.1103/PhysRevD.106.116015} {\bibfield  {journal} {\bibinfo  {journal}
  {Phys. Rev. D}\ }\textbf {\bibinfo {volume} {106}},\ \bibinfo {pages}
  {116015} (\bibinfo {year} {2022})},\ \Eprint
  {http://arxiv.org/abs/2209.12908} {arXiv:2209.12908 [hep-ph]} \BibitemShut
  {NoStop}%
\bibitem [{\citenamefont {Golub}\ \emph {et~al.}(2022)\citenamefont {Golub},
  \citenamefont {Villalba-Ch\'avez},\ and\ \citenamefont
  {M\"uller}}]{Golub:2022cvd}%
  \BibitemOpen
  \bibfield  {author} {\bibinfo {author} {\bibfnamefont {A.}~\bibnamefont
  {Golub}}, \bibinfo {author} {\bibfnamefont {S.}~\bibnamefont
  {Villalba-Ch\'avez}}, \ and\ \bibinfo {author} {\bibfnamefont
  {C.}~\bibnamefont {M\"uller}},\ }\href {\doibase 10.1103/PhysRevD.105.116016}
  {\bibfield  {journal} {\bibinfo  {journal} {Phys. Rev. D}\ }\textbf {\bibinfo
  {volume} {105}},\ \bibinfo {pages} {116016} (\bibinfo {year} {2022})},\
  \Eprint {http://arxiv.org/abs/2203.14776} {arXiv:2203.14776 [hep-ph]}
  \BibitemShut {NoStop}%
\bibitem [{\citenamefont {Diez}\ \emph {et~al.}(2023)\citenamefont {Diez},
  \citenamefont {Alkofer},\ and\ \citenamefont {Kohlf\"urst}}]{Diez:2022ywi}%
  \BibitemOpen
  \bibfield  {author} {\bibinfo {author} {\bibfnamefont {M.}~\bibnamefont
  {Diez}}, \bibinfo {author} {\bibfnamefont {R.}~\bibnamefont {Alkofer}}, \
  and\ \bibinfo {author} {\bibfnamefont {C.}~\bibnamefont {Kohlf\"urst}},\
  }\href {\doibase 10.1016/j.physletb.2023.138063} {\bibfield  {journal}
  {\bibinfo  {journal} {Phys. Lett. B}\ }\textbf {\bibinfo {volume} {844}},\
  \bibinfo {pages} {138063} (\bibinfo {year} {2023})},\ \Eprint
  {http://arxiv.org/abs/2211.07510} {arXiv:2211.07510 [hep-ph]} \BibitemShut
  {NoStop}%
\bibitem [{\citenamefont {Peskin}\ and\ \citenamefont
  {Schroeder}(1995)}]{Peskin:1995ev}%
  \BibitemOpen
  \bibfield  {author} {\bibinfo {author} {\bibfnamefont {M.~E.}\ \bibnamefont
  {Peskin}}\ and\ \bibinfo {author} {\bibfnamefont {D.~V.}\ \bibnamefont
  {Schroeder}},\ }\href@noop {} {\emph {\bibinfo {title} {{An Introduction to
  quantum field theory}}}}\ (\bibinfo  {publisher} {Addison-Wesley},\ \bibinfo
  {address} {Reading, USA},\ \bibinfo {year} {1995})\BibitemShut {NoStop}%
\bibitem [{\citenamefont {Itzykson}\ and\ \citenamefont
  {Zuber}(1980)}]{Itzykson:1980rh}%
  \BibitemOpen
  \bibfield  {author} {\bibinfo {author} {\bibfnamefont {C.}~\bibnamefont
  {Itzykson}}\ and\ \bibinfo {author} {\bibfnamefont {J.~B.}\ \bibnamefont
  {Zuber}},\ }\href@noop {} {\emph {\bibinfo {title} {{Quantum Field
  Theory}}}},\ International Series In Pure and Applied Physics\ (\bibinfo
  {publisher} {McGraw-Hill},\ \bibinfo {address} {New York},\ \bibinfo {year}
  {1980})\BibitemShut {NoStop}%
\bibitem [{\citenamefont {Alkofer}\ and\ \citenamefont
  {Greensite}(2007)}]{Alkofer:2006fu}%
  \BibitemOpen
  \bibfield  {author} {\bibinfo {author} {\bibfnamefont {R.}~\bibnamefont
  {Alkofer}}\ and\ \bibinfo {author} {\bibfnamefont {J.}~\bibnamefont
  {Greensite}},\ }\href {\doibase 10.1088/0954-3899/34/7/S02} {\bibfield
  {journal} {\bibinfo  {journal} {J. Phys. G}\ }\textbf {\bibinfo {volume}
  {34}},\ \bibinfo {pages} {S3} (\bibinfo {year} {2007})},\ \Eprint
  {http://arxiv.org/abs/hep-ph/0610365} {arXiv:hep-ph/0610365} \BibitemShut
  {NoStop}%
\bibitem [{\citenamefont {Bowman}\ and\ \citenamefont
  {Szczepaniak}(2004)}]{Bowman:2004xd}%
  \BibitemOpen
  \bibfield  {author} {\bibinfo {author} {\bibfnamefont {P.~O.}\ \bibnamefont
  {Bowman}}\ and\ \bibinfo {author} {\bibfnamefont {A.~P.}\ \bibnamefont
  {Szczepaniak}},\ }\href {\doibase 10.1103/PhysRevD.70.016002} {\bibfield
  {journal} {\bibinfo  {journal} {Phys. Rev. D}\ }\textbf {\bibinfo {volume}
  {70}},\ \bibinfo {pages} {016002} (\bibinfo {year} {2004})},\ \Eprint
  {http://arxiv.org/abs/hep-ph/0403075} {arXiv:hep-ph/0403075} \BibitemShut
  {NoStop}%
\bibitem [{\citenamefont {Bali}\ \emph {et~al.}(1998)\citenamefont {Bali},
  \citenamefont {Schlichter},\ and\ \citenamefont {Schilling}}]{Bali:1997cp}%
  \BibitemOpen
  \bibfield  {author} {\bibinfo {author} {\bibfnamefont {G.~S.}\ \bibnamefont
  {Bali}}, \bibinfo {author} {\bibfnamefont {C.}~\bibnamefont {Schlichter}}, \
  and\ \bibinfo {author} {\bibfnamefont {K.}~\bibnamefont {Schilling}},\ }\href
  {\doibase 10.1143/PTPS.131.645} {\bibfield  {journal} {\bibinfo  {journal}
  {Prog. Theor. Phys. Suppl.}\ }\textbf {\bibinfo {volume} {131}},\ \bibinfo
  {pages} {645} (\bibinfo {year} {1998})},\ \Eprint
  {http://arxiv.org/abs/hep-lat/9802005} {arXiv:hep-lat/9802005} \BibitemShut
  {NoStop}%
\bibitem [{\citenamefont {Nayak}(2013)}]{Nayak:2012in}%
  \BibitemOpen
  \bibfield  {author} {\bibinfo {author} {\bibfnamefont {G.~C.}\ \bibnamefont
  {Nayak}},\ }\href {\doibase 10.1007/JHEP03(2013)001} {\bibfield  {journal}
  {\bibinfo  {journal} {JHEP}\ }\textbf {\bibinfo {volume} {03}},\ \bibinfo
  {pages} {001} (\bibinfo {year} {2013})},\ \Eprint
  {http://arxiv.org/abs/1201.2666} {arXiv:1201.2666 [hep-ph]} \BibitemShut
  {NoStop}%
\bibitem [{\citenamefont {Isgur}\ and\ \citenamefont
  {Paton}(1983)}]{Isgur:1983wj}%
  \BibitemOpen
  \bibfield  {author} {\bibinfo {author} {\bibfnamefont {N.}~\bibnamefont
  {Isgur}}\ and\ \bibinfo {author} {\bibfnamefont {J.~E.}\ \bibnamefont
  {Paton}},\ }\href {\doibase 10.1016/0370-2693(83)91445-4} {\bibfield
  {journal} {\bibinfo  {journal} {Phys. Lett. B}\ }\textbf {\bibinfo {volume}
  {124}},\ \bibinfo {pages} {247} (\bibinfo {year} {1983})}\BibitemShut
  {NoStop}%
\bibitem [{\citenamefont {Casher}\ \emph {et~al.}(1979)\citenamefont {Casher},
  \citenamefont {Neuberger},\ and\ \citenamefont {Nussinov}}]{Casher:1978wy}%
  \BibitemOpen
  \bibfield  {author} {\bibinfo {author} {\bibfnamefont {A.}~\bibnamefont
  {Casher}}, \bibinfo {author} {\bibfnamefont {H.}~\bibnamefont {Neuberger}}, \
  and\ \bibinfo {author} {\bibfnamefont {S.}~\bibnamefont {Nussinov}},\ }\href
  {\doibase 10.1103/PhysRevD.20.179} {\bibfield  {journal} {\bibinfo  {journal}
  {Phys. Rev. D}\ }\textbf {\bibinfo {volume} {20}},\ \bibinfo {pages} {179}
  (\bibinfo {year} {1979})}\BibitemShut {NoStop}%
\bibitem [{\citenamefont {Andersson}\ \emph {et~al.}(1983)\citenamefont
  {Andersson}, \citenamefont {Gustafson}, \citenamefont {Ingelman},\ and\
  \citenamefont {Sjostrand}}]{Andersson:1983ia}%
  \BibitemOpen
  \bibfield  {author} {\bibinfo {author} {\bibfnamefont {B.}~\bibnamefont
  {Andersson}}, \bibinfo {author} {\bibfnamefont {G.}~\bibnamefont
  {Gustafson}}, \bibinfo {author} {\bibfnamefont {G.}~\bibnamefont {Ingelman}},
  \ and\ \bibinfo {author} {\bibfnamefont {T.}~\bibnamefont {Sjostrand}},\
  }\href {\doibase 10.1016/0370-1573(83)90080-7} {\bibfield  {journal}
  {\bibinfo  {journal} {Phys. Rept.}\ }\textbf {\bibinfo {volume} {97}},\
  \bibinfo {pages} {31} (\bibinfo {year} {1983})}\BibitemShut {NoStop}%
\bibitem [{\citenamefont {Cea}\ \emph {et~al.}(2012)\citenamefont {Cea},
  \citenamefont {Cosmai},\ and\ \citenamefont {Papa}}]{Cea:2012qw}%
  \BibitemOpen
  \bibfield  {author} {\bibinfo {author} {\bibfnamefont {P.}~\bibnamefont
  {Cea}}, \bibinfo {author} {\bibfnamefont {L.}~\bibnamefont {Cosmai}}, \ and\
  \bibinfo {author} {\bibfnamefont {A.}~\bibnamefont {Papa}},\ }\href {\doibase
  10.1103/PhysRevD.86.054501} {\bibfield  {journal} {\bibinfo  {journal} {Phys.
  Rev. D}\ }\textbf {\bibinfo {volume} {86}},\ \bibinfo {pages} {054501}
  (\bibinfo {year} {2012})},\ \Eprint {http://arxiv.org/abs/1208.1362}
  {arXiv:1208.1362 [hep-lat]} \BibitemShut {NoStop}%
\bibitem [{\citenamefont {Buisseret}\ and\ \citenamefont
  {Semay}(2005)}]{Buisseret:2004wm}%
  \BibitemOpen
  \bibfield  {author} {\bibinfo {author} {\bibfnamefont {F.}~\bibnamefont
  {Buisseret}}\ and\ \bibinfo {author} {\bibfnamefont {C.}~\bibnamefont
  {Semay}},\ }\href {\doibase 10.1103/PhysRevD.71.034019} {\bibfield  {journal}
  {\bibinfo  {journal} {Phys. Rev. D}\ }\textbf {\bibinfo {volume} {71}},\
  \bibinfo {pages} {034019} (\bibinfo {year} {2005})},\ \Eprint
  {http://arxiv.org/abs/hep-ph/0412361} {arXiv:hep-ph/0412361} \BibitemShut
  {NoStop}%
\bibitem [{\citenamefont {Oliveira}\ \emph {et~al.}(2016)\citenamefont
  {Oliveira}, \citenamefont {K\i{}z\i{}lersu}, \citenamefont {Silva},
  \citenamefont {Skullerud}, \citenamefont {Sternbeck},\ and\ \citenamefont
  {Williams}}]{Oliveira:2016muq}%
  \BibitemOpen
  \bibfield  {author} {\bibinfo {author} {\bibfnamefont {O.}~\bibnamefont
  {Oliveira}}, \bibinfo {author} {\bibfnamefont {A.}~\bibnamefont
  {K\i{}z\i{}lersu}}, \bibinfo {author} {\bibfnamefont {P.~J.}\ \bibnamefont
  {Silva}}, \bibinfo {author} {\bibfnamefont {J.-I.}\ \bibnamefont
  {Skullerud}}, \bibinfo {author} {\bibfnamefont {A.}~\bibnamefont
  {Sternbeck}}, \ and\ \bibinfo {author} {\bibfnamefont {A.~G.}\ \bibnamefont
  {Williams}},\ }\href {\doibase 10.5506/APhysPolBSupp.9.363} {\bibfield
  {journal} {\bibinfo  {journal} {Acta Phys. Polon. Supp.}\ }\textbf {\bibinfo
  {volume} {9}},\ \bibinfo {pages} {363} (\bibinfo {year} {2016})},\ \Eprint
  {http://arxiv.org/abs/1605.09632} {arXiv:1605.09632 [hep-lat]} \BibitemShut
  {NoStop}%
\bibitem [{\citenamefont {Cucchieri}\ and\ \citenamefont
  {Mendes}(2008)}]{Cucchieri:2008fc}%
  \BibitemOpen
  \bibfield  {author} {\bibinfo {author} {\bibfnamefont {A.}~\bibnamefont
  {Cucchieri}}\ and\ \bibinfo {author} {\bibfnamefont {T.}~\bibnamefont
  {Mendes}},\ }\href {\doibase 10.1103/PhysRevD.78.094503} {\bibfield
  {journal} {\bibinfo  {journal} {Phys. Rev. D}\ }\textbf {\bibinfo {volume}
  {78}},\ \bibinfo {pages} {094503} (\bibinfo {year} {2008})},\ \Eprint
  {http://arxiv.org/abs/0804.2371} {arXiv:0804.2371 [hep-lat]} \BibitemShut
  {NoStop}%
\bibitem [{\citenamefont {Aguilar}\ and\ \citenamefont
  {Papavassiliou}(2008)}]{Aguilar:2007nf}%
  \BibitemOpen
  \bibfield  {author} {\bibinfo {author} {\bibfnamefont {A.~C.}\ \bibnamefont
  {Aguilar}}\ and\ \bibinfo {author} {\bibfnamefont {J.}~\bibnamefont
  {Papavassiliou}},\ }\href {\doibase 10.1103/PhysRevD.77.125022} {\bibfield
  {journal} {\bibinfo  {journal} {Phys. Rev. D}\ }\textbf {\bibinfo {volume}
  {77}},\ \bibinfo {pages} {125022} (\bibinfo {year} {2008})},\ \Eprint
  {http://arxiv.org/abs/0712.0780} {arXiv:0712.0780 [hep-ph]} \BibitemShut
  {NoStop}%
\bibitem [{\citenamefont {Aguilar}\ \emph {et~al.}(2013)\citenamefont
  {Aguilar}, \citenamefont {Ib\'a\~nez},\ and\ \citenamefont
  {Papavassiliou}}]{Aguilar:2013xqa}%
  \BibitemOpen
  \bibfield  {author} {\bibinfo {author} {\bibfnamefont {A.~C.}\ \bibnamefont
  {Aguilar}}, \bibinfo {author} {\bibfnamefont {D.}~\bibnamefont {Ib\'a\~nez}},
  \ and\ \bibinfo {author} {\bibfnamefont {J.}~\bibnamefont {Papavassiliou}},\
  }\href {\doibase 10.1103/PhysRevD.87.114020} {\bibfield  {journal} {\bibinfo
  {journal} {Phys. Rev. D}\ }\textbf {\bibinfo {volume} {87}},\ \bibinfo
  {pages} {114020} (\bibinfo {year} {2013})},\ \Eprint
  {http://arxiv.org/abs/1303.3609} {arXiv:1303.3609 [hep-ph]} \BibitemShut
  {NoStop}%
\bibitem [{\citenamefont {Eichmann}\ \emph {et~al.}(2014)\citenamefont
  {Eichmann}, \citenamefont {Williams}, \citenamefont {Alkofer},\ and\
  \citenamefont {Vujinovic}}]{Eichmann:2014xya}%
  \BibitemOpen
  \bibfield  {author} {\bibinfo {author} {\bibfnamefont {G.}~\bibnamefont
  {Eichmann}}, \bibinfo {author} {\bibfnamefont {R.}~\bibnamefont {Williams}},
  \bibinfo {author} {\bibfnamefont {R.}~\bibnamefont {Alkofer}}, \ and\
  \bibinfo {author} {\bibfnamefont {M.}~\bibnamefont {Vujinovic}},\ }\href
  {\doibase 10.1103/PhysRevD.89.105014} {\bibfield  {journal} {\bibinfo
  {journal} {Phys. Rev. D}\ }\textbf {\bibinfo {volume} {89}},\ \bibinfo
  {pages} {105014} (\bibinfo {year} {2014})},\ \Eprint
  {http://arxiv.org/abs/1402.1365} {arXiv:1402.1365 [hep-ph]} \BibitemShut
  {NoStop}%
\bibitem [{\citenamefont {Mintz}\ \emph {et~al.}(2018)\citenamefont {Mintz},
  \citenamefont {Palhares}, \citenamefont {Sorella},\ and\ \citenamefont
  {Pereira}}]{Mintz:2017qri}%
  \BibitemOpen
  \bibfield  {author} {\bibinfo {author} {\bibfnamefont {B.~W.}\ \bibnamefont
  {Mintz}}, \bibinfo {author} {\bibfnamefont {L.~F.}\ \bibnamefont {Palhares}},
  \bibinfo {author} {\bibfnamefont {S.~P.}\ \bibnamefont {Sorella}}, \ and\
  \bibinfo {author} {\bibfnamefont {A.~D.}\ \bibnamefont {Pereira}},\ }\href
  {\doibase 10.1103/PhysRevD.97.034020} {\bibfield  {journal} {\bibinfo
  {journal} {Phys. Rev. D}\ }\textbf {\bibinfo {volume} {97}},\ \bibinfo
  {pages} {034020} (\bibinfo {year} {2018})},\ \Eprint
  {http://arxiv.org/abs/1712.09633} {arXiv:1712.09633 [hep-th]} \BibitemShut
  {NoStop}%
\bibitem [{\citenamefont {Dudal}\ \emph {et~al.}(2012)\citenamefont {Dudal},
  \citenamefont {Oliveira},\ and\ \citenamefont
  {Rodriguez-Quintero}}]{Dudal:2012zx}%
  \BibitemOpen
  \bibfield  {author} {\bibinfo {author} {\bibfnamefont {D.}~\bibnamefont
  {Dudal}}, \bibinfo {author} {\bibfnamefont {O.}~\bibnamefont {Oliveira}}, \
  and\ \bibinfo {author} {\bibfnamefont {J.}~\bibnamefont
  {Rodriguez-Quintero}},\ }\href {\doibase 10.1103/PhysRevD.86.105005}
  {\bibfield  {journal} {\bibinfo  {journal} {Phys. Rev. D}\ }\textbf {\bibinfo
  {volume} {86}},\ \bibinfo {pages} {105005} (\bibinfo {year} {2012})},\
  \Eprint {http://arxiv.org/abs/1207.5118} {arXiv:1207.5118 [hep-ph]}
  \BibitemShut {NoStop}%
\bibitem [{\citenamefont {Cyrol}\ \emph {et~al.}(2015)\citenamefont {Cyrol},
  \citenamefont {Huber},\ and\ \citenamefont {von Smekal}}]{Cyrol:2014kca}%
  \BibitemOpen
  \bibfield  {author} {\bibinfo {author} {\bibfnamefont {A.~K.}\ \bibnamefont
  {Cyrol}}, \bibinfo {author} {\bibfnamefont {M.~Q.}\ \bibnamefont {Huber}}, \
  and\ \bibinfo {author} {\bibfnamefont {L.}~\bibnamefont {von Smekal}},\
  }\href {\doibase 10.1140/epjc/s10052-015-3312-1} {\bibfield  {journal}
  {\bibinfo  {journal} {Eur. Phys. J. C}\ }\textbf {\bibinfo {volume} {75}},\
  \bibinfo {pages} {102} (\bibinfo {year} {2015})},\ \Eprint
  {http://arxiv.org/abs/1408.5409} {arXiv:1408.5409 [hep-ph]} \BibitemShut
  {NoStop}%
\bibitem [{\citenamefont {Huber}(2020)}]{Huber:2018ned}%
  \BibitemOpen
  \bibfield  {author} {\bibinfo {author} {\bibfnamefont {M.~Q.}\ \bibnamefont
  {Huber}},\ }\href {\doibase 10.1016/j.physrep.2020.04.004} {\bibfield
  {journal} {\bibinfo  {journal} {Phys. Rept.}\ }\textbf {\bibinfo {volume}
  {879}},\ \bibinfo {pages} {1} (\bibinfo {year} {2020})},\ \Eprint
  {http://arxiv.org/abs/1808.05227} {arXiv:1808.05227 [hep-ph]} \BibitemShut
  {NoStop}%
\bibitem [{\citenamefont {Pinto-G\'omez}\ \emph {et~al.}(2023)\citenamefont
  {Pinto-G\'omez}, \citenamefont {De~Soto}, \citenamefont {Ferreira},
  \citenamefont {Papavassiliou},\ and\ \citenamefont
  {Rodr\'\i{}guez-Quintero}}]{Pinto-Gomez:2022brg}%
  \BibitemOpen
  \bibfield  {author} {\bibinfo {author} {\bibfnamefont {F.}~\bibnamefont
  {Pinto-G\'omez}}, \bibinfo {author} {\bibfnamefont {F.}~\bibnamefont
  {De~Soto}}, \bibinfo {author} {\bibfnamefont {M.~N.}\ \bibnamefont
  {Ferreira}}, \bibinfo {author} {\bibfnamefont {J.}~\bibnamefont
  {Papavassiliou}}, \ and\ \bibinfo {author} {\bibfnamefont {J.}~\bibnamefont
  {Rodr\'\i{}guez-Quintero}},\ }\href {\doibase 10.1016/j.physletb.2023.137737}
  {\bibfield  {journal} {\bibinfo  {journal} {Phys. Lett. B}\ }\textbf
  {\bibinfo {volume} {838}},\ \bibinfo {pages} {137737} (\bibinfo {year}
  {2023})},\ \Eprint {http://arxiv.org/abs/2208.01020} {arXiv:2208.01020
  [hep-ph]} \BibitemShut {NoStop}%
\bibitem [{\citenamefont {Maris}\ \emph {et~al.}(2003)\citenamefont {Maris},
  \citenamefont {Raya}, \citenamefont {Roberts},\ and\ \citenamefont
  {Schmidt}}]{Maris:2002mt}%
  \BibitemOpen
  \bibfield  {author} {\bibinfo {author} {\bibfnamefont {P.}~\bibnamefont
  {Maris}}, \bibinfo {author} {\bibfnamefont {A.}~\bibnamefont {Raya}},
  \bibinfo {author} {\bibfnamefont {C.~D.}\ \bibnamefont {Roberts}}, \ and\
  \bibinfo {author} {\bibfnamefont {S.~M.}\ \bibnamefont {Schmidt}},\ }\href
  {\doibase 10.1140/epja/i2002-10206-6} {\bibfield  {journal} {\bibinfo
  {journal} {Eur. Phys. J. A}\ }\textbf {\bibinfo {volume} {18}},\ \bibinfo
  {pages} {231} (\bibinfo {year} {2003})},\ \Eprint
  {http://arxiv.org/abs/nucl-th/0208071} {arXiv:nucl-th/0208071} \BibitemShut
  {NoStop}%
\bibitem [{\citenamefont {Alkofer}\ \emph {et~al.}(2004)\citenamefont
  {Alkofer}, \citenamefont {Detmold}, \citenamefont {Fischer},\ and\
  \citenamefont {Maris}}]{Alkofer:2003jj}%
  \BibitemOpen
  \bibfield  {author} {\bibinfo {author} {\bibfnamefont {R.}~\bibnamefont
  {Alkofer}}, \bibinfo {author} {\bibfnamefont {W.}~\bibnamefont {Detmold}},
  \bibinfo {author} {\bibfnamefont {C.~S.}\ \bibnamefont {Fischer}}, \ and\
  \bibinfo {author} {\bibfnamefont {P.}~\bibnamefont {Maris}},\ }\href
  {\doibase 10.1103/PhysRevD.70.014014} {\bibfield  {journal} {\bibinfo
  {journal} {Phys. Rev. D}\ }\textbf {\bibinfo {volume} {70}},\ \bibinfo
  {pages} {014014} (\bibinfo {year} {2004})},\ \Eprint
  {http://arxiv.org/abs/hep-ph/0309077} {arXiv:hep-ph/0309077} \BibitemShut
  {NoStop}%
\bibitem [{\citenamefont {Leinweber}\ \emph {et~al.}(2022)\citenamefont
  {Leinweber}, \citenamefont {Biddle}, \citenamefont {Kamleh},\ and\
  \citenamefont {Virgili}}]{Leinweber:2022ukj}%
  \BibitemOpen
  \bibfield  {author} {\bibinfo {author} {\bibfnamefont {D.}~\bibnamefont
  {Leinweber}}, \bibinfo {author} {\bibfnamefont {J.}~\bibnamefont {Biddle}},
  \bibinfo {author} {\bibfnamefont {W.}~\bibnamefont {Kamleh}}, \ and\ \bibinfo
  {author} {\bibfnamefont {A.}~\bibnamefont {Virgili}},\ }\href {\doibase
  10.1051/epjconf/202227401002} {\bibfield  {journal} {\bibinfo  {journal} {EPJ
  Web Conf.}\ }\textbf {\bibinfo {volume} {274}},\ \bibinfo {pages} {01002}
  (\bibinfo {year} {2022})},\ \Eprint {http://arxiv.org/abs/2211.13421}
  {arXiv:2211.13421 [hep-lat]} \BibitemShut {NoStop}%
\bibitem [{\citenamefont {K\i{}z\i{}lers\"u}\ \emph {et~al.}(2021)\citenamefont
  {K\i{}z\i{}lers\"u}, \citenamefont {Oliveira}, \citenamefont {Silva},
  \citenamefont {Skullerud},\ and\ \citenamefont
  {Sternbeck}}]{Kizilersu:2021jen}%
  \BibitemOpen
  \bibfield  {author} {\bibinfo {author} {\bibfnamefont {A.}~\bibnamefont
  {K\i{}z\i{}lers\"u}}, \bibinfo {author} {\bibfnamefont {O.}~\bibnamefont
  {Oliveira}}, \bibinfo {author} {\bibfnamefont {P.~J.}\ \bibnamefont {Silva}},
  \bibinfo {author} {\bibfnamefont {J.-I.}\ \bibnamefont {Skullerud}}, \ and\
  \bibinfo {author} {\bibfnamefont {A.}~\bibnamefont {Sternbeck}},\ }\href
  {\doibase 10.1103/PhysRevD.103.114515} {\bibfield  {journal} {\bibinfo
  {journal} {Phys. Rev. D}\ }\textbf {\bibinfo {volume} {103}},\ \bibinfo
  {pages} {114515} (\bibinfo {year} {2021})},\ \Eprint
  {http://arxiv.org/abs/2103.02945} {arXiv:2103.02945 [hep-lat]} \BibitemShut
  {NoStop}%
\bibitem [{\citenamefont {Kizilersu}\ \emph {et~al.}(2007)\citenamefont
  {Kizilersu}, \citenamefont {Leinweber}, \citenamefont {Skullerud},\ and\
  \citenamefont {Williams}}]{Kizilersu:2006et}%
  \BibitemOpen
  \bibfield  {author} {\bibinfo {author} {\bibfnamefont {A.}~\bibnamefont
  {Kizilersu}}, \bibinfo {author} {\bibfnamefont {D.~B.}\ \bibnamefont
  {Leinweber}}, \bibinfo {author} {\bibfnamefont {J.-I.}\ \bibnamefont
  {Skullerud}}, \ and\ \bibinfo {author} {\bibfnamefont {A.~G.}\ \bibnamefont
  {Williams}},\ }\href {\doibase 10.1140/epjc/s10052-007-0250-6} {\bibfield
  {journal} {\bibinfo  {journal} {Eur. Phys. J. C}\ }\textbf {\bibinfo {volume}
  {50}},\ \bibinfo {pages} {871} (\bibinfo {year} {2007})},\ \Eprint
  {http://arxiv.org/abs/hep-lat/0610078} {arXiv:hep-lat/0610078} \BibitemShut
  {NoStop}%
\bibitem [{\citenamefont {Alkofer}\ \emph {et~al.}(2009)\citenamefont
  {Alkofer}, \citenamefont {Fischer}, \citenamefont {Llanes-Estrada},\ and\
  \citenamefont {Schwenzer}}]{Alkofer:2008tt}%
  \BibitemOpen
  \bibfield  {author} {\bibinfo {author} {\bibfnamefont {R.}~\bibnamefont
  {Alkofer}}, \bibinfo {author} {\bibfnamefont {C.~S.}\ \bibnamefont
  {Fischer}}, \bibinfo {author} {\bibfnamefont {F.~J.}\ \bibnamefont
  {Llanes-Estrada}}, \ and\ \bibinfo {author} {\bibfnamefont {K.}~\bibnamefont
  {Schwenzer}},\ }\href {\doibase 10.1016/j.aop.2008.07.001} {\bibfield
  {journal} {\bibinfo  {journal} {Annals Phys.}\ }\textbf {\bibinfo {volume}
  {324}},\ \bibinfo {pages} {106} (\bibinfo {year} {2009})},\ \Eprint
  {http://arxiv.org/abs/0804.3042} {arXiv:0804.3042 [hep-ph]} \BibitemShut
  {NoStop}%
\bibitem [{\citenamefont {Aguilar}\ \emph {et~al.}(2023)\citenamefont
  {Aguilar}, \citenamefont {Ferreira}, \citenamefont {Iba\~nez},\ and\
  \citenamefont {Papavassiliou}}]{Aguilar:2023mam}%
  \BibitemOpen
  \bibfield  {author} {\bibinfo {author} {\bibfnamefont {A.~C.}\ \bibnamefont
  {Aguilar}}, \bibinfo {author} {\bibfnamefont {M.~N.}\ \bibnamefont
  {Ferreira}}, \bibinfo {author} {\bibfnamefont {D.}~\bibnamefont {Iba\~nez}},
  \ and\ \bibinfo {author} {\bibfnamefont {J.}~\bibnamefont {Papavassiliou}},\
  }\href {\doibase 10.1140/epjc/s10052-023-12103-8} {\bibfield  {journal}
  {\bibinfo  {journal} {Eur. Phys. J. C}\ }\textbf {\bibinfo {volume} {83}},\
  \bibinfo {pages} {967} (\bibinfo {year} {2023})},\ \Eprint
  {http://arxiv.org/abs/2308.16297} {arXiv:2308.16297 [hep-ph]} \BibitemShut
  {NoStop}%
\bibitem [{\citenamefont {Windisch}\ \emph {et~al.}(2013)\citenamefont
  {Windisch}, \citenamefont {Hopfer},\ and\ \citenamefont
  {Alkofer}}]{Windisch:2012de}%
  \BibitemOpen
  \bibfield  {author} {\bibinfo {author} {\bibfnamefont {A.}~\bibnamefont
  {Windisch}}, \bibinfo {author} {\bibfnamefont {M.}~\bibnamefont {Hopfer}}, \
  and\ \bibinfo {author} {\bibfnamefont {R.}~\bibnamefont {Alkofer}},\ }\href
  {\doibase 10.5506/APhysPolBSupp.6.347} {\bibfield  {journal} {\bibinfo
  {journal} {Acta Phys. Polon. Supp.}\ }\textbf {\bibinfo {volume} {6}},\
  \bibinfo {pages} {347} (\bibinfo {year} {2013})},\ \Eprint
  {http://arxiv.org/abs/1210.8428} {arXiv:1210.8428 [hep-ph]} \BibitemShut
  {NoStop}%
\bibitem [{\citenamefont {Lessa}\ \emph {et~al.}(2023)\citenamefont {Lessa},
  \citenamefont {Serna}, \citenamefont {El-Bennich}, \citenamefont {Bashir},\
  and\ \citenamefont {Oliveira}}]{Lessa:2022wqc}%
  \BibitemOpen
  \bibfield  {author} {\bibinfo {author} {\bibfnamefont {J.~R.}\ \bibnamefont
  {Lessa}}, \bibinfo {author} {\bibfnamefont {F.~E.}\ \bibnamefont {Serna}},
  \bibinfo {author} {\bibfnamefont {B.}~\bibnamefont {El-Bennich}}, \bibinfo
  {author} {\bibfnamefont {A.}~\bibnamefont {Bashir}}, \ and\ \bibinfo {author}
  {\bibfnamefont {O.}~\bibnamefont {Oliveira}},\ }\href {\doibase
  10.1103/PhysRevD.107.074017} {\bibfield  {journal} {\bibinfo  {journal}
  {Phys. Rev. D}\ }\textbf {\bibinfo {volume} {107}},\ \bibinfo {pages}
  {074017} (\bibinfo {year} {2023})},\ \Eprint
  {http://arxiv.org/abs/2202.12313} {arXiv:2202.12313 [hep-ph]} \BibitemShut
  {NoStop}%
\bibitem [{\citenamefont {Llanes-Estrada}\ \emph {et~al.}(2004)\citenamefont
  {Llanes-Estrada}, \citenamefont {Cotanch}, \citenamefont {Szczepaniak},\ and\
  \citenamefont {Swanson}}]{Llanes-Estrada:2004edu}%
  \BibitemOpen
  \bibfield  {author} {\bibinfo {author} {\bibfnamefont {F.~J.}\ \bibnamefont
  {Llanes-Estrada}}, \bibinfo {author} {\bibfnamefont {S.~R.}\ \bibnamefont
  {Cotanch}}, \bibinfo {author} {\bibfnamefont {A.~P.}\ \bibnamefont
  {Szczepaniak}}, \ and\ \bibinfo {author} {\bibfnamefont {E.~S.}\ \bibnamefont
  {Swanson}},\ }\href {\doibase 10.1103/PhysRevC.70.035202} {\bibfield
  {journal} {\bibinfo  {journal} {Phys. Rev. C}\ }\textbf {\bibinfo {volume}
  {70}},\ \bibinfo {pages} {035202} (\bibinfo {year} {2004})},\ \Eprint
  {http://arxiv.org/abs/hep-ph/0402253} {arXiv:hep-ph/0402253} \BibitemShut
  {NoStop}%
\bibitem [{\citenamefont {Hopfer}\ \emph {et~al.}(2012)\citenamefont {Hopfer},
  \citenamefont {Windisch},\ and\ \citenamefont {Alkofer}}]{Hopfer:2012cnq}%
  \BibitemOpen
  \bibfield  {author} {\bibinfo {author} {\bibfnamefont {M.}~\bibnamefont
  {Hopfer}}, \bibinfo {author} {\bibfnamefont {A.}~\bibnamefont {Windisch}}, \
  and\ \bibinfo {author} {\bibfnamefont {R.}~\bibnamefont {Alkofer}},\ }\href
  {\doibase 10.22323/1.171.0073} {\bibfield  {journal} {\bibinfo  {journal}
  {PoS}\ }\textbf {\bibinfo {volume} {ConfinementX}},\ \bibinfo {pages} {073}
  (\bibinfo {year} {2012})},\ \Eprint {http://arxiv.org/abs/1301.3672}
  {arXiv:1301.3672 [hep-ph]} \BibitemShut {NoStop}%
\bibitem [{\citenamefont {Windisch}(2014)}]{Windisch:2014lce}%
  \BibitemOpen
  \bibfield  {author} {\bibinfo {author} {\bibfnamefont {A.}~\bibnamefont
  {Windisch}},\ }\emph {\bibinfo {title} {{Features of strong quark
  correlations at vanishing and non-vanishing density}}},\ \href@noop {} {Ph.D.
  thesis},\ \bibinfo  {school} {Graz U.} (\bibinfo {year} {2014})\BibitemShut
  {NoStop}%
\bibitem [{\citenamefont {Alkofer}(2023)}]{Alkofer:2023lrl}%
  \BibitemOpen
  \bibfield  {author} {\bibinfo {author} {\bibfnamefont {R.}~\bibnamefont
  {Alkofer}},\ }\href {\doibase 10.3390/sym15091787} {\bibfield  {journal}
  {\bibinfo  {journal} {Symmetry}\ }\textbf {\bibinfo {volume} {15}},\ \bibinfo
  {pages} {1787} (\bibinfo {year} {2023})},\ \Eprint
  {http://arxiv.org/abs/2309.09679} {arXiv:2309.09679 [hep-ph]} \BibitemShut
  {NoStop}%
\bibitem [{\citenamefont {Albino}\ \emph {et~al.}(2019)\citenamefont {Albino},
  \citenamefont {Bashir}, \citenamefont {Guerrero}, \citenamefont {Bennich},\
  and\ \citenamefont {Rojas}}]{Albino:2018ncl}%
  \BibitemOpen
  \bibfield  {author} {\bibinfo {author} {\bibfnamefont {L.}~\bibnamefont
  {Albino}}, \bibinfo {author} {\bibfnamefont {A.}~\bibnamefont {Bashir}},
  \bibinfo {author} {\bibfnamefont {L.~X.~G.}\ \bibnamefont {Guerrero}},
  \bibinfo {author} {\bibfnamefont {B.~E.}\ \bibnamefont {Bennich}}, \ and\
  \bibinfo {author} {\bibfnamefont {E.}~\bibnamefont {Rojas}},\ }\href
  {\doibase 10.1103/PhysRevD.100.054028} {\bibfield  {journal} {\bibinfo
  {journal} {Phys. Rev. D}\ }\textbf {\bibinfo {volume} {100}},\ \bibinfo
  {pages} {054028} (\bibinfo {year} {2019})},\ \Eprint
  {http://arxiv.org/abs/1812.02280} {arXiv:1812.02280 [nucl-th]} \BibitemShut
  {NoStop}%
\bibitem [{\citenamefont {Gimeno-Segovia}\ and\ \citenamefont
  {Llanes-Estrada}(2008)}]{Gimeno-Segovia:2008fqu}%
  \BibitemOpen
  \bibfield  {author} {\bibinfo {author} {\bibfnamefont {M.}~\bibnamefont
  {Gimeno-Segovia}}\ and\ \bibinfo {author} {\bibfnamefont {F.~J.}\
  \bibnamefont {Llanes-Estrada}},\ }\href {\doibase
  10.1140/epjc/s10052-008-0676-5} {\bibfield  {journal} {\bibinfo  {journal}
  {Eur. Phys. J. C}\ }\textbf {\bibinfo {volume} {56}},\ \bibinfo {pages} {557}
  (\bibinfo {year} {2008})},\ \Eprint {http://arxiv.org/abs/0805.4145}
  {arXiv:0805.4145 [hep-th]} \BibitemShut {NoStop}%
\bibitem [{\citenamefont {Escudero-Pedrosa}\ \emph {et~al.}(2021)\citenamefont
  {Escudero-Pedrosa}, \citenamefont {Llanes-Estrada}, \citenamefont {Oller},\
  and\ \citenamefont {Salas-Bern\'ardez}}]{Escudero-Pedrosa:2020rwb}%
  \BibitemOpen
  \bibfield  {author} {\bibinfo {author} {\bibfnamefont {J.}~\bibnamefont
  {Escudero-Pedrosa}}, \bibinfo {author} {\bibfnamefont {F.~J.}\ \bibnamefont
  {Llanes-Estrada}}, \bibinfo {author} {\bibfnamefont {J.~A.}\ \bibnamefont
  {Oller}}, \ and\ \bibinfo {author} {\bibfnamefont {A.}~\bibnamefont
  {Salas-Bern\'ardez}},\ }\href {\doibase 10.1016/j.nuclphysbps.2021.05.022}
  {\bibfield  {journal} {\bibinfo  {journal} {Nucl. Part. Phys. Proc.}\
  }\textbf {\bibinfo {volume} {312-317}},\ \bibinfo {pages} {82} (\bibinfo
  {year} {2021})},\ \Eprint {http://arxiv.org/abs/2012.02616} {arXiv:2012.02616
  [hep-ph]} \BibitemShut {NoStop}%
\bibitem [{\citenamefont {Horak}\ \emph {et~al.}(2022)\citenamefont {Horak},
  \citenamefont {Pawlowski}, \citenamefont {Rodr\'\i{}guez-Quintero},
  \citenamefont {Turnwald}, \citenamefont {Urban}, \citenamefont {Wink},\ and\
  \citenamefont {Zafeiropoulos}}]{Horak:2021syv}%
  \BibitemOpen
  \bibfield  {author} {\bibinfo {author} {\bibfnamefont {J.}~\bibnamefont
  {Horak}}, \bibinfo {author} {\bibfnamefont {J.~M.}\ \bibnamefont
  {Pawlowski}}, \bibinfo {author} {\bibfnamefont {J.}~\bibnamefont
  {Rodr\'\i{}guez-Quintero}}, \bibinfo {author} {\bibfnamefont
  {J.}~\bibnamefont {Turnwald}}, \bibinfo {author} {\bibfnamefont {J.~M.}\
  \bibnamefont {Urban}}, \bibinfo {author} {\bibfnamefont {N.}~\bibnamefont
  {Wink}}, \ and\ \bibinfo {author} {\bibfnamefont {S.}~\bibnamefont
  {Zafeiropoulos}},\ }\href {\doibase 10.1103/PhysRevD.105.036014} {\bibfield
  {journal} {\bibinfo  {journal} {Phys. Rev. D}\ }\textbf {\bibinfo {volume}
  {105}},\ \bibinfo {pages} {036014} (\bibinfo {year} {2022})},\ \Eprint
  {http://arxiv.org/abs/2107.13464} {arXiv:2107.13464 [hep-ph]} \BibitemShut
  {NoStop}%
\bibitem [{\citenamefont {Horak}\ \emph {et~al.}(2023)\citenamefont {Horak},
  \citenamefont {Pawlowski}, \citenamefont {Turnwald}, \citenamefont {Urban},
  \citenamefont {Wink},\ and\ \citenamefont {Zafeiropoulos}}]{Horak:2023xfb}%
  \BibitemOpen
  \bibfield  {author} {\bibinfo {author} {\bibfnamefont {J.}~\bibnamefont
  {Horak}}, \bibinfo {author} {\bibfnamefont {J.~M.}\ \bibnamefont
  {Pawlowski}}, \bibinfo {author} {\bibfnamefont {J.}~\bibnamefont {Turnwald}},
  \bibinfo {author} {\bibfnamefont {J.~M.}\ \bibnamefont {Urban}}, \bibinfo
  {author} {\bibfnamefont {N.}~\bibnamefont {Wink}}, \ and\ \bibinfo {author}
  {\bibfnamefont {S.}~\bibnamefont {Zafeiropoulos}},\ }\href {\doibase
  10.1103/PhysRevD.107.076019} {\bibfield  {journal} {\bibinfo  {journal}
  {Phys. Rev. D}\ }\textbf {\bibinfo {volume} {107}},\ \bibinfo {pages}
  {076019} (\bibinfo {year} {2023})},\ \Eprint
  {http://arxiv.org/abs/2301.07785} {arXiv:2301.07785 [hep-ph]} \BibitemShut
  {NoStop}%
\bibitem [{\citenamefont {Glendenning}\ and\ \citenamefont
  {Matsui}(1983)}]{Glendenning:1983qq}%
  \BibitemOpen
  \bibfield  {author} {\bibinfo {author} {\bibfnamefont {N.~K.}\ \bibnamefont
  {Glendenning}}\ and\ \bibinfo {author} {\bibfnamefont {T.}~\bibnamefont
  {Matsui}},\ }\href {\doibase 10.1103/PhysRevD.28.2890} {\bibfield  {journal}
  {\bibinfo  {journal} {Phys. Rev. D}\ }\textbf {\bibinfo {volume} {28}},\
  \bibinfo {pages} {2890} (\bibinfo {year} {1983})}\BibitemShut {NoStop}%
\bibitem [{\citenamefont {Hebenstreit}\ \emph {et~al.}(2013)\citenamefont
  {Hebenstreit}, \citenamefont {Berges},\ and\ \citenamefont
  {Gelfand}}]{Hebenstreit:2013baa}%
  \BibitemOpen
  \bibfield  {author} {\bibinfo {author} {\bibfnamefont {F.}~\bibnamefont
  {Hebenstreit}}, \bibinfo {author} {\bibfnamefont {J.}~\bibnamefont {Berges}},
  \ and\ \bibinfo {author} {\bibfnamefont {D.}~\bibnamefont {Gelfand}},\ }\href
  {\doibase 10.1103/PhysRevLett.111.201601} {\bibfield  {journal} {\bibinfo
  {journal} {Phys. Rev. Lett.}\ }\textbf {\bibinfo {volume} {111}},\ \bibinfo
  {pages} {201601} (\bibinfo {year} {2013})},\ \Eprint
  {http://arxiv.org/abs/1307.4619} {arXiv:1307.4619 [hep-ph]} \BibitemShut
  {NoStop}%
\bibitem [{\citenamefont {Kasper}\ \emph {et~al.}(2014)\citenamefont {Kasper},
  \citenamefont {Hebenstreit},\ and\ \citenamefont {Berges}}]{Kasper:2014uaa}%
  \BibitemOpen
  \bibfield  {author} {\bibinfo {author} {\bibfnamefont {V.}~\bibnamefont
  {Kasper}}, \bibinfo {author} {\bibfnamefont {F.}~\bibnamefont {Hebenstreit}},
  \ and\ \bibinfo {author} {\bibfnamefont {J.}~\bibnamefont {Berges}},\ }\href
  {\doibase 10.1103/PhysRevD.90.025016} {\bibfield  {journal} {\bibinfo
  {journal} {Phys. Rev. D}\ }\textbf {\bibinfo {volume} {90}},\ \bibinfo
  {pages} {025016} (\bibinfo {year} {2014})},\ \Eprint
  {http://arxiv.org/abs/1403.4849} {arXiv:1403.4849 [hep-ph]} \BibitemShut
  {NoStop}%
\bibitem [{\citenamefont {Llanes-Estrada}\ \emph {et~al.}(2006)\citenamefont
  {Llanes-Estrada}, \citenamefont {Fischer},\ and\ \citenamefont
  {Alkofer}}]{Llanes-Estrada:2004hnb}%
  \BibitemOpen
  \bibfield  {author} {\bibinfo {author} {\bibfnamefont {F.~J.}\ \bibnamefont
  {Llanes-Estrada}}, \bibinfo {author} {\bibfnamefont {C.~S.}\ \bibnamefont
  {Fischer}}, \ and\ \bibinfo {author} {\bibfnamefont {R.}~\bibnamefont
  {Alkofer}},\ }\href {\doibase 10.1016/j.nuclphysbps.2005.08.008} {\bibfield
  {journal} {\bibinfo  {journal} {Nucl. Phys. B Proc. Suppl.}\ }\textbf
  {\bibinfo {volume} {152}},\ \bibinfo {pages} {43} (\bibinfo {year} {2006})},\
  \Eprint {http://arxiv.org/abs/hep-ph/0407332} {arXiv:hep-ph/0407332}
  \BibitemShut {NoStop}%
\end{thebibliography}%
\end{document}